\documentclass[fleqn,usenatbib]{mnras}

\usepackage{newtxtext,newtxmath}

\usepackage[T1]{fontenc}

\DeclareRobustCommand{\VAN}[3]{#2}
\let\VANthebibliography\thebibliography
\def\thebibliography{\DeclareRobustCommand{\VAN}[3]{##3}\VANthebibliography}

\usepackage{graphicx}	
\usepackage{amsmath}	
\usepackage{orcidlink}  

\usepackage{here} 

\defcitealias{Getman_2022}{G22}

\title[Evolution of rotation--activity relation]{Modelling the emergence and evolution of the rotation--activity relation}

\author[K. A. Stuart et al.]{
Kieran A. Stuart,$^{1}$\thanks{E-mail: 2476691@dundee.ac.uk (KS); sgregory001@dundee.ac.uk (SG)}\orcidlink{https://orcid.org/0009-0005-8084-0759}
Scott. G. Gregory,$^{1}$\footnotemark[1]\orcidlink{https://orcid.org/0000-0003-3674-5568}
\\
$^{1}$School of Science and Engineering, University of Dundee, Nethergate, Dundee, DD1 4HN\\
}

\date{Accepted by MNRAS 8$^{\rm{th}}$ April 2025 }

\pubyear{2024}

\begin{document}
\label{firstpage}
\pagerange{\pageref{firstpage}--\pageref{lastpage}}
\maketitle

\begin{abstract}
Main-sequence stars follow a well-defined rotation--activity relation. There are two primary regimes: saturated, where the fractional X-ray luminosity $\log(L_{\rm X}/L_*)$ is approximately constant, and unsaturated, where the fractional X-ray luminosity decreases with increasing Rossby number (or decreasing rotation rate). Pre-main sequence (PMS) stars have a larger scatter in $\log(L_{\rm X}/L_*)$ than main-sequence stars, are observed to have saturated levels of X-ray emission, and do not follow the rotation--activity relation. We investigate how PMS stars evolve in the rotation--activity plane and the timescale over which the X-ray rotation--activity relation emerges. Using observational data of $\sim$600 stars from four PMS clusters, stellar internal structure models, a rotational evolution model, and observed X-ray luminosity trends with age, we simulate the evolution of the PMS stars in the rotation--activity plane up to ages of 100\,Myr. Our model reproduces the rotation--activity relation found for main-sequence stars, with higher-mass stars beginning to form the unsaturated regime from around 10 Myr. After $\sim$25\,Myr, the gradient of the unsaturated regime matches that found for main-sequence stars. For stars of mass greater than 0.6\,M$_{\sun}$, the maximum age by which a star has left the saturated regime correlates with when the star leaves the PMS. We find that an intra-cluster age spread is a key factor in contributing to the observed scatter in $\log(L_{\rm X}/L_*)$, particularly for ages < 10\,Myr. 
\end{abstract}




\section{Introduction}
Coronal X-ray emission from low-mass stars arises from plasma confined along magnetic loops 
\citep{Vaiana_1981}. Magnetic loops erupt into the stellar atmosphere having been generated by the interior dynamo of a star. For main-sequence, solar-like stars, magnetic loops are generated in the shear layer at the interface of the inner radiative core and the outer convective zone due to differential rotation \citep{Parker_1993}. For fully convective stars, a turbulent magnetic dynamo can generate strong, often dipole-dominant, large-scale magnetic fields \citep{Durney_1993, Morin_2010, Yadav_2015}.

The relation between coronal activity and stellar rotation is well-established for main-sequence stars. X-ray luminosity $L_{\rm{X}}$ decays with decreasing rotation rate \citep{Pallavicini_1981b}. A star's fractional X-ray luminosity ($L_\textrm{X}/L_*$) is related to the Rossby number $Ro$, the dimensionless ratio of the stellar rotation period $P_{\rm{rot}}$ and the convective turnover time $\tau_c$, $Ro$  = $P_{\rm{rot}}/\tau_c$ \citep{Mangeney_1984}. The rotation--activity relation has three regimes. Low Rossby number stars are in the saturated regime, where the fractional X-ray luminosity is approximately constant (or has a weak decrease with increasing $Ro$) at around $\log (L_\textrm{X}/L_*) = -3$. High Rossby number, slower rotating stars are in the unsaturated regime, where the fractional X-ray luminosity shows a characteristic sharp decay with decreasing rotation rate \citep{Nunez_2024,Magaudda_2022,Pizzolato_2003,Wright_2011,Wright_2018}. A third supersaturated regime exists for stars with the highest rotation rates, which have lower $L_\textrm{X}/L_*$ compared to saturated regime stars \citep{Jeffries_2011,Randich_1996,Prosser_1996}. Centrifugal stripping may create a limit to the X-ray luminosity \citep[see][]{Jardine_1999}, by constraining the size of X-ray emitting coronal loops \citep{Argiroffi_2016,Nunez_2017}. Similar rotation--activity relations with Rossby number have also been found for the surface average magnetic field strength \citep{Reiners_2022,Vidotto_2014} and H$\alpha$ emission line luminosity - a tracer of chromospheric activity \citep{Noyes_1984,Nunez_2024}.

Observations of young PMS clusters (age < 5\,Myr) find almost all stars have Rossby numbers that place them in the saturated regime of the rotation--activity relation. Stars in young PMS clusters are also observed to have a far greater scatter of fractional X-ray luminosities than found for main sequence stars \citep{Briggs_2007,Preibisch_2005}. The scatter in $\log(L_\textrm{X}/L_*)$  -- quantified by the median absolute deviation (MAD) -- decreases with cluster age, with \citet{Alexander_2012} reporting that it reaches the level found for main-sequence stars by $\sim$30\,Myr -- the age where a solar mass star reaches the zero-age main-sequence (ZAMS). By such ages, and as early as $\sim$13\,Myr, there is evidence of higher-mass stars in PMS clusters having evolved in the rotation--activity plane to create the unsaturated regime \citep{Argiroffi_2016,Jeffries_2011}. In the cluster $\alpha$~Persei, of age at least 50\,Myr, the unsaturated regime of the rotation--activity relation is apparent and is similar to that found for field stars \citep{Randich_1996}. 

How PMS stars evolve from the saturated regime to form the unsaturated regime of the rotation--activity relation is not well understood. The transition may be related to a change in the interior dynamo process - not only for stars which transition from fully to partially convective interiors, but for fully convective stars for which the dynamo process is found to be dependent on rotation rate \citep{Warnecke_2020}. While a fully convective star is rapidly rotating, the dynamo can generate large-scale, dipole-like field loops that can enclose hot ($\sim$30\,MK), X-ray-emitting plasma heated from mega-flares that dominate the X-ray emission \citep{Cohen_2017,Getman_2021}. As stars spin-down, and become slower rotators, emission may be dominated from plasma at temperatures of $\lesssim$10\,MK heated from flaring in smaller scale field loops similar to the contemporary Sun \citep{Gudel_2004}. 

For main sequence stars, there is an observed decay of X-ray luminosity with age, the form of which varies with stellar mass \citep{Gudel_2004,Maggio_1987,Nunez_2016}. The decay in X-ray luminosity is expected given that as stars age on the main sequence, the angular momentum loss from magnetised stellar wind leads to an approximately Skumanich decay ($\propto t^{-1/2})$ in rotation rate  \citep{skumanich_1972}, which in turn leads to a reduction in X-ray luminosity with increasing age.

Analysis of \textit{Chandra} satellite observations has given insight into PMS star X-ray evolution. Initially, the median X-ray luminosities are insensitive to age and begin to decay after 7\,Myr \citep{Getman_2022}. However, this study focused on stars of mass $M_{*}$ > 0.7\,M$_{\sun}$ due to the lack of detections for lower-mass stars that are typically less X-ray luminous. The observed decline in X-ray luminosity is likely influenced by the development of large radiative cores after stars leave the Hayashi track. PMS stars that have evolved onto Henyey tracks are known to have typically lower fractional X-ray luminosities than those on Hayashi tracks \citep{Gregory_2016,Rebull_2006}.

Previous studies have used established relations between X-ray luminosity and rotation period / Rossby number to study the change of X-ray emission using rotation rate evolution models \citep[e.g.][] {Gondoin_2018,Johnstone_2021,Magaudda_2020}. In this paper we explore the timescales over which the X-ray rotation-activity relation forms. We combine rotational evolution models with our understanding of X-ray luminosity trends with age to evolve the position of stars across the rotation--activity plane. We start with observational data of PMS stars where the multi-regime rotation--activity relation is not yet established and where the scatter in fractional X-ray luminosities is orders-of-magnitude higher than for main sequence stars. We investigate if our evolutionary model can reproduce the rotation--activity relation -- with regimes of saturation and unsaturation -- along with investigating the behaviour in the scatter of fractional X-ray luminosities as PMS stars evolve. 

In Sec.~\ref{sec: PMS data collection}, we outline the data collected from four PMS clusters to place and evolve the position of stars on the rotation--activity plane. Data from field stars and open clusters are selected to construct the main-sequence rotation--activity relation for comparison (Sec.~\ref{sec: MS data collection}). We determine stellar parameters from the position of stars in the Hertzsprung--Russell (H--R) diagram as described in Sec.~\ref{sec: ARR interpolation}, to establish rotation--activity plots with $\sim$600 PMS stars and the main-sequence sample for comparison (Sec.~\ref{sec: init ARR results}). By using observed trends between X-ray luminosity and age (Sec.~\ref{sec: X-ray evo}), a coupled core-envelope rotational evolution model for stars (Sec.~\ref{sec: Rot Evo}) and stellar evolutionary mass tracks, we model the emergence and evolution of the rotation--activity relation. Our results are described in Sec~\ref{sec: rot eve results} and Sec~\ref{sec: ARR results}. In Sec~\ref{sec: Rx spread} we examine the change in the median fractional X-ray luminosity for our sample and its scatter. In Sec.~\ref{sec: discussion} we discuss the potential limitations of our model, and we conclude in Sec.~\ref{sec: conclusions}.


\section{Pre-main sequence cluster data}
\label{sec: PMS data collection}

In this section, we describe the observational data collected as input for our models. To position a star on the rotation-activity plane we require the bolometric luminosity, X-ray luminosity, convective turnover time, and stellar rotation period. To determine the current age and mass of PMS stars -- and corresponding parameters such as stellar radius and convective turnover time -- we position stars on the H--R diagram using their effective temperatures $T_{\rm{eff}}$ and bolometric luminosities $L_*$. For our rotational evolution models, we also collect information on disc status, if available. This allows us to determine how to evolve the stellar rotation rate from the observed age (see Section \ref{sec: Rot Evo}). We only select stars with X-ray luminosities calculated from X-ray detections and ignore published upper limits.

The PMS clusters we study in this paper and the data sources used to collect the parameters required to position a star on the rotation--activity plane are outlined in the following subsections. We begin with the data from \cite{Getman_2022}, hereafter G22, where possible, which provides X-ray luminosities for PMS stars, along with (in many cases) calculated values of $T_{\text{eff}}$ and $L_{*}$. Data from other literature is cross-matched using stellar coordinates, matched to within an error of 2 arcseconds. X-ray and bolometric luminosities not taken from G22 are corrected using the cluster distances listed in table 1 of \citet{Getman_2021}. 

\subsection{Orion} 
The Orion Nebula Cluster and the surrounding star-forming regions, at a distance of $\sim$400\,pc \citep{Kuhn_2019,Getman_2019}, are amongst the best-studied PMS clusters. The ages of stars in these regions fall between $\sim$0.1--10\,Myr with median ages around 2\,Myr \citep{Briceno_2019,Getman_2014}. The X-ray luminosities for Orion stars are from the MYStIX (Massive Young Star-Forming Complex Study in Infrared and X-Ray) and SFiNCs (Star Formation In Nearby Clouds) surveys as outlined in \citetalias{Getman_2022}. The regions included in Orion, as defined in \cite{Getman_2021}, are the Orion Nebula Cluster (ONC), the Orion flanking fields (north and south), the Orion molecular clouds 2--3 (OMC23), and the reflection nebula NGC~2068. 

The observed rotation periods are taken from \cite{Serna_2021}, which uses the latest \textit{TESS} data to build upon the catalogue of rotation periods for stars identified in the Orion star-forming complex. We prioritise use of the \textit{TESS} periods when available. Then, we use earlier literature periods as available in tables from \cite{Serna_2021} (taking the first listed period in cases of multiple previously reported periods). This study also includes the classification of PMS stars as outlined by \cite{Briceno_2019}, where the equivalent width of the H$\alpha$ line is used to indicate the accretion status. For stars where $L_{*}$ and $T_{\text{eff}}$ are not available in \citetalias{Getman_2022}, we use values from \cite{Serna_2021}. 396 Orion stars are identified with X-ray luminosities and H--R position data, of which 197 have published rotation periods.

\subsection{NGC 2264} 
At a distance of $\sim$730\,pc, NGC~2264 is a well-studied star-forming region due to low extinction from foreground sources \citep{Flaccomio_2023,Kuhn_2019}.
The X-ray luminosity values for PMS stars in this cluster are obtained from the MYStIX / SFiNCs survey as presented in \citetalias{Getman_2022}. Additional values of  $L_{\rm{X}}$ are based on \textit{XMM-Newton} observations sourced from tables in \cite{Dahm_2007}. Stellar rotation periods are taken from \cite{Venuti_2017}, who analysed the periodicity of NGC~2264 members from \textit{CoRoT} observations. We exclude rotation periods from the sample that are classed as eclipsing binaries, multi-periodic/beating, or undefined. Data from \cite{Venuti_2018} provides H--R diagram positions (for use where not available from \citetalias{Getman_2022}) and a classification of the circumstellar disc status based on the infrared excess in \textit{Spitzer} observations  -- see \cite{Cody_2014} for classification details. For NGC~2264, 502 stars are identified with X-ray luminosities and H--R position data, of which 171 have published rotation periods.

\subsection{IC~348}
IC~348 is another nearby star-forming region at a distance of 324\,pc \citep{Kuhn_2019}. X-ray luminosities for IC~348 are also obtained from the MYStIX / SFiNCs survey results in \citetalias{Getman_2022}. We cross-match these values with data from \cite{Alexander_2012}, who collated rotation periods from previous studies, taking mean values in cases of multiple sources, and matched these to \textit{Chandra} X-ray observations for which we use their values of X-ray luminosity when not available from \citetalias{Getman_2022}. H--R diagram positions are obtained from \citet{Luhman_2003} when not available from \citetalias{Getman_2022}. \cite{Lada_2006} provides data on disc status based on excess infrared emission from \textit{Spitzer} observations. Stars classed with `anaemic' discs are given indeterminate disc status in our models, as many weak disc stars are fast rotators. We collate 243 IC~348 stars with X-ray luminosities and H--R position data, of which 88 have published rotation periods.

\subsection{h Persei} 

h~Persei (NGC~869) is another PMS cluster with previous work investigating the rotation--activity relation. At a distance of $\sim$2.5\,kpc, stars in the h~Persei cluster have an average age of $\sim$13\,Myr, making them, on average, older than our other clusters. h~Persei has a larger distribution of rotation rates since most of the stars no longer have discs and have been able to spin up. \citetalias{Getman_2022} provides X-ray luminosities for h~Persei stars as a part of their investigations into star clusters in the age range of 7-25\,Myr. For rotation periods, we use data provided in  \cite{Argiroffi_2016}, who matched rotation periods of stars from the Monitor project -- see \cite{Moraux_2013} -- with \textit{Chandra} X-ray observations. Where \citetalias{Getman_2022} does not have available X-ray information for a star, we use the values from \cite{Argiroffi_2016}. 830 stars in h~Persei are identified with X-ray luminosities and H--R position data, of which 193 have published rotation periods.

\begin{figure}

 \includegraphics[width=0.5\textwidth]{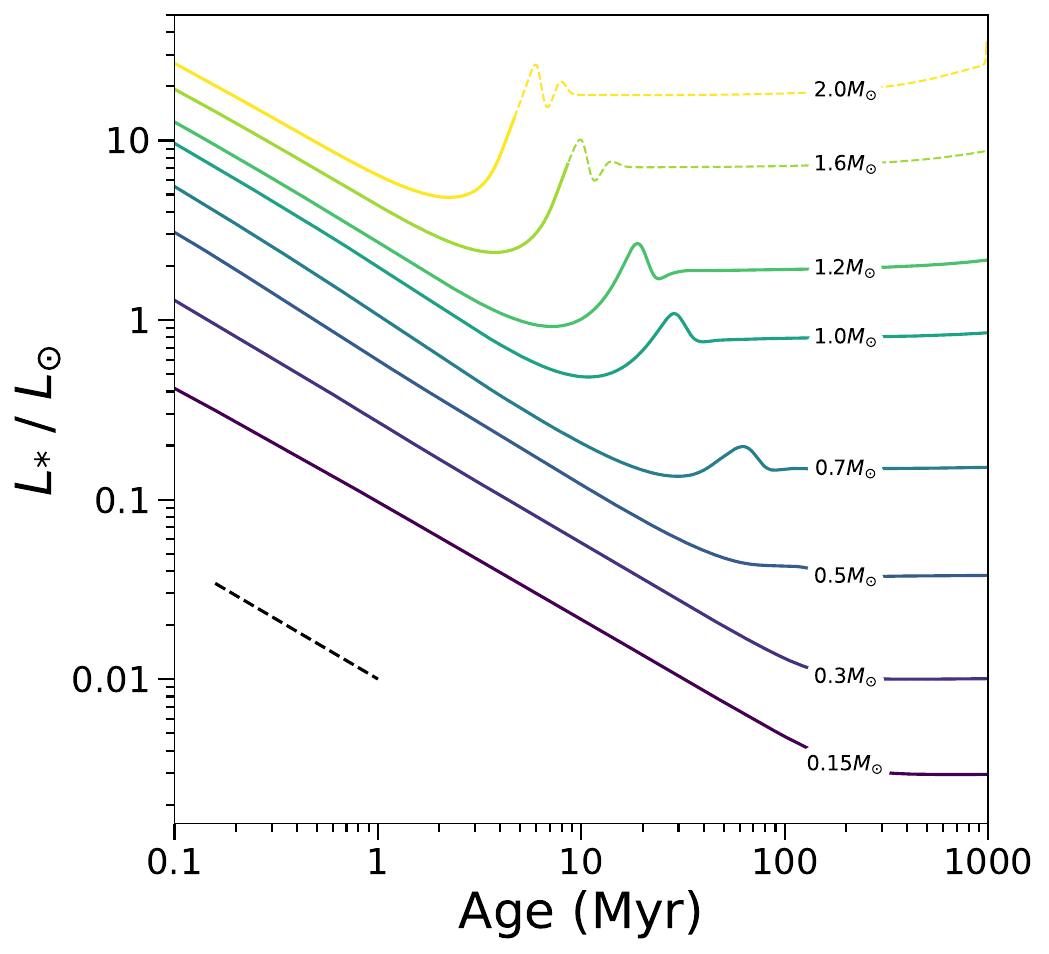}
 \caption{Bolometric luminosity versus age for stars of different mass from the YaPSI stellar evolution models. The dashed regions of the tracks indicate where the star no longer has an outer convective envelope. The black dashed line represents the expected bolometric luminosity decrease with age for stars evolving along Hayashi tracks, $L_{*}\propto t^{-2/3}$.}
 \label{fig: LstarVsAge}

\end{figure}
%

\begin{figure}

 \includegraphics[width=\columnwidth]{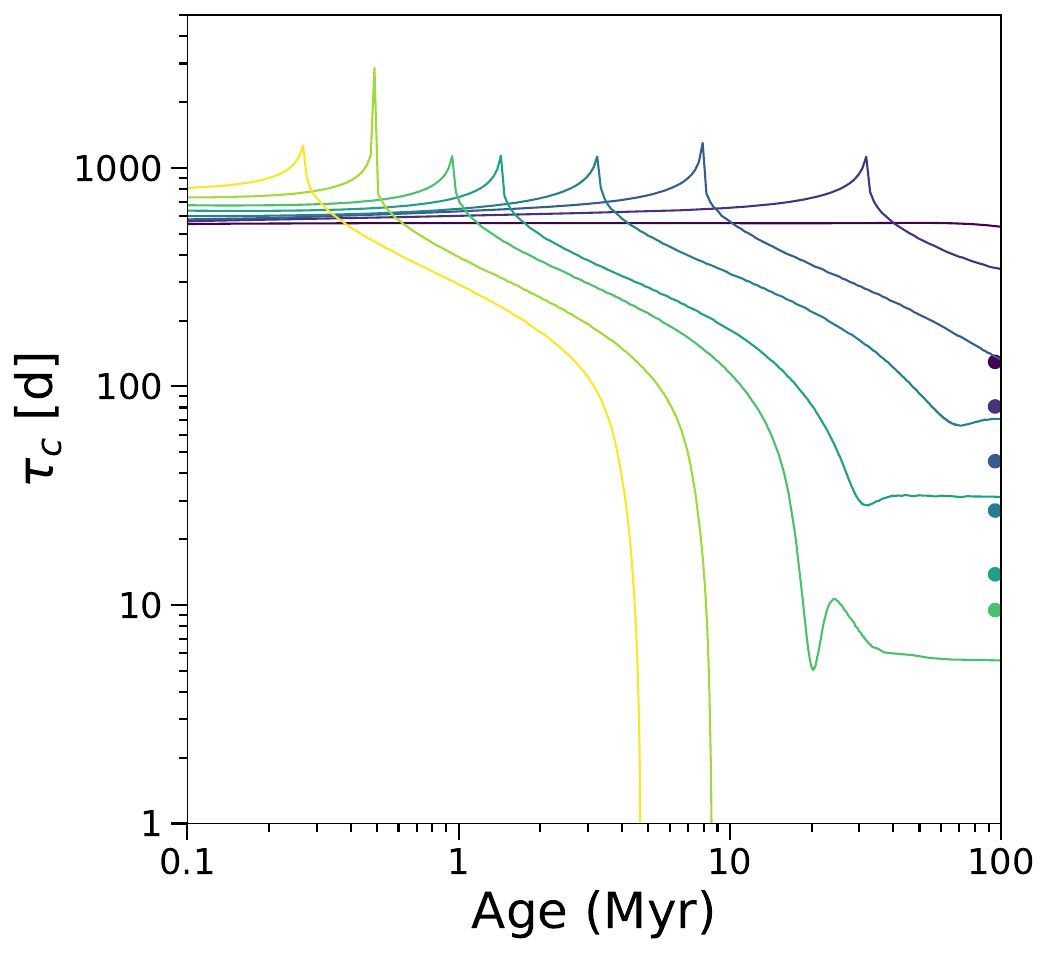}
 \caption{Convective turnover time versus age for stars of different mass from the YaPSI stellar evolution models. The colours match the stellar masses from Fig.~\protect\ref{fig: LstarVsAge}, ranging from 0.15 (top right) to 2$\,{\rm M}_\odot$ (bottom left). The filled circles on the right with a colour corresponding to the mass tracks represent the empirically determined values of convective turnover time from \protect\cite{Wright_2018}.}
 \label{fig: TaucVsAge}
\end{figure}
%

\begin{figure}
 \includegraphics[width=\columnwidth]{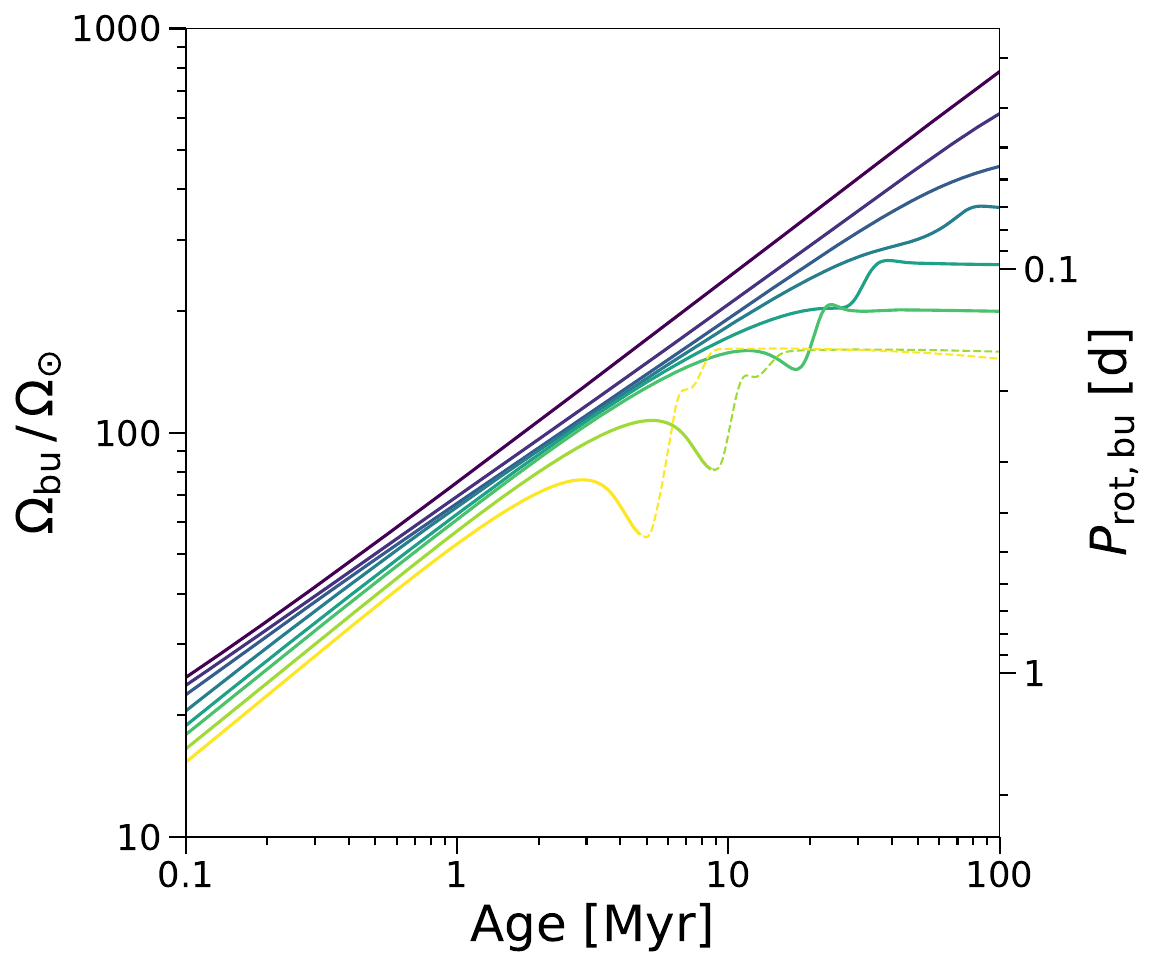}
 \caption{Break-up rotation rate relative to the solar rotation rate versus age. The right axis indicates the stellar rotation period at break-up $P_{\rm{rot,bu}} = 2\pi / \Omega_{\rm{bu}}$. The colours match the stellar masses from Fig.~\protect\ref{fig: LstarVsAge}, ranging from 0.15 (top left) to 2$\,{\rm M}_\odot$ (bottom right).}
 \label{fig: OmegabuVsAge}
\end{figure}

\section{Main sequence star data}
\label{sec: MS data collection}
For main sequence stars, we use the data from \cite{Wright_2011} that has been a standard in previous rotation--activity relation studies; and provides all relevant data (X-ray luminosity, rotational period, and H--R diagram positions) from several open cluster stars, ages 40--700\,Myr, and field stars. The main sequence sample consists of 824 stars.

\begin{figure*}
\includegraphics[width=0.4\textwidth]{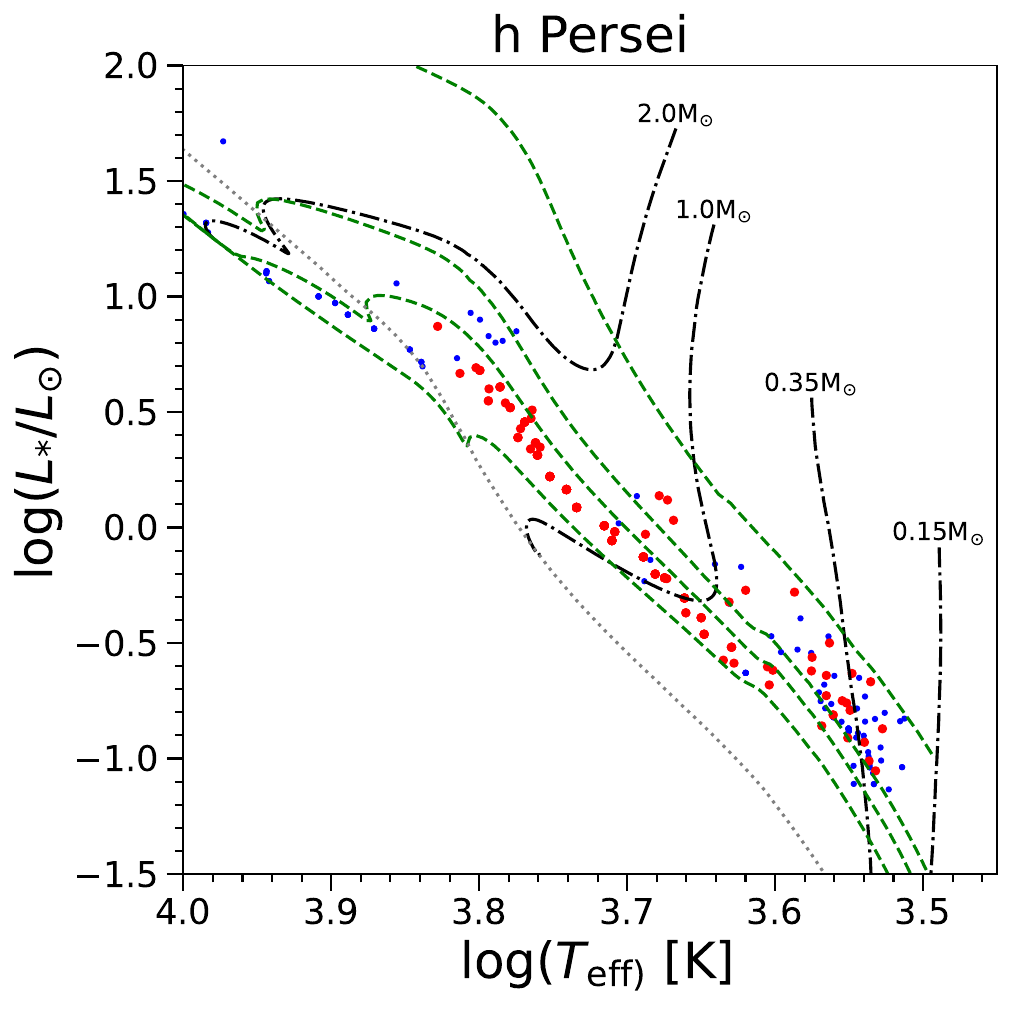}
\includegraphics[width=0.4\textwidth]{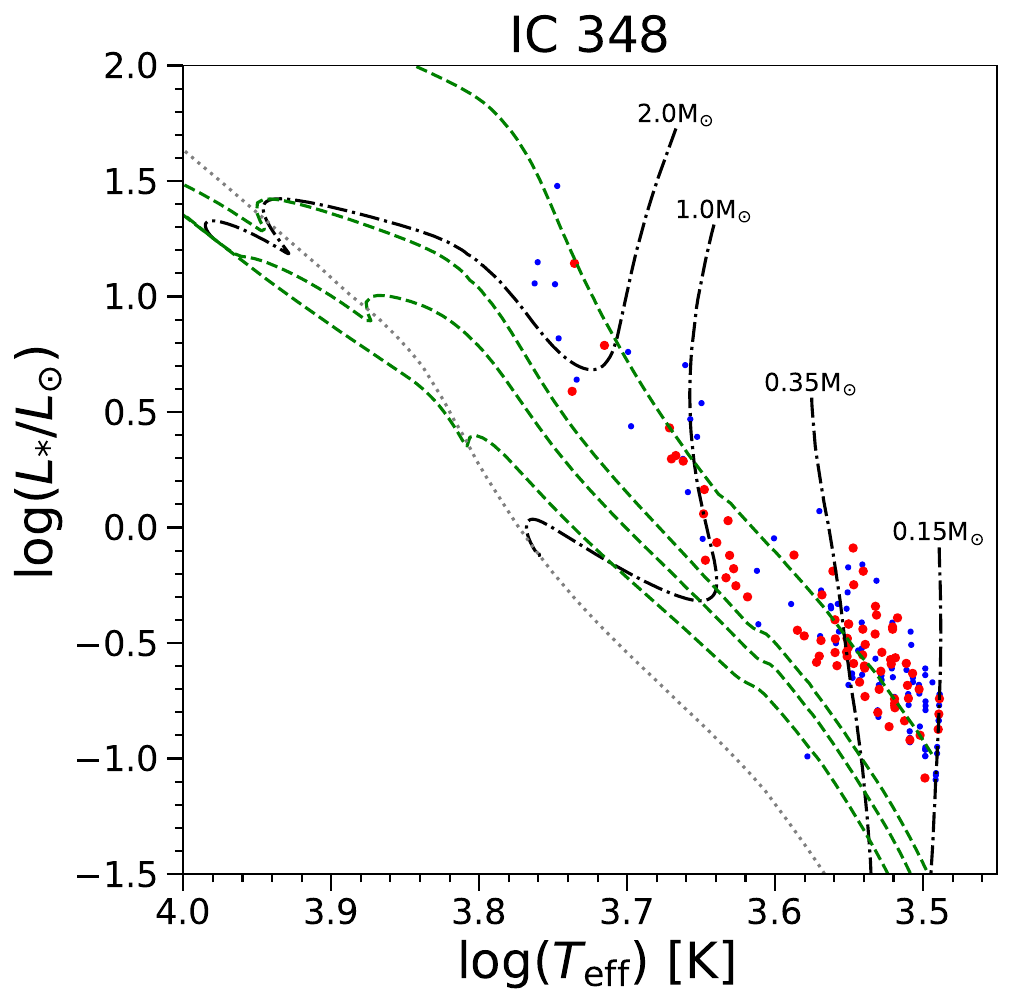}
\includegraphics[width=0.4\textwidth]{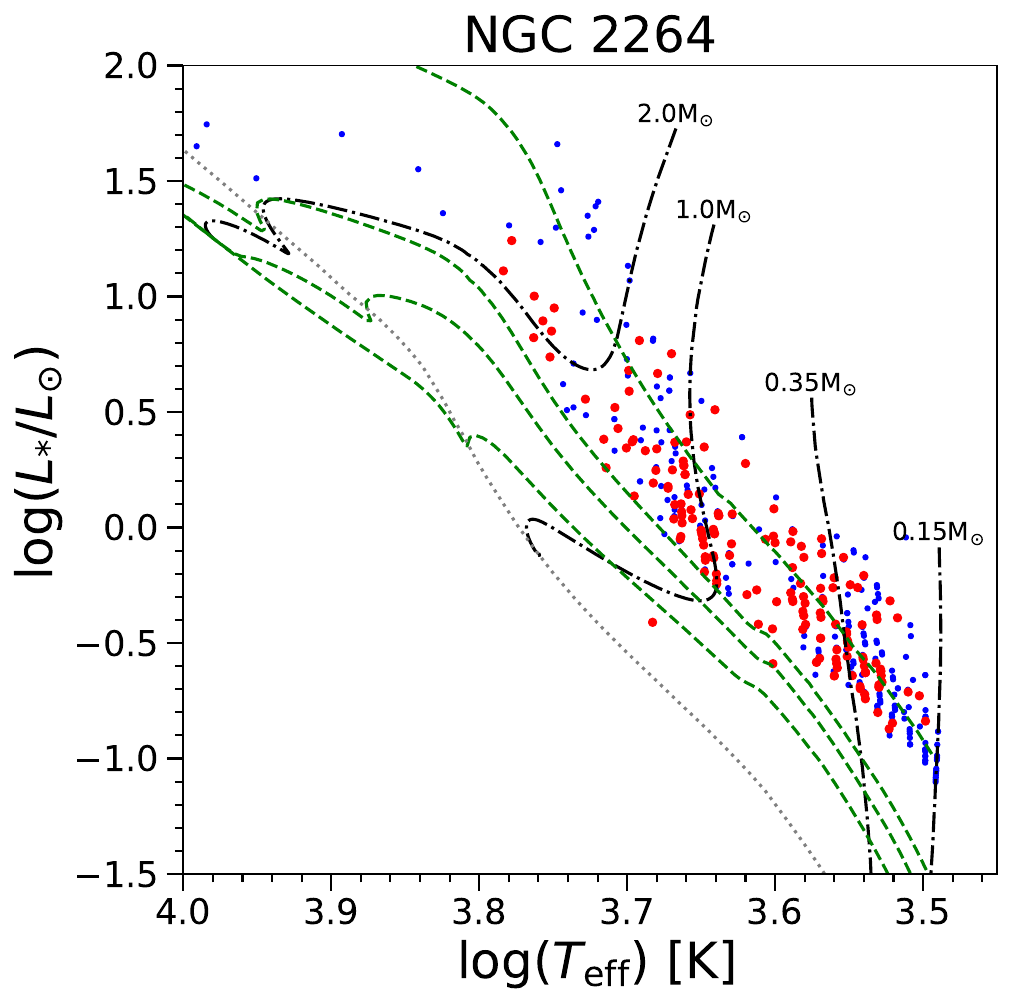}
\includegraphics[width=0.4\textwidth]{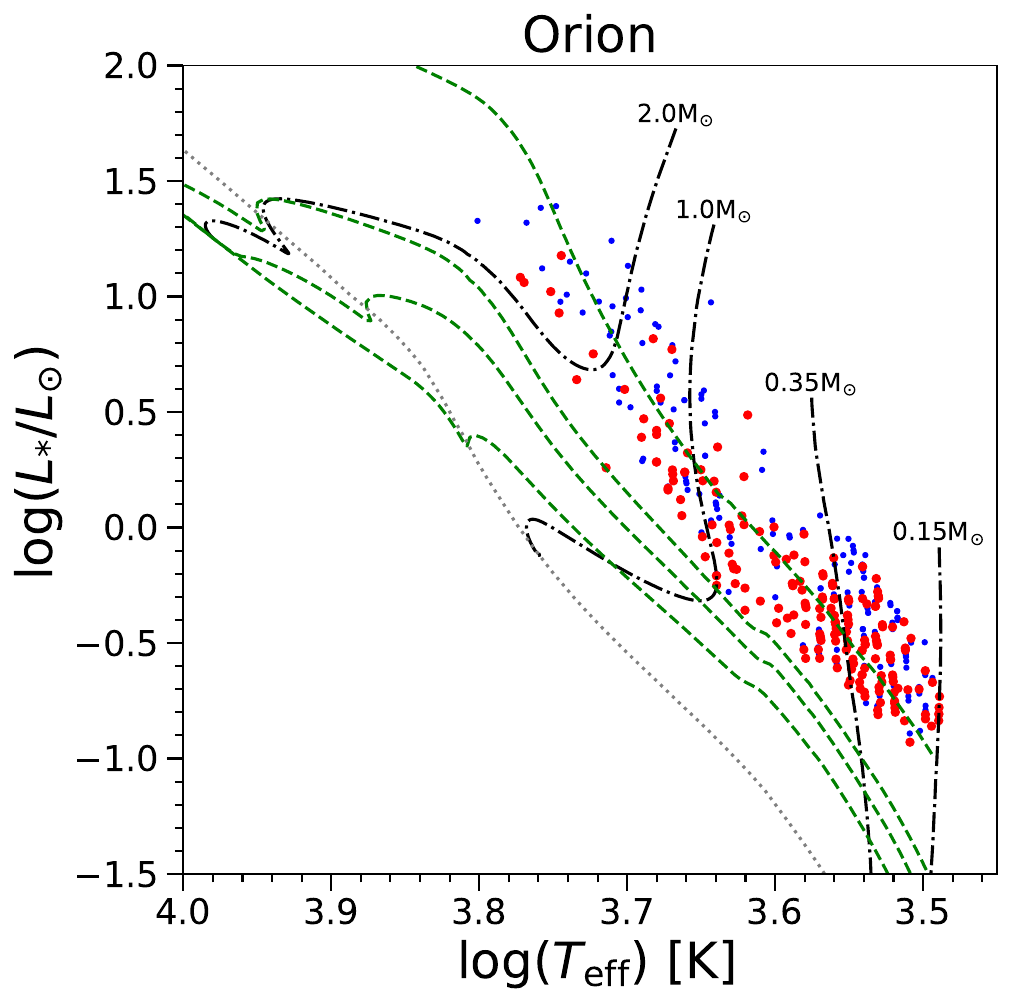}
\caption{H--R diagrams for stars in four PMS clusters. Large red-filled circles represent stars where X-ray luminosity and rotation period data are available. Small blue-filled circles represent stars with only X-ray luminosity data. The green dashed lines represent YaPSI isochrones for 1, 6, 10 and 20\,Myr from top to bottom. The dotted grey line represents the zero-age-main sequence. The black dot-dashed lines are selected mass tracks.}
\label{fig: cluster HR diagrams}

\end{figure*}
%

\begin{figure}
\includegraphics[width=\columnwidth]{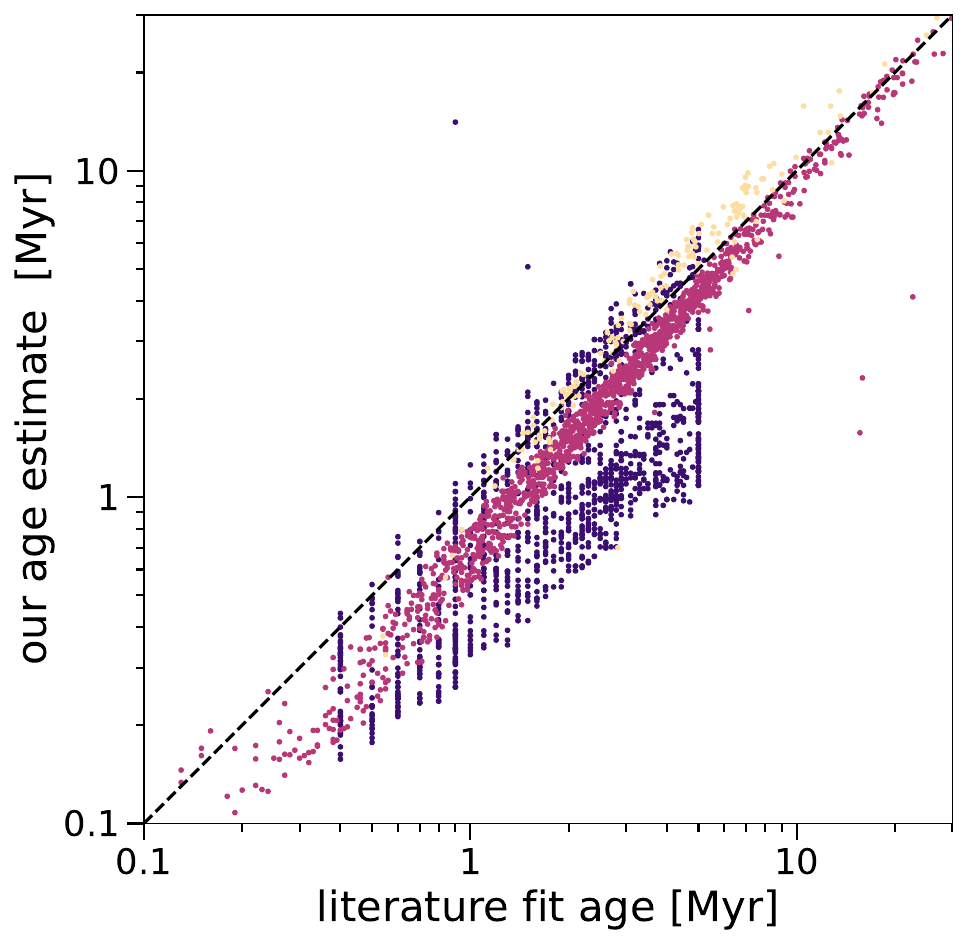}
\caption{Comparison of stellar ages derived from the YaPSI models versus age estimates from the literature for stars in PMS clusters. The dark purple, maroon and yellow circles indicate the age estimates from \protect\citetalias{Getman_2022}, \protect\cite{Serna_2021} and \protect\cite{Venuti_2018} respectively.}
\label{fig: Age comparisons}

\end{figure}
%

\begin{figure*}
\includegraphics[width=0.3\textwidth]{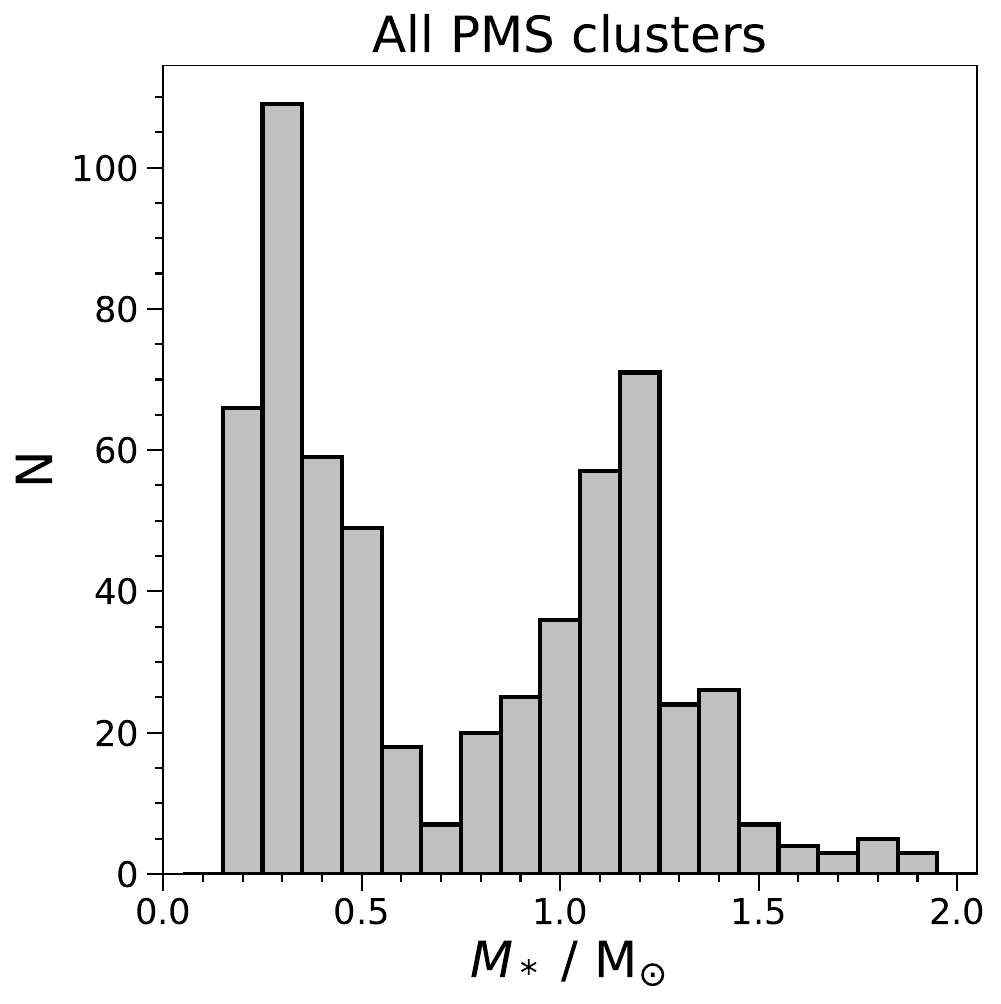}
\includegraphics[width=0.3\textwidth]{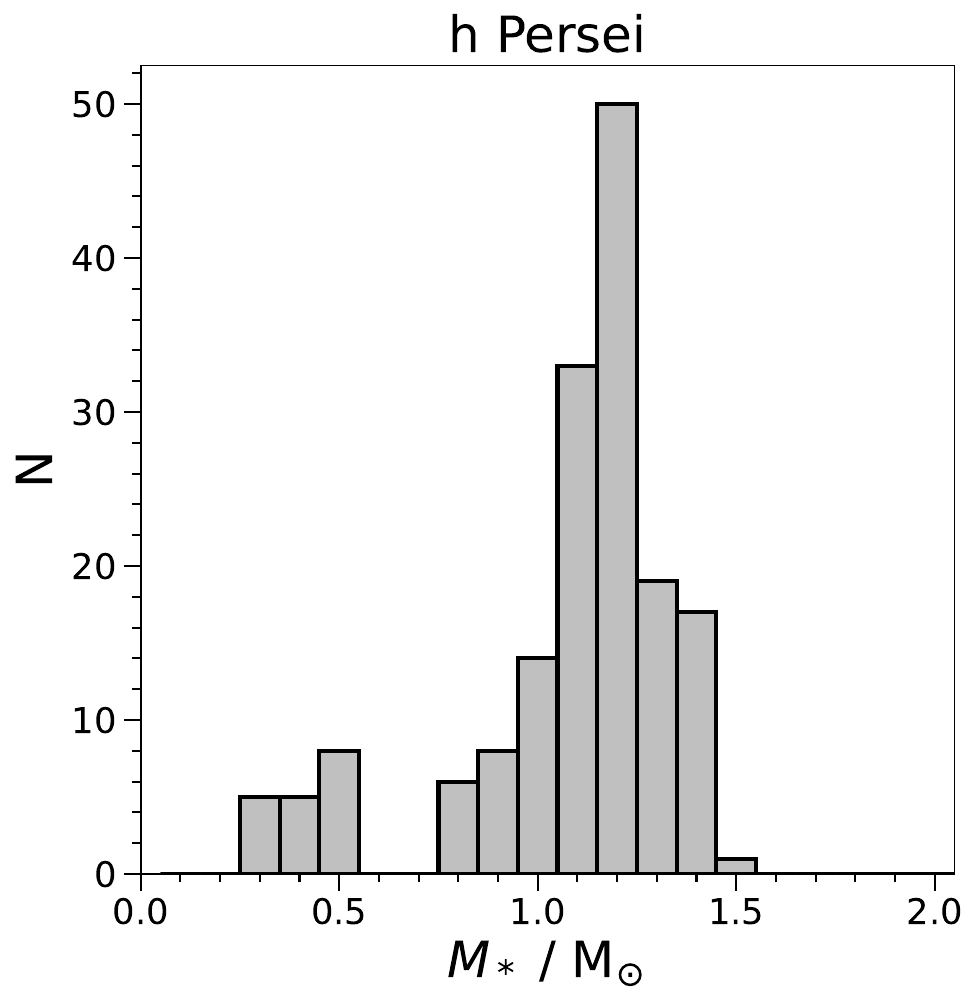}
\includegraphics[width=0.3\textwidth]{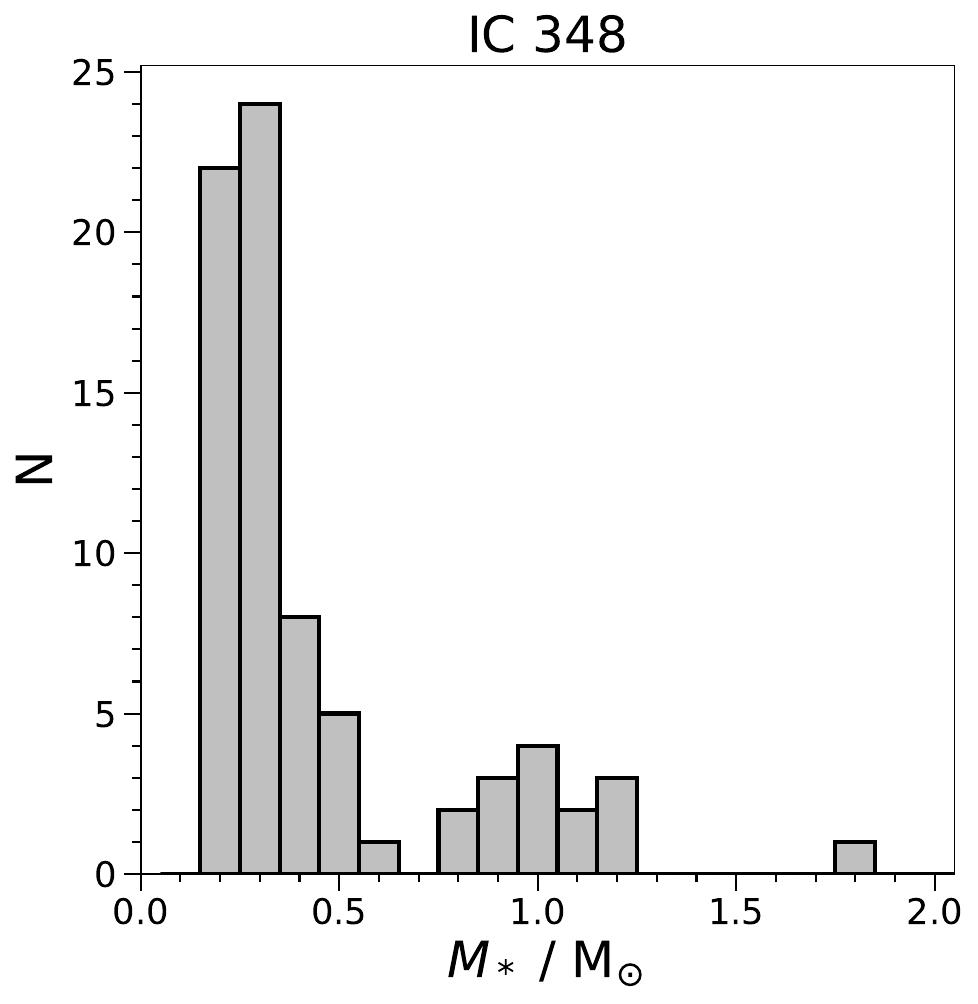}
\includegraphics[width=0.3\textwidth]{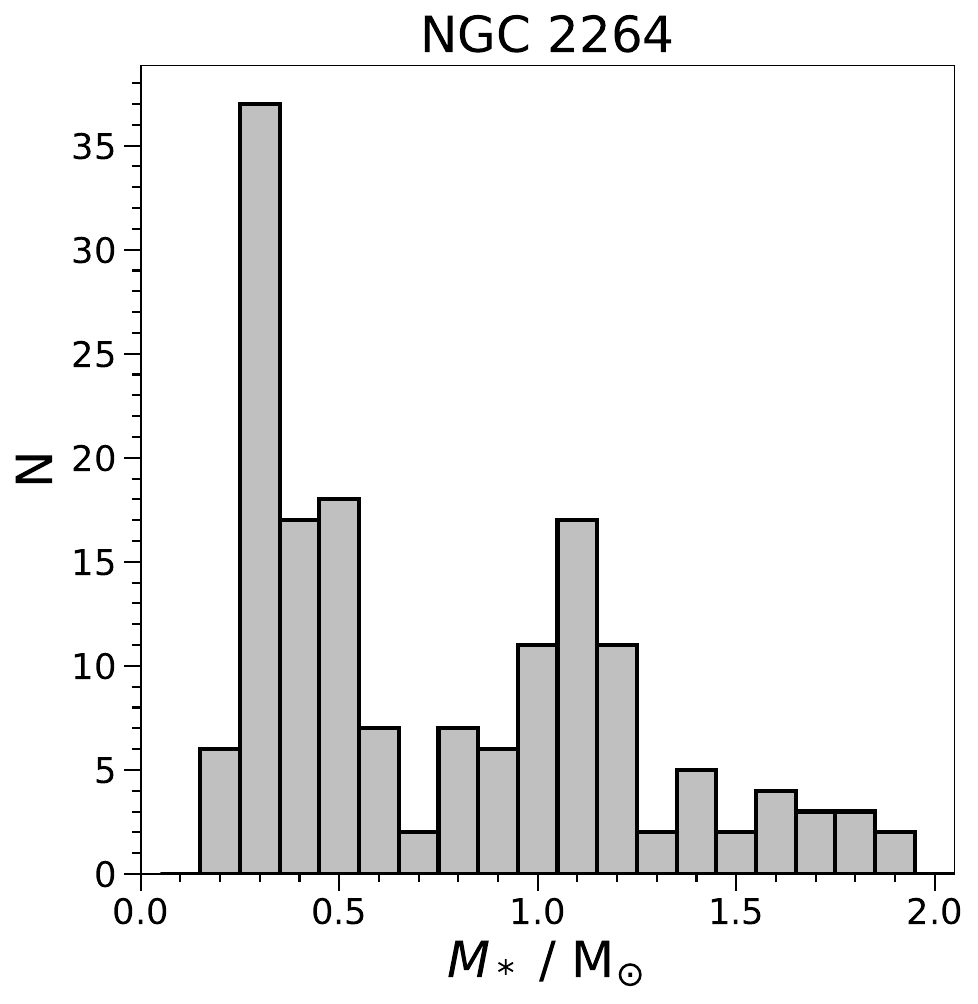}
\includegraphics[width=0.3\textwidth]{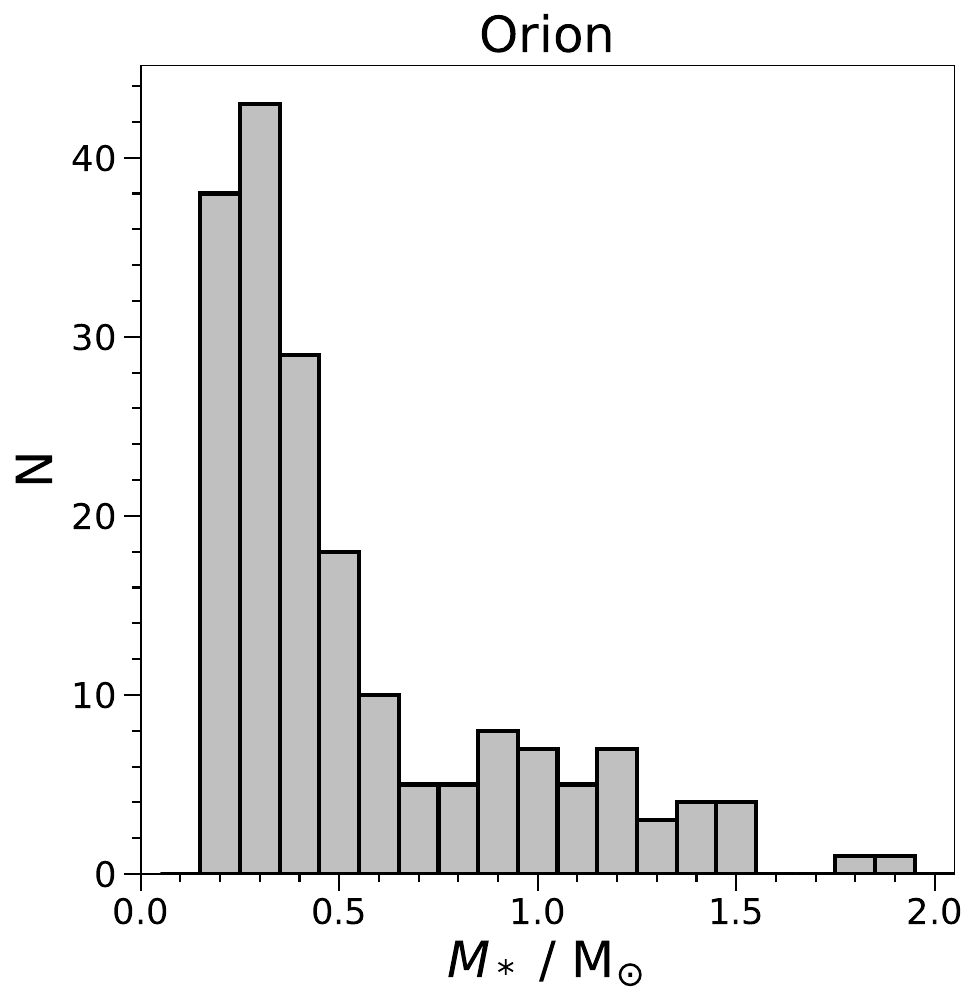}
\caption{Histograms of stellar mass for PMS stars used in our simulations (those with $L_{\rm{X}}$ and rotation period data) for the whole PMS sample and the individual clusters as labelled.}
\label{fig: init Mass hists}

\end{figure*}
%


\section{Stellar parameters from the H--R diagram}
\label{sec: ARR interpolation}

We can position stars with $T_{\rm{eff}}$, $L_{*}$, $L_{\rm{X}}$ and $P_{\text{rot}}$ on the rotation activity plane. To do so, we also need the convective turnover time $\tau_c$ which can be derived from a star's H--R diagram position, as well as the mass and age of the star. Using $T_{\rm{eff}}$ and $L_{*}$, we determine $\tau_c$, mass, and age, of each star by interpolating to the closest-matching point on an evolutionary mass track. This then allows the star's parameters to be evolved in time along the appropriate mass track. We use the `Yale--Potsdam Stellar Isochrones' (YaPSI) evolutionary tracks with solar metallicities from \cite{Spada_2017}.\footnote{The YaPSI mass tracks and isochrones are available at \href{http://www.astro.yale.edu/yapsi/}{http://www.astro.yale.edu/yapsi/}.} The YaPSI evolutionary models provide self-consistent calculations of the convective turnover times, meaning this does not need to be derived from other models [see, e.g. \cite{Kim_1996}, \cite{Landin_2023}] or empirical calculations [see, e.g. \cite{Pizzolato_2003}, \cite{Wright_2018}]. The bolometric luminosity and convective turnover time evolution for selected mass tracks from the YaPSI models are shown in Fig.~\ref{fig: LstarVsAge} and Fig.~\ref{fig: TaucVsAge}, respectively. The lower mass limit in the YaPSI models ($M_{*}  \geq 0.15\,\rm{M}_{\sun}$) is satisfactory as the primary focus of our study is on stars around a solar mass. The YaPSI models use two surface boundary conditions to determine calibrated mixing length parameters $\alpha_{\rm{MLT}}$. As a result, stars of mass $0.6\rm\,{M}_{\sun} \leq M_* \leq 1.1\,\rm{M}_{\sun}$ have two separate evolutionary models corresponding to the two different mixing length parameters. This paper uses the tracks with $\alpha_{\rm{MLT}} = 1.91804$ where possible.

All stars start the PMS phase with a fully convective internal structure. The lowest masses stars ($M_{*}  \leq 0.25\rm{M}_{\sun}$) remain fully convective on the PMS and on the main-sequence. As a higher-mass star contracts during the PMS, the opacity in the centre of the star decreases sufficiently as the temperature increases to become stable against convection.
For a range of stellar masses, the emerging radiative cores are only temporary but last beyond the first 100\,Myr of the lifetime of a star. For stars in the range 0.26--0.30\,$\rm{M}_{\sun}$, the temporary core is stable. For stars in the range of 0.31--0.34\,$\rm{M}_{\sun}$, the temporary core is unstable. The instability results from the behaviour of non-equilibrium $^3\rm{He}$ fusion \citep{Baraffe_2018, Van_Saders_2012}. At ages of 100\,Myr, these temporary radiative cores will be present. Stars with radiative cores can exhibit inner radiative shells briefly where the inner core is convective. For partially convective stars, this inner convective core mass lasts briefly and is a relatively low proportion of the mass of the whole star -- we do not consider this in our rotational evolution. For stars of mass $M_{*} > 1.2\,\rm{M}_{\sun}$, by the main sequence, the outer convective envelope of a star will cease to exist. At that point, the star no longer has a classifiable convective turnover time or position on the rotation--activity plane. The above mass conditions on the stellar interior are based on the YaPSI mass tracks. The exact values vary from model to model but are broadly in agreement \citep[see for example][]{Baraffe_2015}. 

When interpolating the observed position of a star on the H--R diagram to a point on a YaPSI mass track we impose a fitting criteria. We require that the values of $T_{\rm{eff}}$ and $L_*$ are within 2\% of the observed values. This criteria avoids fitting ages and masses to stars that fall in a position of the H--R diagram not covered by mass tracks. Primarily, this cuts off stars with low values of effective temperature that fall below the limit of the YaPSI tracks as these stars are likely lower mass than the $0.15\,\rm{M}_{\sun}$ limit for the tracks. Additionally, this cuts off stars that fall below the ZAMS (whose observed position is likely erroneous). We also use an upper age limit of 50\,Myr. This upper age limit is more than adequate to encapsulate the ages we would expect for our stars in PMS clusters (see Figures \ref{fig: cluster HR diagrams} and $\ref{fig: Age comparisons}$). For fitting main-sequence/field stars of the \cite{Wright_2011} sample, the upper age limit imposed is set by the age when fusing hydrogen is depleted in the inner core, defined using the age a helium core forms. 

We only investigate stars up to a mass limit of 2\,$\rm{M}_{\sun}$. Many higher-mass stars have no detected X-ray emission, so the statistics describing the decay in $L_{\rm{X}}$ with time are affected in the analysis of \citetalias{Getman_2022}. Furthermore, stars with $M_* > 2\,\rm{M}_{\sun}$ lose their outer convective envelopes within 5\,Myr. We also do not model stars observed near the fully radiative stage of their evolution (defined if their core mass fraction is > 99.5\% of the total stellar mass). 

Finally, we look at the break-up rotation rate $\Omega_{\rm{bu}}$ of the stars in our sample. This is the rotation rate $\Omega_* = 2\pi / P_{\rm{rot}}$ where the centrifugal force overcomes the gravitational force at the surface, which is given as
\begin{equation}
\Omega_{\rm{bu}} = \left( \frac{ G M_{*} }{ R_*^3 }\right)^{1/2}.
\label{Eq: breakup velocity}
\end{equation}
Stars in PMS clusters are observed near break-up rotation rates \citep{Rebull_2018}. The calculated values of break-up rotation rates for different stellar masses from the YaPSI mass tracks are shown in Fig.~\ref{fig: OmegabuVsAge}. One star in our sample exceeds the break-up rotation rate at its observed age. This star is rejected from our sample, as this indicates its published rotation period is likely an alias.

\subsection{Stellar mass and age}
\label{sec: interpolation results}

For our sample of 649 PMS stars with all the relevant data to position them on the rotation--activity plane, 589 stars are retained after the interpolation and rejection criteria described in the previous section. In total, 1488 PMS cluster stars that met the criteria have $L_{\rm{X}}$ data available, but only a subset have rotation period data. For the sample of 824 main-sequence stars, 753 pass our interpolation and rejection criteria and will be used to determine the slopes of the saturated and unsaturated regime and the typical scatter in fractional X-ray luminosities for saturate regime stars.

In Fig.~\ref{fig: cluster HR diagrams}, we display the individual PMS cluster stars on the H--R diagram. Only stars that meet the 2\% matching criteria are displayed. Selected YaPSI mass tracks and isochrones are shown on the H--R diagrams. Any star that falls above the 2\,M$_{\odot}$ mass track is not used in our rotation--activity model. 

When comparing the H--R diagrams, it is clear that h~Persei is an older star-forming cluster. Most h~Persei stars are between the 10 and 20\,Myr isochrone. Meanwhile, almost all of the stars in the other clusters are above the 6\,Myr isochrone, with many at ages less than 1\,Myr. Also evident is a clear bias for the lower mass (lower effective temperature) stars to be typically younger. This result has been reported before for the clusters studied here and is a known effect of biases in PMS evolutionary models \citep{Gregory_2016,Hillenbrand_2008,Smith_2023,Venuti_2018}.

When comparing our age values to previously published age estimates, as shown in Fig.~\ref{fig: Age comparisons}, we see that our age estimates are typically below literature values. Our age estimates are up to a factor of four younger. Such a range of age differences is typical when comparing different methods of age-determination for stars in PMS clusters \citep{Getman_2014,Sung_2010}. Age estimates from \cite{Serna_2021} and \cite{Venuti_2018} agree best with our interpolation, where the masses and ages are also determined from H--R diagram positions derived from effective temperature and photometry. The ages from \citetalias{Getman_2022} agree less strongly. The difference is likely influenced by \citetalias{Getman_2022}'s inclusion of X-ray photometry to determine ages and the maximum cut-off of 5\,Myr in age used in the method \cite[see the chronometer from][]{Getman_2014}. 

Our H--R diagram interpolation also reveals the stellar mass distribution of our PMS cluster sample. Fig.~\ref{fig: init Mass hists} shows mass histograms of our PMS star sample. The mass distribution highlights that a large amount of the sample is below $0.75\rm{M}_{\sun}$, which is the lower mass limit over which \citetalias{Getman_2022} establish their relations between X-ray luminosity and age. The distributions also highlight that h~Persei, unlike the other younger star-forming clusters, is dominated by observations of higher-mass stars. The higher-mass star distribution is heavily influenced by the sensitivity of \textit{Chandra} observations, with h~Persei being considerably further away than the other clusters, making detections of lower-mass stars with typically lower X-ray luminosities more unlikely.

\begin{figure*}

\includegraphics[scale=0.33]{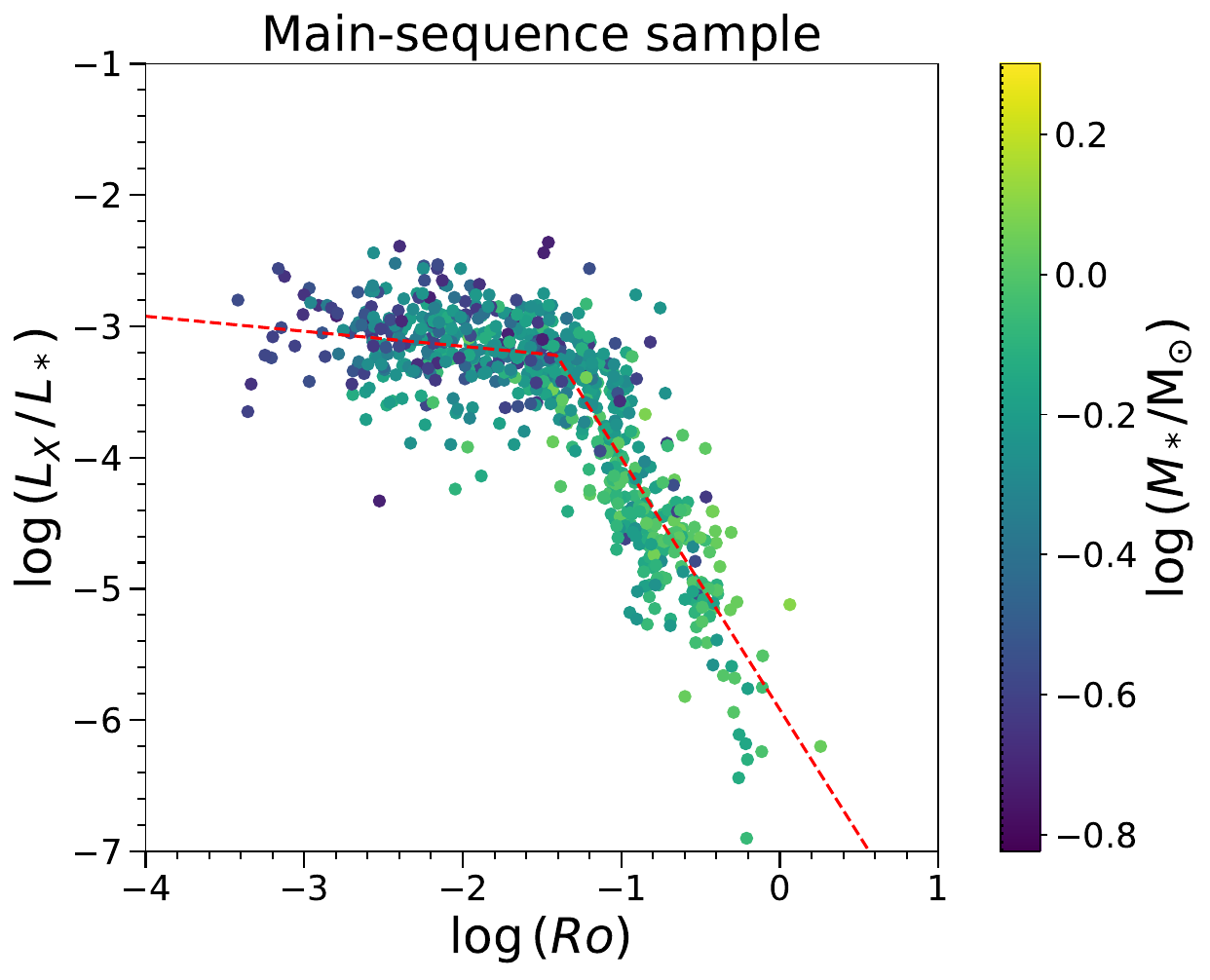} \caption{The rotation--activity relation for open cluster / field stars from \citet{Wright_2011}. Our best fit, a dual power-law (see Equation~\ref{Eq: ARR dual power-law}), is shown as the red dashed line. The colour of the data points represents stellar mass.}
\label{fig: W11_init_ARRplot}

\end{figure*}
%

\begin{figure*}

\includegraphics[scale=0.33]{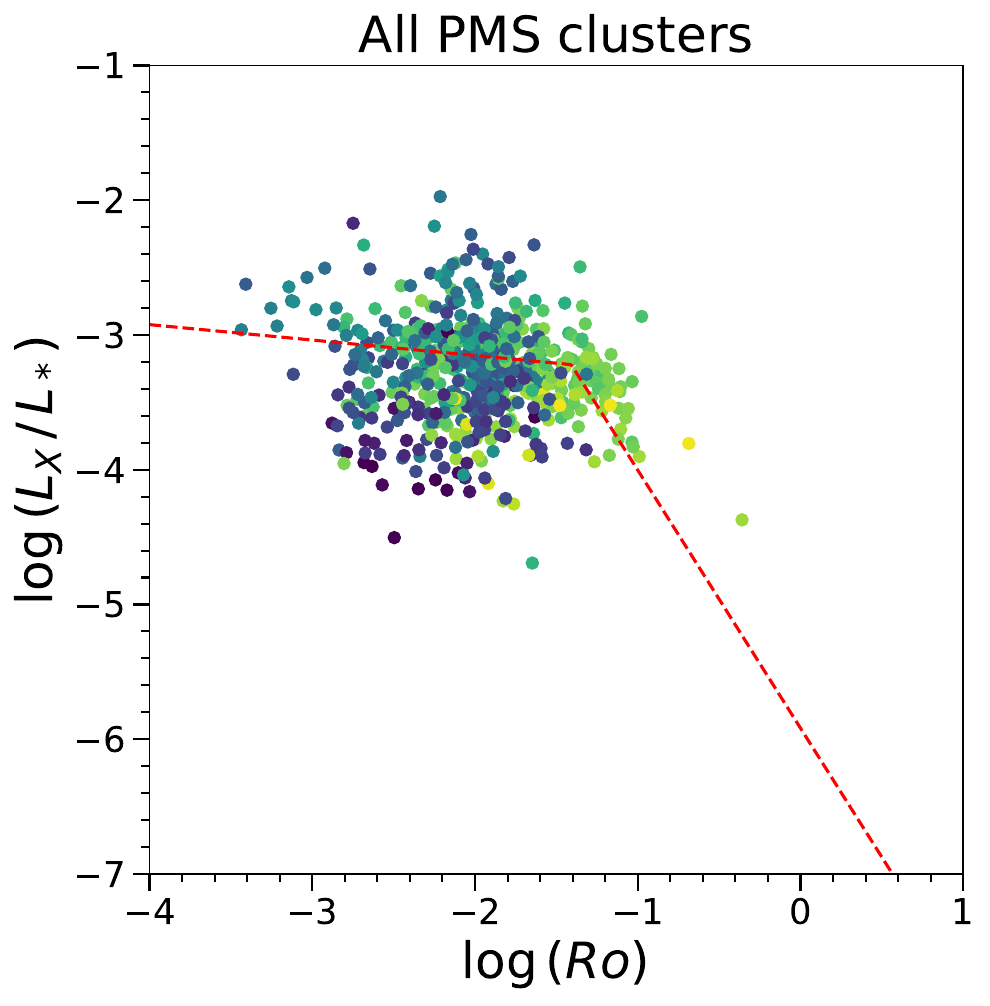}
\includegraphics[scale=0.33]{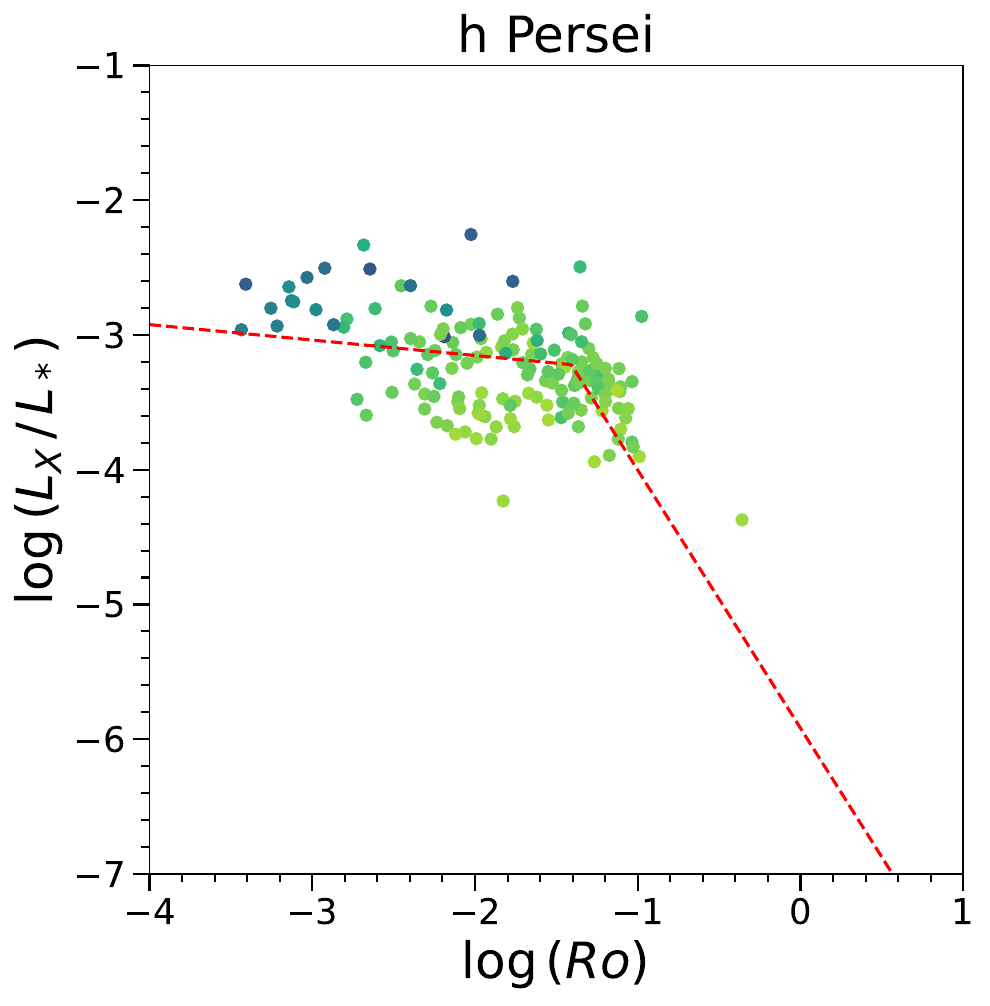}
\includegraphics[scale=0.33]{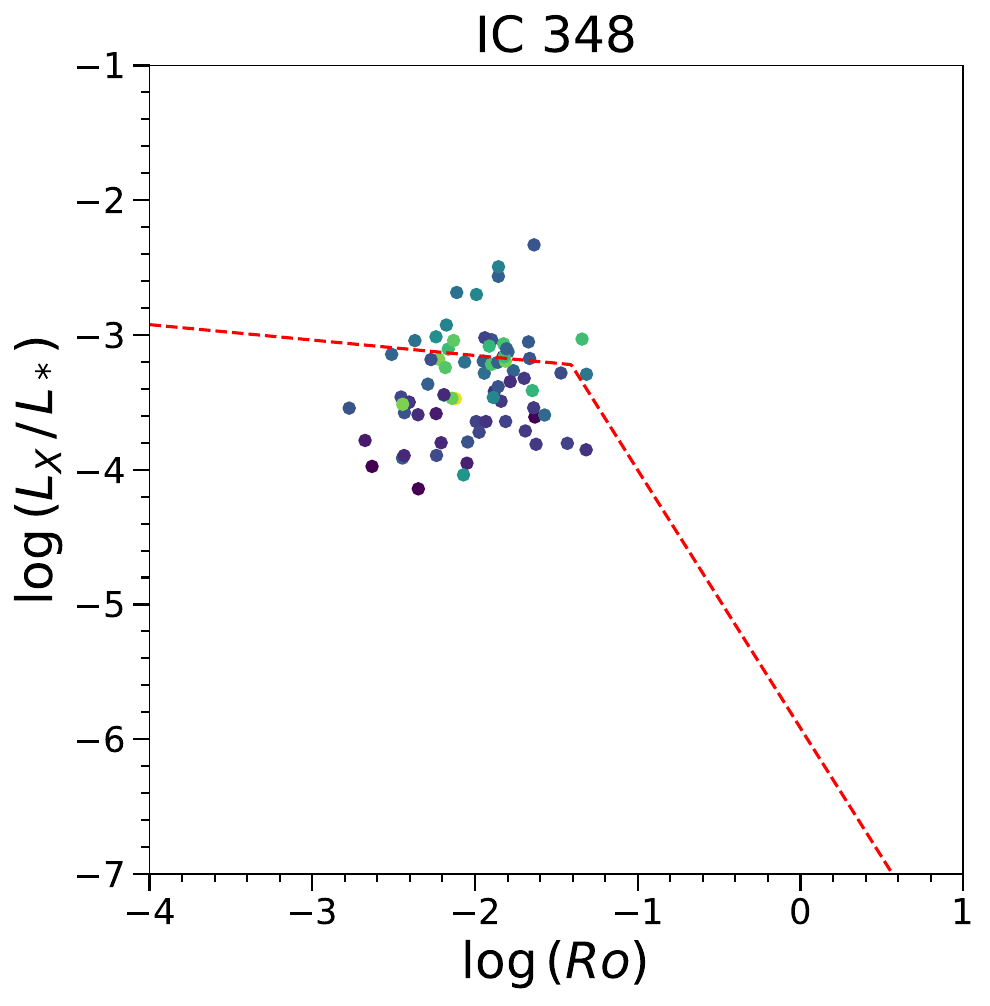}
\includegraphics[scale=0.33]{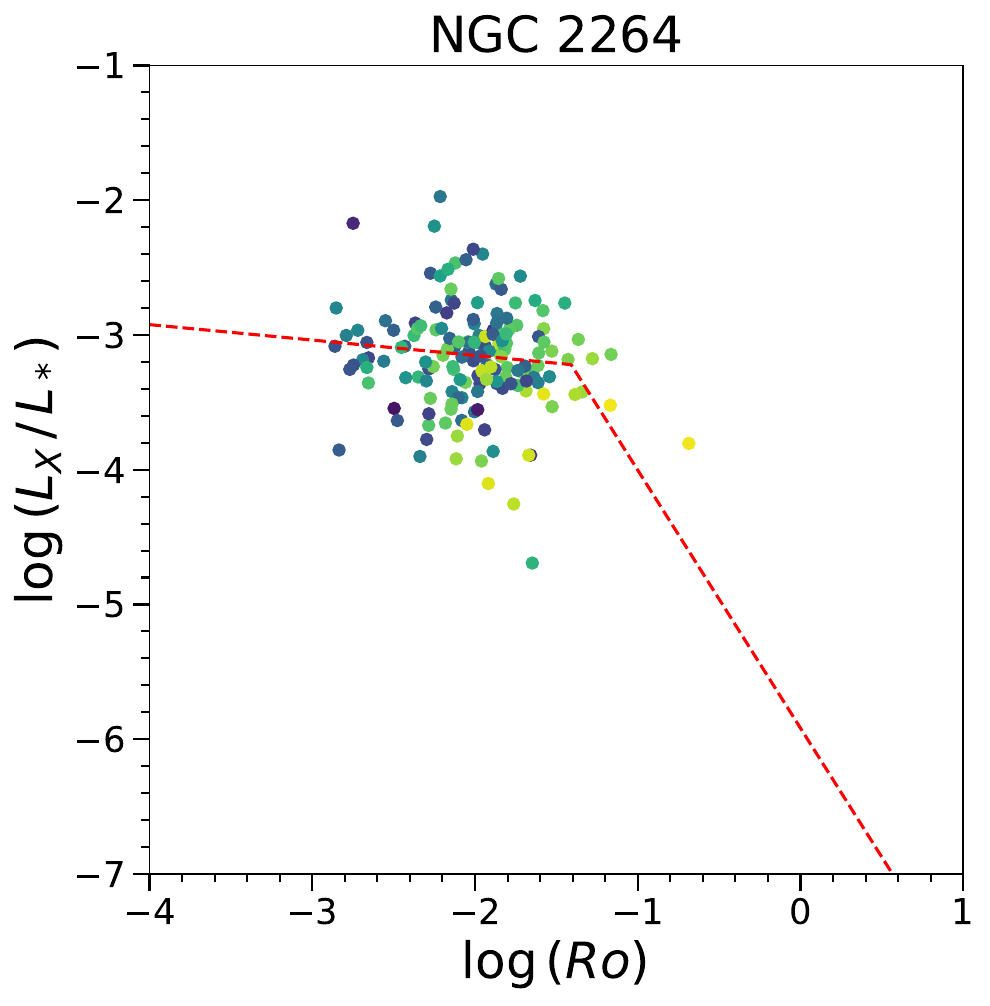}
\includegraphics[scale=0.33]{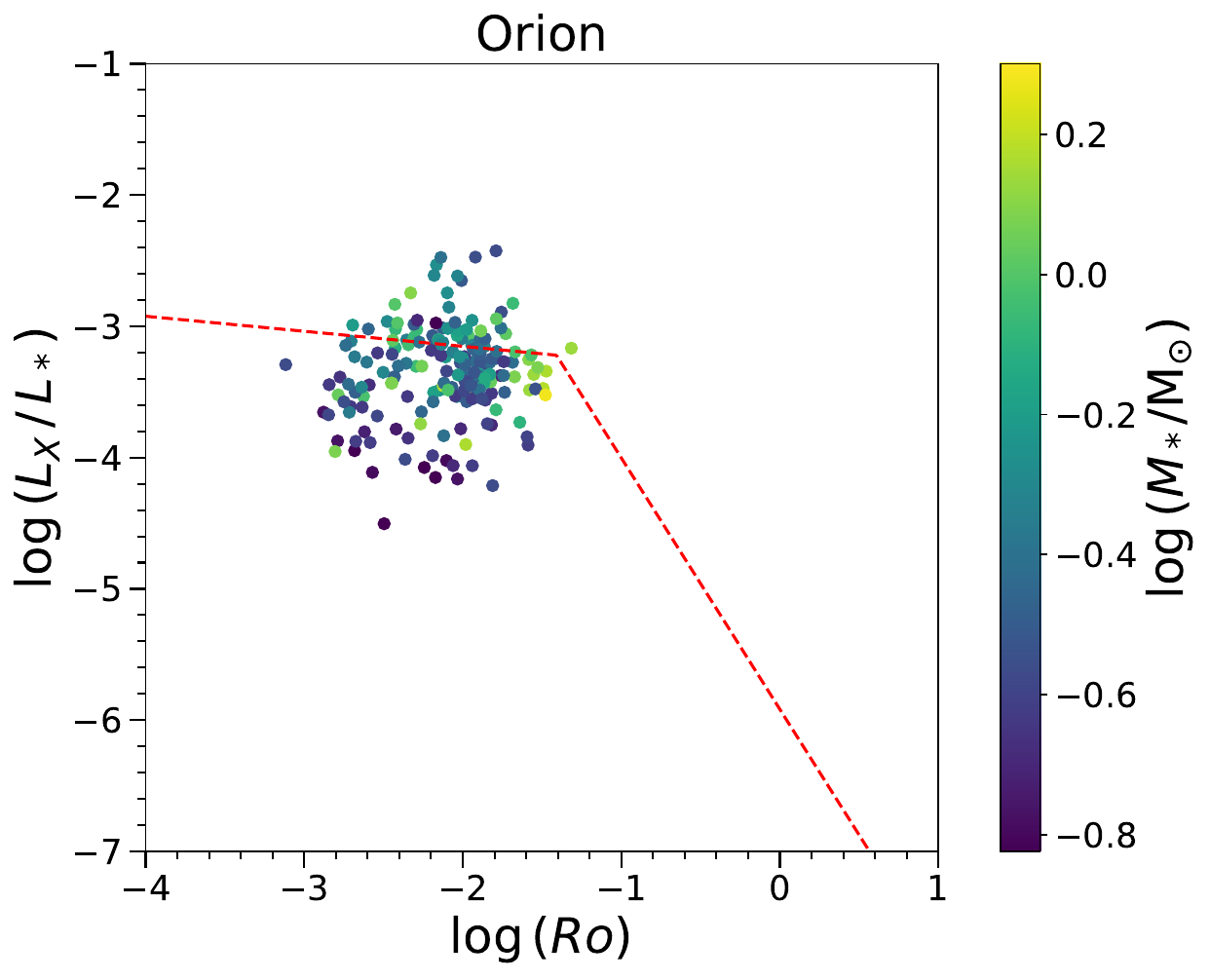}
\caption{Rotation--activity plots for our whole PMS sample (top left) and the individual clusters (as labelled). The data points, fractional X-ray luminosities and Rossby numbers, are derived from the observational data. The red dashed line indicates the dual power-law fit to the main-sequence sample in Fig.~\ref{fig: W11_init_ARRplot}, which is shown for comparison to the PMS data. The colour represents the stellar mass as indicated by the scale at the bottom right.}
\label{fig: PMS_init_ARRplot}

\end{figure*}

\subsection{Observed rotation--activity relations}
\label{sec: init ARR results}
In Fig.~\ref{fig: W11_init_ARRplot}, we display the rotation--activity plot for the open cluster sample of \cite{Wright_2011}. The plot highlights the mass of the individual stars and shows how the higher-mass stars define the slope of the unsaturated slope. 

The rotation--activity relation for main-sequence stars is defined by its shape: the Rossby number below which stars are saturated ($Ro_{\rm{sat}}$) and the slope of the saturated and unsaturated regimes. Such features can be described by a dual power-law relation \citep{Johnstone_2021,Magaudda_2020,Reiners_2014},
\begin{equation}
L_\textrm{X}/L_* = 
\begin{cases}
C_{\rm{unsat}} Ro^{\beta_{\rm{unsat}}} & Ro > Ro_{\rm{sat}} \\
C_{\rm{sat}} Ro^{\beta_{\rm{sat}}} & Ro \leq Ro_{\rm{sat}}, \\
\end{cases}
\label{Eq: ARR dual power-law}
\end{equation}
where $\beta_{\rm{unsat}}$ and $\beta_{\rm{sat}}$ are constants defining the gradient of the unsaturated and saturated regime, respectively, when plotting the rotation-activity-relation as $\log(L_{\rm X}/L_*)$ versus $\log(Ro)$. The constants $C_{\rm{unsat}}$ and $C_{\rm{sat}}$ can be chosen so that the values of both slopes at $Ro_{\rm{sat}}$ are an equivalent value of $\log{(L_\textrm{X}/L_*)}_{\rm{sat}}$. 

The choice of parameterisation/calculation of the convective turnover times can influence the best-fit of equation~(\ref{Eq: ARR dual power-law}). \cite{Magaudda_2020} highlights how the values for fitting the dual power-law change depending on the choice of how the convective turnover time is determined. Thus, to compare the rotation--activity relation of the PMS stars in this work to main-sequence stars, we must fit the dual power-law for main-sequence stars using our chosen parameterisation and interpolation of the convective turnover times.

We fit the dual power-law, equation~(\ref{Eq: ARR dual power-law}), to the main-sequence sample of \citet{Wright_2011}. We obtain the best-fit using the OLS(Y|X) method -- see appendix C of \cite{Johnstone_2021} who provide details of this procedure. We obtain values of $Ro_{\rm{sat}}$ = 0.039, $\log{(L_\textrm{X}/L_*)_{\rm{sat}}}$ = -3.179, $\beta_{\rm{sat}}$ = -0.141, and $\beta_{\rm{unsat}}$ = -2.220. This best-fit is plotted in Fig.~\ref{fig: W11_init_ARRplot}. The gradient we find for the unsaturated regime $\beta_{\rm{unsat}}$ agrees with the typical range in the literature of $\sim$ -2.4 to -1.8. The median value of $\log{(L_\textrm{X}/L_*)}$ for stars in the saturated regime is -3.15, which agrees with the constant fit value for the saturated regime from \cite{Wright_2011}.

Our value of $Ro_{\rm{sat}}$ agrees excellently with the value of 0.04 found in the analysis of \cite{Argiroffi_2016}, who use the models of \cite{Ventura_1998} and consider the convective turnover time as a function of the fraction of the convective envelope depth to stellar radius. Other methods used in the literature, see \citet{Magaudda_2020} and \citet{Wright_2011} for example, find a value of $\sim$0.14. The difference is to be expected given the convective turnover times in these works are around a factor of three times smaller than those calculated in the YaPSI models. This can be seen in Fig.~\ref{fig: TaucVsAge} where we have compared the empirical values for $\tau_c$ found in \cite{Wright_2018} to the YaPSI mass track values -- note one must compare the main-sequence values from the YaPSI models. This factor of three difference has been highlighted by \cite{Argiroffi_2016} where they find the difference is a result of the convective turnover times found in the empirical relations using the local convective turnover time while the YaPSI models and the models of \cite{Ventura_1998} use the global convective turnover time. The global convective turnover time is a radial integration of the convective velocity over the depth of the convective region. In contrast, the local convective turnover time is the ratio of the mixing length over the convective velocity calculated at a point -- conventionally half a mixing length above the bottom of the convective envelope. Such convention comes with the problem that this method is insufficient for fully convective stars where half a mixing length overshoots the radius of a star. This issue is discussed in detail in \cite{Landin_2023}.

\citet{Johnstone_2021}, who also use the YaPSI evolutionary models to infer convective turnover times for stars in the \cite{Wright_2011} sample, found a higher value of $Ro_{\rm{sat}}=0.0605$. Our value of $Ro_{\rm{sat}}$ differing from that found in \cite{Johnstone_2021} will be influenced by our removal of stars from the sample that fail our interpolation criteria. Additionally, we have individual stellar age estimates, while previous works use isochrone fitting based on average cluster age. Suppose stars are found to be younger than the cluster's average age from the literature. This can lead to much higher convective turnover times and thus lower Rossby numbers than previous calculations (see Fig.~\ref{fig: TaucVsAge}). Note that this is not the case if a star on the main-sequence is found to be much older than the average cluster age, as the convective turnover time is relatively invariable on the main sequence. Our approach to age estimation can lower the average Rossby number and, along with it, the fitted value of $Ro_{\rm{sat}}$

In Fig.~\ref{fig: PMS_init_ARRplot} we show the rotation--activity plots for the PMS cluster stars. The majority of our sample is in the saturated regime. PMS stars are mostly expected to be saturated as they have high convective turnover times. Also apparent is the much larger range in $(L_\textrm{X}/L_*)$ when compared to the main-sequence saturated regime. PMS stars can reach up to $\log(L_\textrm{X}/L_*) = -2$ and a considerable number have values as low as $\log(L_\textrm{X}/L_*)= -4$. The lower values of fractional X-ray luminosity are to be expected as while the PMS stars have high X-ray luminosities, many are still on Hayashi tracks in the H--R diagram, with bolometric luminosities considerably higher than what they will be once settled on the main-sequence. We can quantify the scatter of
$\log(L_\textrm{X}/L_*)$ for stars in the saturated regime using the MAD. In comparison to the main-sequence level of MAD[$\log(L_\textrm{X}/L_*)$] = 0.180, we find for our clusters of h~Persei, IC~348, NGC~2264 and Orion, MAD[$\log(L_\textrm{X}/L_*)$] = 0.266, 0.235. 0.196 and 0.197, respectively.

The cluster h~Persei is the oldest cluster and has the most evolved PMS stars within our sample. It is the only cluster that shows a clear mass stratification of the stars on the rotation--activity plot (higher-mass stars have typically higher Rossby numbers). This matches the analysis of \cite{Argiroffi_2016}. h~Persei stars also make up very few of the lower $\log(L_\textrm{X}/L_*)$ values in the whole sample, which is because these stars are older and have typically higher mass than the other clusters. Most h~Persei stars have completed their Hayashi track evolution (note the relative drop in bolometric luminosity by the time a star evolves onto the main-sequence decreases as stellar mass increases, as can be seen in Fig.~\ref{fig: LstarVsAge}). 


\section{Rotation--activity evolution model}
\label{sec: RA evo}

To model the evolution of stars on the rotation--activity plane, we need to evolve the rotation period/rate and X-ray luminosity. In this section, we outline the models used to evolve these parameters. The other stellar parameters can be determined at any age by linearly interpolating between points along the star's mass track, beginning from the star's observed age as determined from its H--R diagram position and as described in Sec.~\ref{sec: ARR interpolation}. The relevant parameters include the stellar radius, the radiative core radius, the radiative core mass, the total moment of inertia, the radiative core moment of inertia, the bolometric luminosity, and the convective turnover time.

\subsection{X-ray luminosity evolution}
\label{sec: X-ray evo}

\begin{figure}

\includegraphics[width=0.5\textwidth]{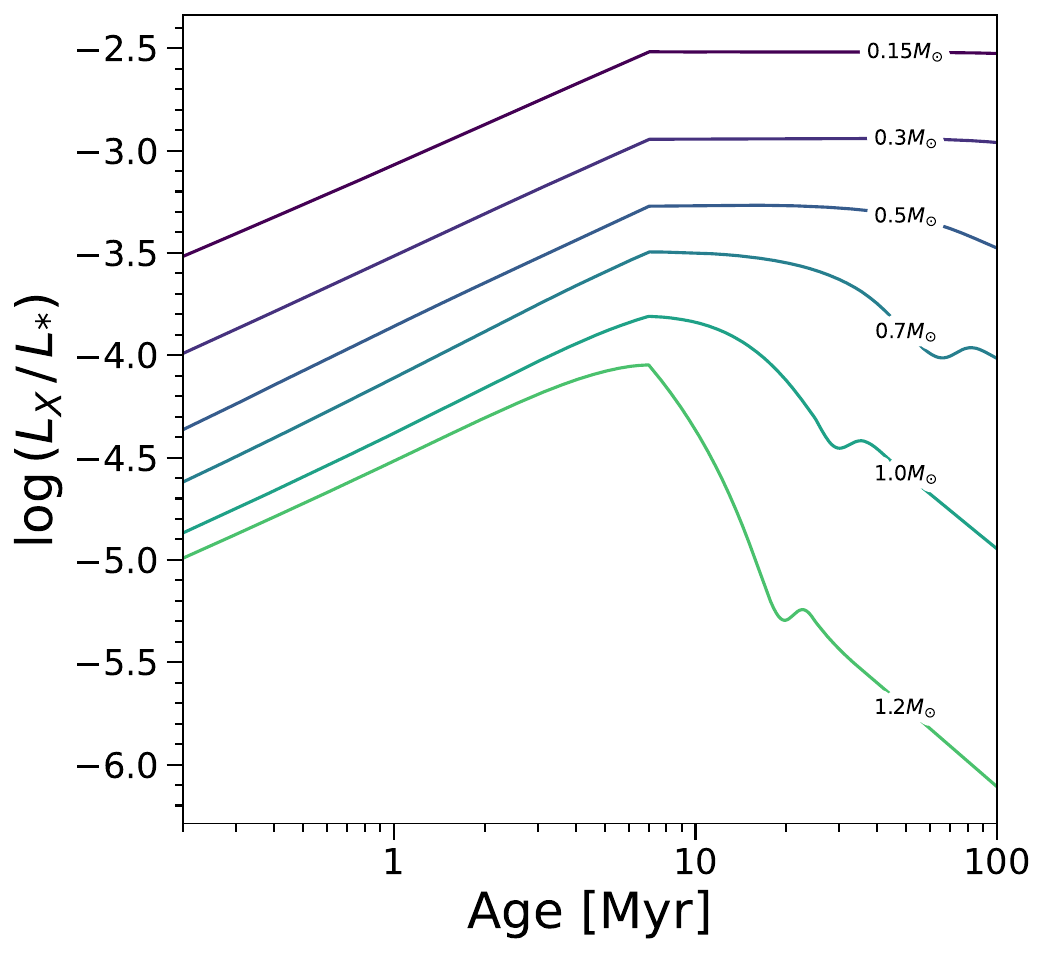}
\caption{Example fractional X-ray luminosity evolution versus age for stars of different mass as indicated. The mass values and their representative colours are the same as in Fig.~\ref{fig: LstarVsAge}. Only stars which retain an outer convective envelope are shown. The X-ray luminosity of each star is chosen to be $\log{L_{X}} = 29.5$ at 1\,Myr. This assumption is made for illustrative purposes to make the tracks distinguishable in this plot and thus the values of $\log{(L_{X}/L_{*})}$ in this plot should not be taken as an indication of the expected average value for a star of given mass; rather, the tracks show the shape/trend of $\log{(L_{X}/L_{*})}$ with increasing age.}
\label{fig: RxVsAge_examples}

\end{figure}

We evolve coronal X-ray luminosities $L_{\rm X}$ by adopting the observed trends with age determined from the median X-ray luminosities of stars in young clusters. The observational trends can be well described by power-laws. Thus, the evolution of $L_{\rm{X}}$ with age can be described by
\begin{equation}
L_{\text{X}} = L_{\text{X},0}\;t^{\beta_{\rm{X}}},
\end{equation}
where $L_{\text{X},0}$ is the initial X-ray luminosity and $\beta_{\rm{X}}$ is a constant that depends on age and the stellar mass range considered.

For our models, we adopt the median $L_{\rm{X}}$ relations with age determined by \citetalias{Getman_2022}. These relation are used to describe the behaviour for the first 25\,Myr of stellar evolution. For the first 7\,Myr, X-ray luminosities are found to be roughly constant,
\begin{equation}
\beta_{\rm{X}} = 0  \quad\quad (\textrm{for } t<7\,\textrm{Myr}).
\end{equation}
For ages 7--25\,Myr, the decay of median X-ray luminosity with age, encoded by the exponent $\beta_{\rm X}$, is dependent on stellar mass,
\begin{equation}
\beta_{\rm{X}} = 
\begin{cases}
-0.66 & M_* / \rm{M}_{\sun} \leq 0.75, \\
-0.46 & 0.75 < M_* / \rm{M}_{\sun} \leq 1, \\
-1.78 & 1. < M_* / \rm{M}_{\sun} \leq 2. \\
\end{cases}
\label{Eq: LXvt Getman} 
\end{equation}
Note that while the analysis \citetalias{Getman_2022} is for stars of mass $\ge0.75$\,M$_{\sun}$, we assume it is also applicable to lower mass stars (we discuss this assumption further in Section~\ref{sec: discussion}). 

For the age range of 25--100\,Myr the trends used are extrapolated from the median values of clusters in this age range listed in \cite{Gudel_2004},
\begin{equation}
\beta_{\rm{X}} = 
\begin{cases}
-0.66  & M_* / \rm{M}_{\sun} \leq 0.7, \\
-1.18  & 0.7 < M_* / \rm{M}_{\sun} \leq 1.1, \\
-1.26  & 1.1 < M_* / \rm{M}_{\sun} \leq 1.4, \\
\end{cases}
\label{Eq: LXvt Gudel}
\end{equation}
where the trend for stars of mass $\rm{M}_{\sun} \geq 1.4$ is not of concern as they have all become fully radiative by 25\,Myr.

To visualise how we expect the fractional X-ray luminosity to evolve with time, Fig.~\ref{fig: RxVsAge_examples} shows how $\log(L_\textrm{X}/L_*)$ evolves for selected stellar masses, with the assumption they have an identical X-ray luminosity of $\log{L_{X}}$ = 29.5 at 1\,Myr (this assumption is used here to compare the relative behaviour of stars of different mass). Apparent is the initial increasing fractional X-ray luminosities as the bolometric luminosities decrease during Hayashi track contraction, while the X-ray luminosities are kept constant until 7\,Myr. Once $L_{\rm{X}}$ starts decreasing, $L_\textrm{X}/L_*$ decreases or, for the lowest-mass stars, we have roughly constant values. The constant values can be explained by considering the evolution of X-ray and bolometric luminosity. The decreasing rate of $L_\textrm{X}$ for low-mass stars is effectively cancelled out by the decreasing rate of bolometric luminosity for the low-mass stars still on Hayashi tracks. During this evolutionary stage, the effective temperature is approximately constant as the star contracts. One can show for a fully convective star that $R_* \propto t^{-1/3}$  \citep{Batygin_2013,Palla_1993}. This gives us $L_* \propto t^{-2/3}$ from Stefan--Boltzmann's law. Hence, as $L_{\rm X}\propto t^{-0.66}$, then  $L_{\rm{X}}/L_{*}$ is approximately constant during Hayashi track evolution once the X-ray luminosity starts decreasing -- see Fig.~\ref{fig: LstarVsAge}.

\subsection{Stellar rotational evolution}
\label{sec: Rot Evo}
To track how the Rossby number evolves forwards and backwards with age for our sample of stars, we construct a model for the evolution of the stellar rotation rate. We use the basis of previous models -- as established in \cite{MacGregor_1991}, \cite{Gallet_2013} and \cite{Johnstone_2021} -- that assume, if a star is not fully convective, that the outer convective envelope and the central radiative core of a star can be treated as solid bodies with individual rotation rates. Angular momentum loss occurs from the convective zone due to thermally driven magnetised stellar wind. The core and convective envelope are coupled so that if there is a difference in rotation rate between the two, angular momentum is exchanged over a prescribed timescale such that their rotation rates will converge. The core is defined as the inner region of the star that does not include the outer convective envelope. Following similar notation to \cite{Johnstone_2021}, the two equations describing the change in the rotation rates of the core $\Omega_{\rm{core}}$ and outer envelope $\Omega_{*}$ are
\begin{equation}
\label{Eq: dOmega_core}
\frac{\mathrm{d}\Omega_{\rm{core}}}{\mathrm{d}t} = \frac{1}{I_{\rm{core}}}\left(- \tau_{\rm{ce}} - \tau_{\rm{cg}} - \Omega_{\rm{core}}\frac{\mathrm{d}I_{\rm{core}}}{\mathrm{d}t} \right),
\end{equation}
\begin{equation}
\label{Eq: dOmega_cz}
\frac{\mathrm{d}\Omega_{*}}{\mathrm{d}t} = \frac{1}{I_{\rm{cz}}}\left( \tau_{\rm{ce}} + \tau_{\rm{cg}} + \tau_{\rm{d}} + \tau_{\rm{w}} - \Omega_{*}\frac{\mathrm{d}I_{\rm{cz}}}{\mathrm{d}t} \right).
\end{equation}
The choice of notation allows the change in rotation rates to be represented in terms of the torques on the bodies and the rates of change of the moments of inertia of the core $I_{\rm{core}}$ and convective envelope $I_{\rm{cz}}$. 

Two torque terms describe the types of angular momentum exchange between the core and envelope. $\tau_{\rm{cg}}$ is the core-growth torque, representing when the core grows (or shrinks), mass is exchanged away from (or to) the convective envelope and, with it, the angular momentum. The angular momentum transferred is equivalent to the angular momentum carried by the spherical shell of the convective envelope surrounding the core that is exchanged to the core, which gives us the equation
\begin{equation}
\label{Eq: core mass growth torque}
\tau_{\rm{cg}} = -\frac{2}{3}R_{\rm{core}}^2\Omega_{*} \frac{\mathrm{d}M_{\rm{core}}}{\mathrm{d}t}.
\end{equation}
A negative $\tau_{\rm{cg}}$ represents a loss of angular momentum from the convective envelope to the core --  the case for when the core mass is increasing. If the core mass is decreasing ($\mathrm{d}M_{\rm{core}}/\mathrm{d}t < 0$) the core is exchanging angular momentum to the convective envelope from its outer shell of mass and so the $\Omega_{*}$ term in this equation becomes $\Omega_{\rm{core}}$.

The other torque term describing angular momentum exchange between the core and envelope is $\tau_{\rm{ce}}$. This term represents the coupling timescale of the core and the convective envelope such that the two rotation rates converge after a desired timescale. $\tau_{\rm{ce}}$ can be expressed as
\begin{equation}
\label{Eq: core-envelope torque}
\tau_{\rm{ce}} = \frac{\Delta J}{t_{\rm{ce}}}.
\end{equation}
$\Delta J$ is the angular momentum that must be exchanged to make the core and convective envelope rotation rates equivalent, where
\begin{equation}
\label{Eq: Angular momentum equator}
\Delta J = \frac{I_{\rm{cz}}I_{\rm{core}}}{I_{\rm{cz}}+I_{\rm{core}}} (\Omega_{\rm{core}} -\Omega_{*}).
\end{equation}
$t_{\rm{ce}}$ is the coupling timescale over which the angular momentum exchange occurs. We adopt the relation for this timescale with rotation rate found in \cite{Spada_2011} that is supported by findings that the coupling timescale is longer in slower initial rotators \citep{Bouvier_2008}. The fitted coupling timescale can be expressed as
\begin{equation}
\label{Eq: core-envelope coupling timescale}
t_{\rm{ce}} = t_0 \left(\frac{\beta \Omega_{\sun}}{|\Omega_{\rm{core}} -\Omega_{*}|} \right)^{\alpha}.
\end{equation}
We use the constants from \citet{Spada_2011}: $t_0 = 57.7$\,Myr, $\alpha=0.076$ and $\beta = 0.2$.

One can retrieve the simplified equation for a fully convective star from equation (\ref{Eq: dOmega_cz}) by removing the torques that involve core-envelope interaction ($\tau_{\rm{ce}}$ and $\tau_{\rm{cg}}$). Equating the convective zone moment of inertia to the total moment of inertia ($I_{\rm{cz}} = I$), the simplified equation becomes
\begin{equation}
\frac{\mathrm{d}\Omega_{*}}{\mathrm{d}t} = \frac{1}{I}\left(\tau_{\rm{d}} + \tau_{\rm{w}} - \Omega_{*}\frac{\mathrm{d}I}{\mathrm{d}t} \right).
\end{equation}

The next torque term $\tau_{\rm{w}}$ represents the wind torque that describes the angular momentum lost from a star due to a thermally driven, magnetised stellar wind. We use the model established in \cite{Bouvier_1997} that builds on the \cite{Kawaler_1988} wind model, which includes two different wind regimes for unsaturated and saturated stars such that,
\begin{equation}
\tau_w = 
\begin{cases}
 -K_{\rm{w}} \Omega_*^3 \left(\dfrac{R_*}{\rm{R}_{\sun}}\right)^{1/2} \left(\dfrac{M_*}{\rm{M}_{\sun}}\right)^{-1/2} 
 & (\Omega_* < \Omega_{\rm{sat}}), \\
 \\
 -K_{\rm{w}} \Omega_*\,\Omega_{\rm{sat}}^2 \left(\dfrac{R_*}{\rm{R}_{\sun}}\right)^{1/2} \left(\dfrac{M_*}{\rm{M}_{\sun}}\right)^{-1/2} 
 & (\Omega_* \geq \Omega_{\rm{sat}}). \\
 
\end{cases}
\label{Eq: wind torque tauw}
\end{equation}
$\Omega_{\rm{sat}}$ is the rotation rate that gives us the saturated Rossby number $\Omega_{\rm{sat}} = Ro_{\rm{sat}} / \tau_c$. $K_{\rm{w}}$ is a constant, a free parameter that we fit for different mass stars (see Appendix \ref{App: Kw calibration} for details)

The final torque term $\tau_d$ represents the disc-locking torque acting on a star. When a young star has an accretion disc and strong magnetic fields on the order of a kilogauss on average, the magnetic star-disc coupling influences the torque exerted on the star \citep{Koenigl_1991}. The model assumes, while an accretion disc is present, the disc torque cancels out any other torques acting on the convective envelope such that the envelope's rotation rate stays constant ($\mathrm{d}\Omega_{*}/\mathrm{d}t = 0$). This implementation in rotation models matches stars with accretion discs having, on average, lower rotation periods than disc-free stars \citep{Serna_2021, Venuti_2017}. 

\begin{figure}

 \includegraphics[width=\columnwidth]{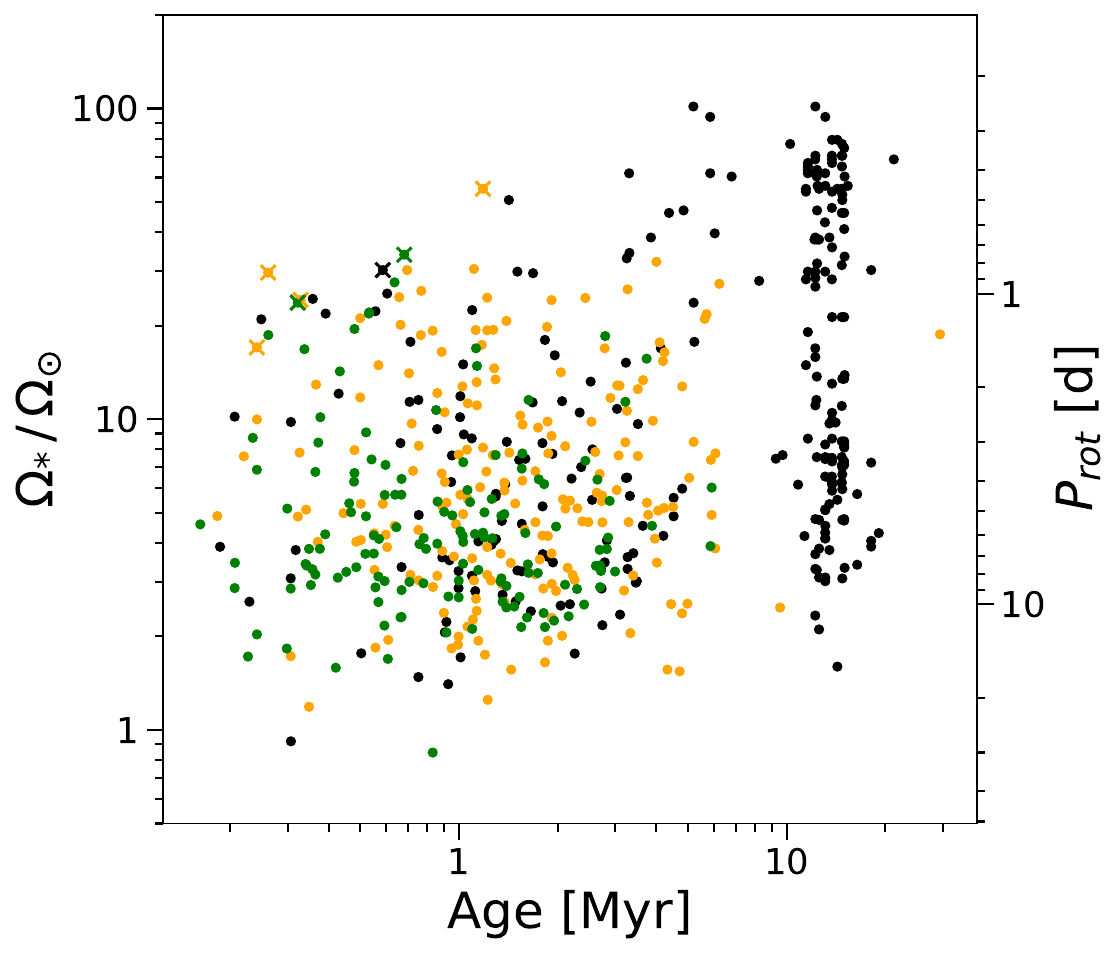}
 \caption{Distribution of stellar rotation rate/period versus age for the observational sample of PMS stars, for which we subsequently model their rotational evolution. Green points are stars with observed signatures of a circumstellar disc; orange points are stars without discs; and black points with indeterminate disc status. Crosses indicate stars whose rotation rate always exceeds the break-up rotation rate when their rotational evolution is modelled.}
 \label{fig: init ProtvAge}
\end{figure}
%

\begin{figure*}
 \includegraphics[scale=0.5]{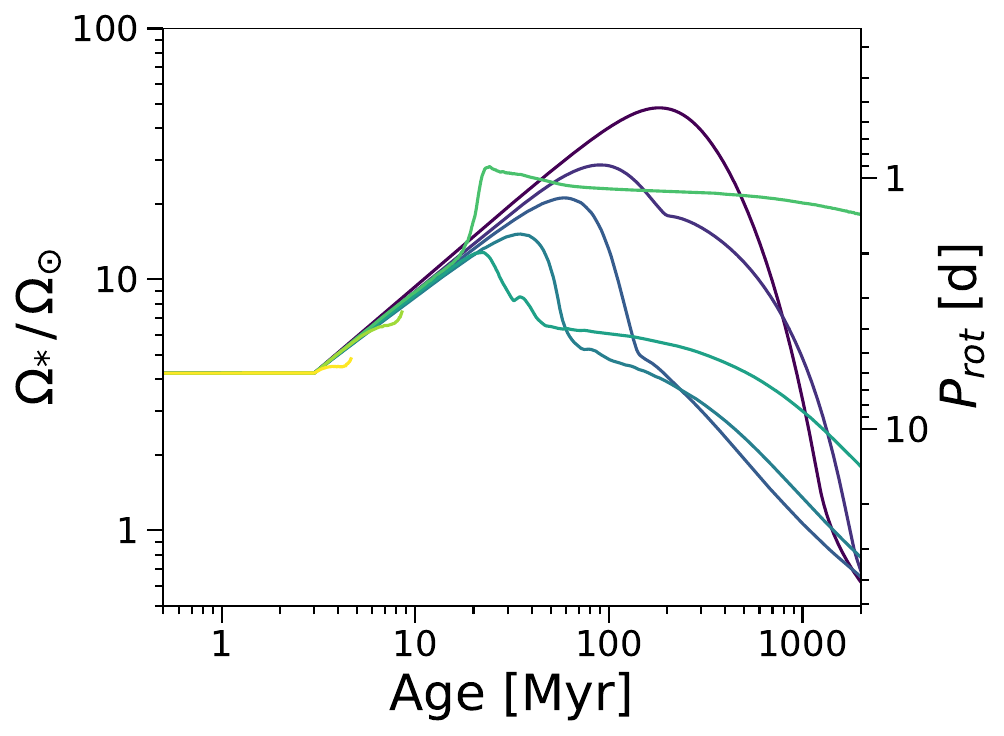}
 \includegraphics[scale=0.5]{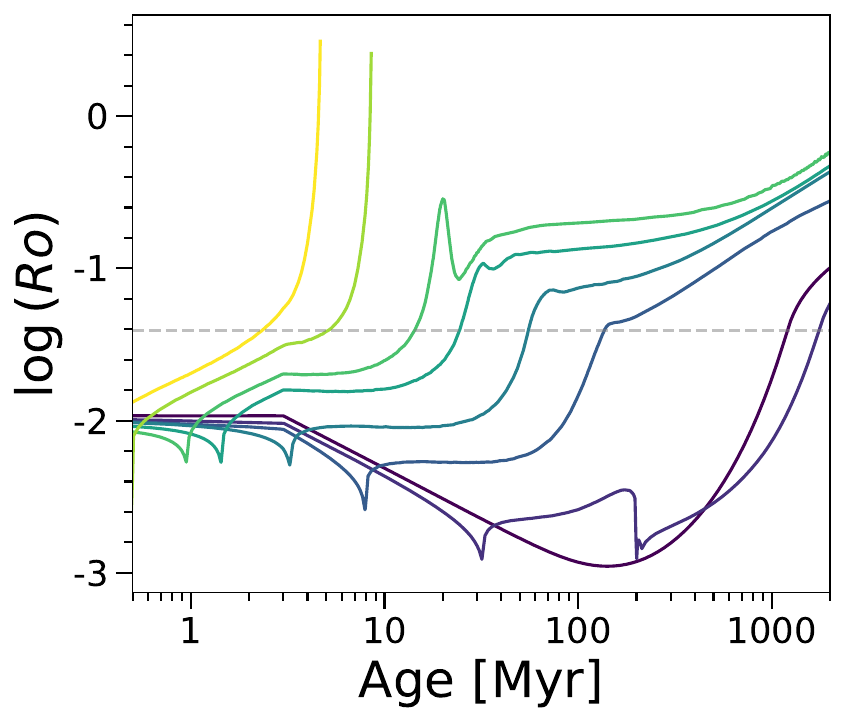}
 \caption{Illustrative examples of rotation rate (left) and the corresponding Rossby number (right) evolution for stars of different mass where the disc lifetime is 3\,Myr and the initial rotation period is 6\,d. The tracks stop if a star no longer has an outer convective envelope. Each line on the plots corresponds to a particular stellar mass, ranging from 0.15 to 2$\,{\rm M}_\odot$, with colours matching those used in Fig.~\ref{fig: LstarVsAge}. The dashed horizontal line marks the saturated Rossby number $Ro_{\rm{sat}}$, below which a star is in the saturated regime of the rotation-activity relation.} 
 \label{fig: Omega&Ro evo}

\end{figure*}

In the literature that our rotational evolution model is founded on, model rotation rates of stars are evolved forward in time assuming initial rotation rates, often based on the rotation rate distributions of young PMS clusters like the ONC. The timescale over this initial period where a disc is present and a disc--locking torque is active is also assumed and, in some cases, this disc--locking timescale has a dependence on the initial rotation rate \citep{Gallet_2013,Johnstone_2021}. 

In our model, we consider the rotation rate evolution of each star, beginning from an initial age of 0.15\,Myr, and determine the initial rotation rate that ensures that we match the observed rotation rate at the observed age of the star. However, from the observations, we only know the rotation rate of a star's outer convective zone. A rotation rate for the star's radiative core must also be determined. We achieve this using the fact at our initial model age a radiative core has not formed for any star in our considered mass range ($0.15 - 2.0$\,M$_{\sun}$), and therefore once the core forms we can set its initial rotation rate to match that of the outer convective envelope.

We consider the evolution of our entire PMS sample of stars up to an age of 100\,Myr. We run our simulations, with the entire sample of stars, 100 times. We assign an individual disc lifetime to each star that varies with each simulation run. Each disc lifetime is randomly chosen, but we must ensure all the random disc lifetimes give us an accurate distribution of stars with discs at any given age. To determine the disc lifetimes we use a Monte--Carlo sampling method. This requires knowledge of the cumulative distribution function describing disc lifetimes $P_{\rm{disc}}(t)$. This can be derived from the function describing the fraction of disc-bearing stars at a given age $f_{\rm{disc}}(t)$. This function has been fit to observations of disc fractions using the form of an exponential decay relation \cite[see for example][]{Briceno_2019, Mamajek_2009, Ribas_2014}. Assuming all stars initially have a disc, then
\begin{equation}
f_{\rm{disc}}(t) = e^{-t/\tau_{\rm{disc}}}.
\label{Eq: Disc fraction function}
\end{equation}
$\tau_{\rm{disc}}$ is the average disc lifetime, which, from fits described in the cited studies, a value of 3\,Myr is appropriate. It is also consistent with the typical disc lifetimes determined from rotation rate distributions \citep{Gallet_2015,Serna_2021}. We do not attempt to assign average disc lifetime as a function of stellar mass. However, some evidence does point to higher mass stars having typically shorter disc lifetimes \citep{Pfalzner_2022,Ribas_2015,Venuti_2024}. If we restrict the possible disc lifetimes to an age range between $t_{\rm{min}}$ and $t_{\rm{max}}$, then the cumulative distribution function is given by the integral of our disc fraction function
\begin{equation}
P_{\rm{disc}}(t) = \int^{t}_{t_{\rm{min}}} Af_{\rm{disc}}(t') \, dt'. 
\end{equation}
The constant A takes a value to ensure the function is normalised such that
\begin{equation}
P_{\rm{disc}}(t = t_{\rm{max}}) = \int^{t_{\rm{max}}}_{t_{\rm{min}}} Ae^{-t'/\tau_{\rm{disc}}}\, dt' = 1. 
\end{equation}
One can now solve for a random disc lifetime $t_{\rm{disc}}$ by assigning the cumulative distribution function a random number $\xi$ in the range of $0-1$. When there is no restriction on disc lifetime (range of $0-\infty$), the disc lifetime is given by
\begin{equation}
t_{\rm{disc}} = - \tau_{\rm{disc}} \ln{(1-\xi)}.      
\end{equation}
If a star in our observed sample is disc-bearing, then the minimum disc lifetime is the observed age of the star. If a star is observed to be disc-less, the maximum disc lifetime is the observed age of the star. We implement a general upper limit for disc lifetime of $t_{\rm{max}}$ = 15\,Myr, by which age the disc fraction is very small, to prevent random disc lifetimes from being unphysically old. A general lower limit is also placed on the disc lifetime at the initial age of our model of 0.15\,Myr, ensuring stars start the simulation in a disc-locked state.

Furthermore, if the random disc lifetime leads to a rotation rate that exceeds break-up (see equation \ref{Eq: breakup velocity}) before its observed age, the disc lifetime is discarded, and another is chosen. This follows the assumption that a star cannot reach the break-up rotation rate before its observed age.

If a star reaches an age where it no longer has an outer convective envelope, we can no longer apply our model. At this point, we cannot position the star in the rotation--activity plane, and its rotational evolution is suspected to be no longer controlled by a magnetised stellar wind \citep{Aerts_2019,Wolff_1997}.


\section{Results}

\subsection{Stellar rotational evolution }
\label{sec: rot eve results}

\begin{figure*}

 \includegraphics[scale=0.4]{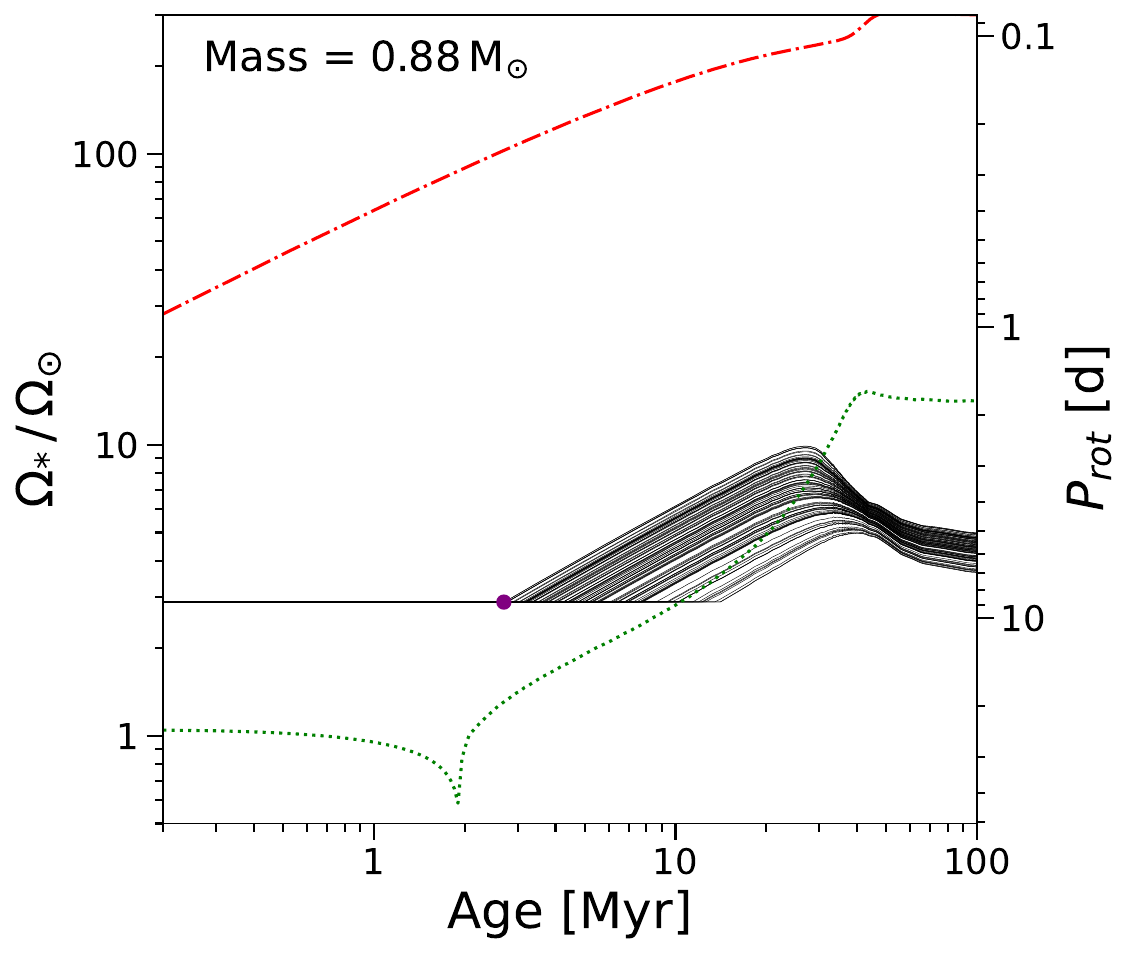}
 \includegraphics[scale=0.4]{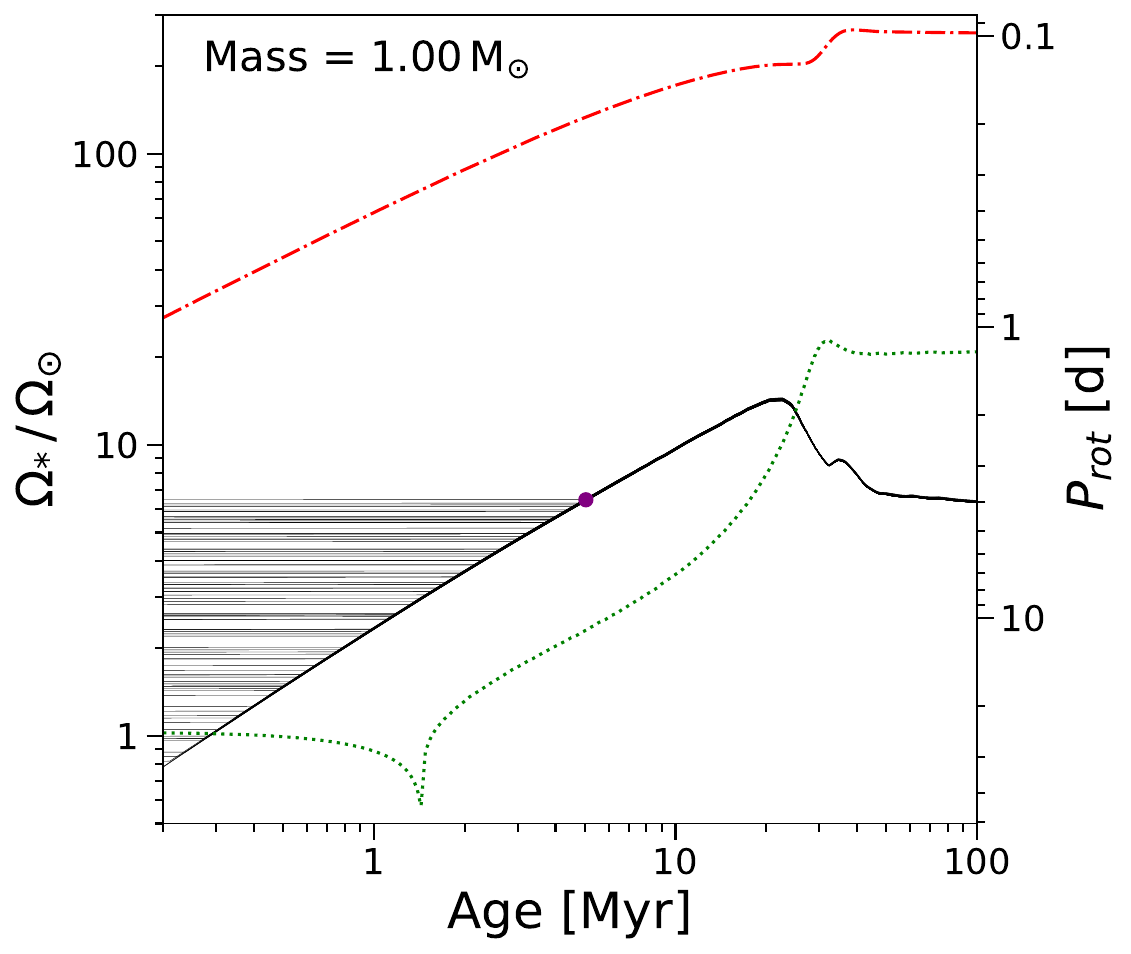}
 \includegraphics[scale=0.4]{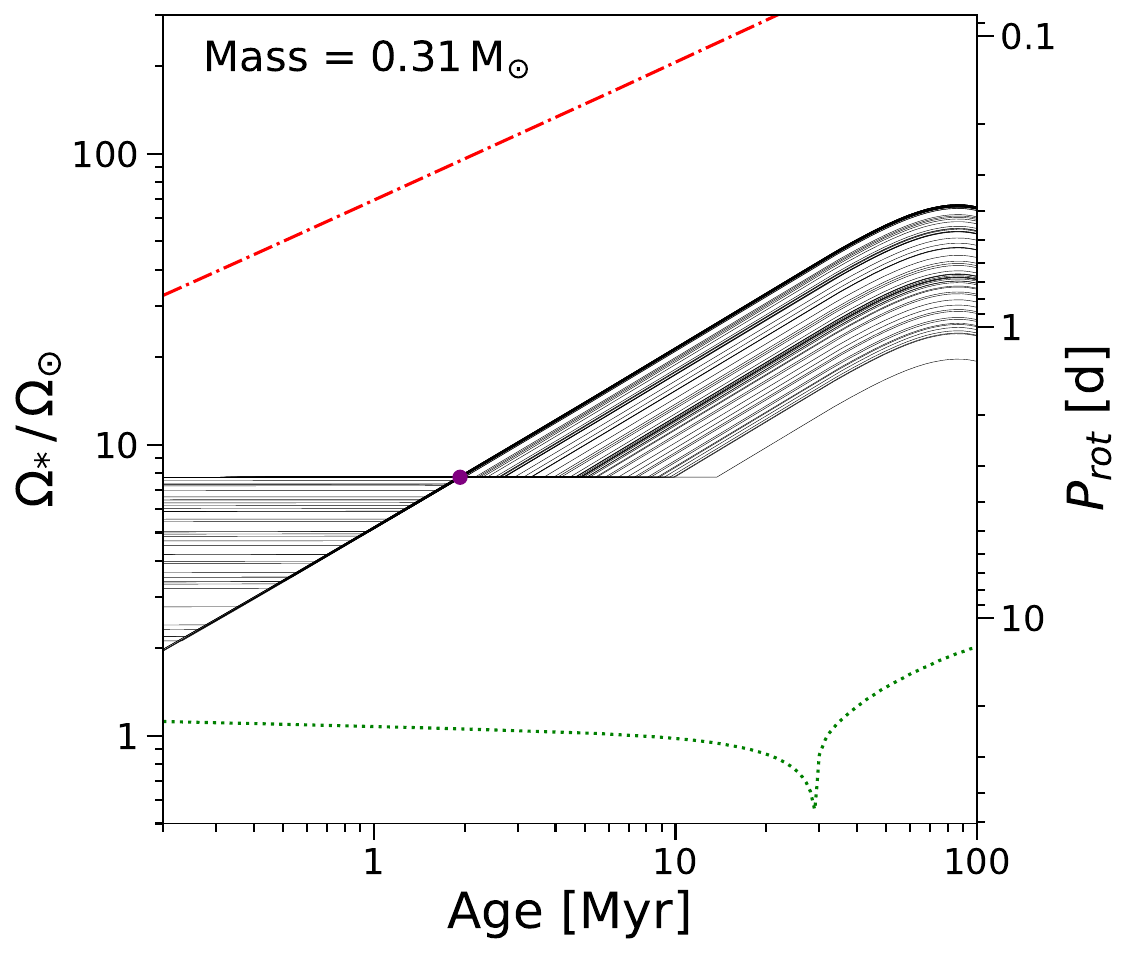}
 \includegraphics[scale=0.4]{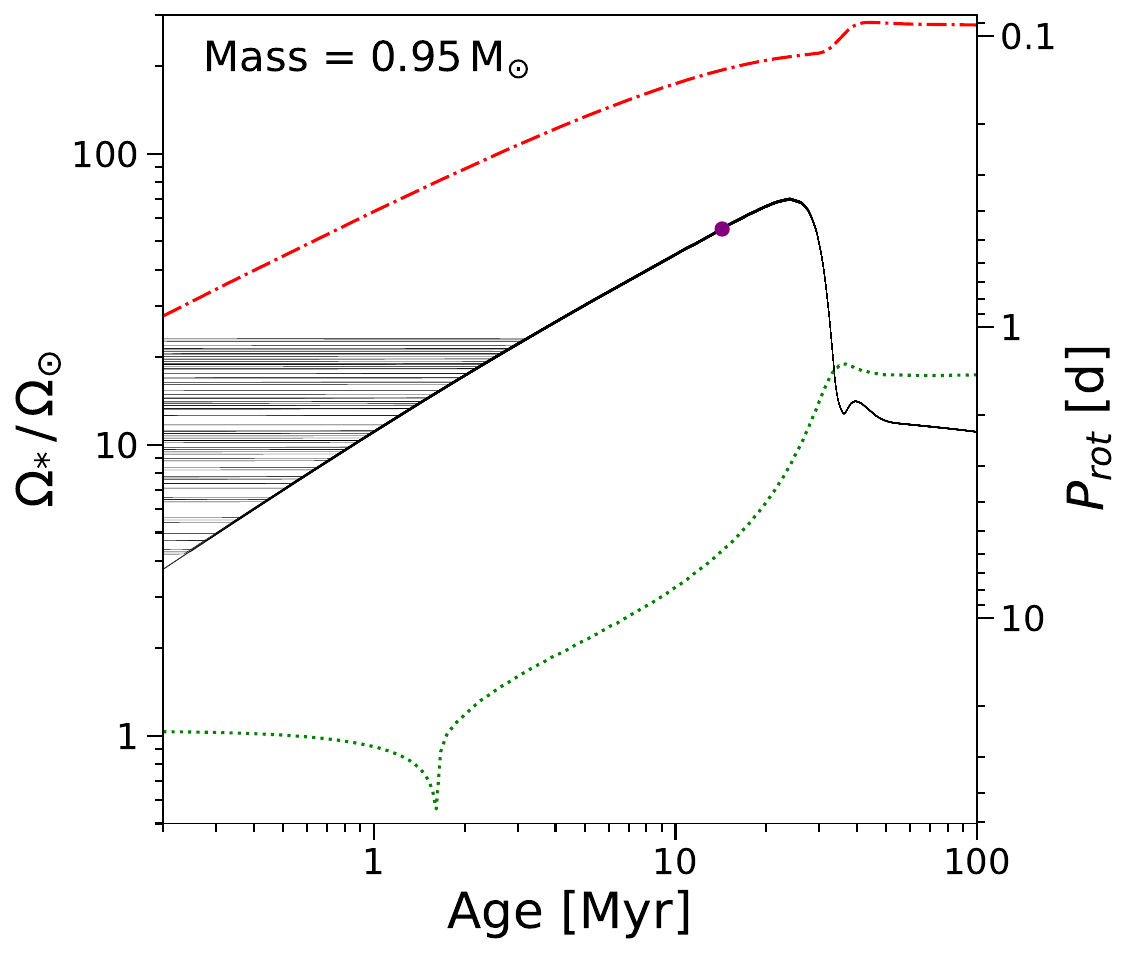}
 \caption{Plots of rotation rate evolution tracks of all 100 simulations for four individual PMS stars, chosen as illustrative examples. The purple circles indicate the observed values of the rotation rate. The red dot-dash line indicates the break-up rotation rate. The green dotted line denotes the separation between the saturated and unsaturated regime, where stars above this line fall in the saturated regime of the rotation--activity relation. Top left and right show examples of stars observed with and without a disc, respectively. Bottom left shows the case of a star with an undetermined disc status. Bottom right shows the case of a star with an undetermined disc status where the maximum disc lifetime is limited by the break-up rotation rate.}
 \label{fig: 4 star Rotevo examples}

\end{figure*}
%

\begin{figure*}
 \includegraphics[scale=0.4]{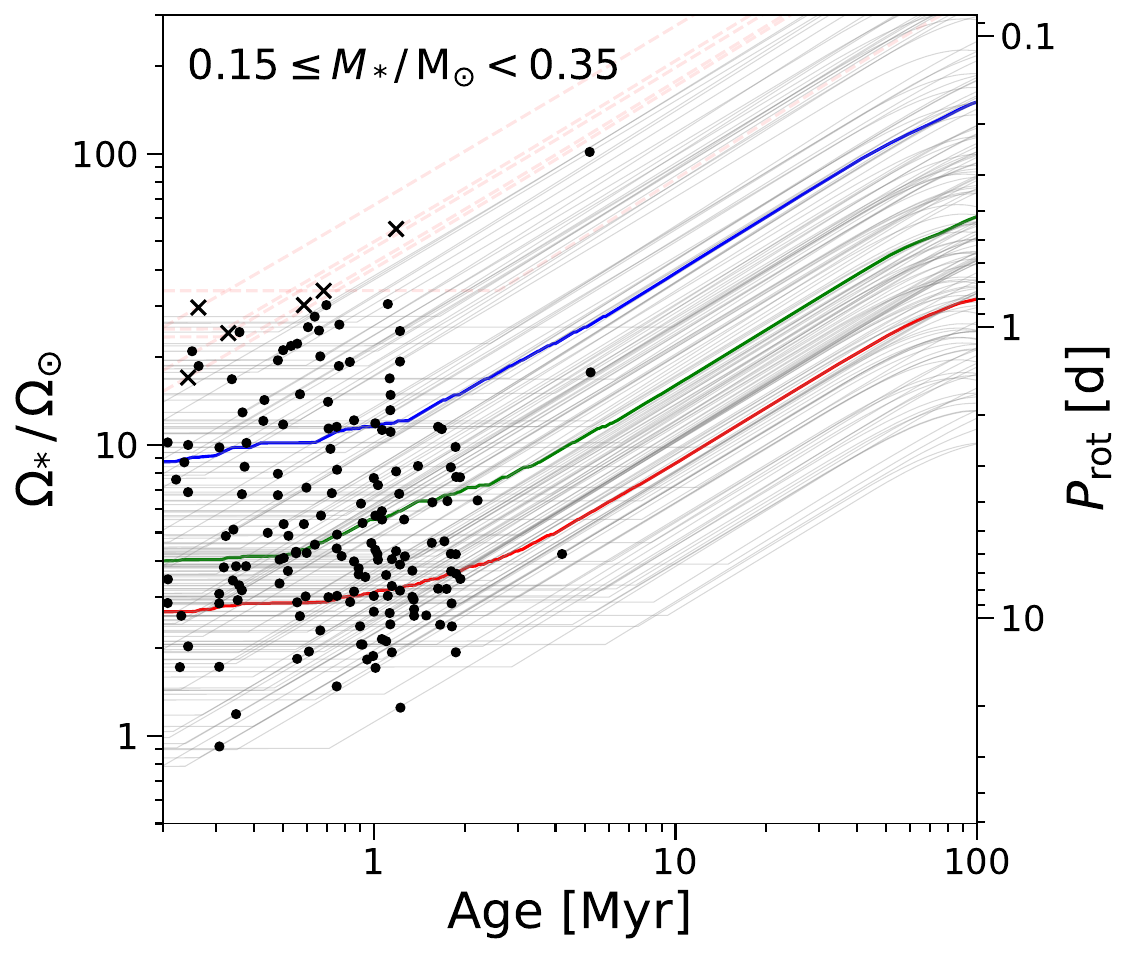}
 \includegraphics[scale=0.4]{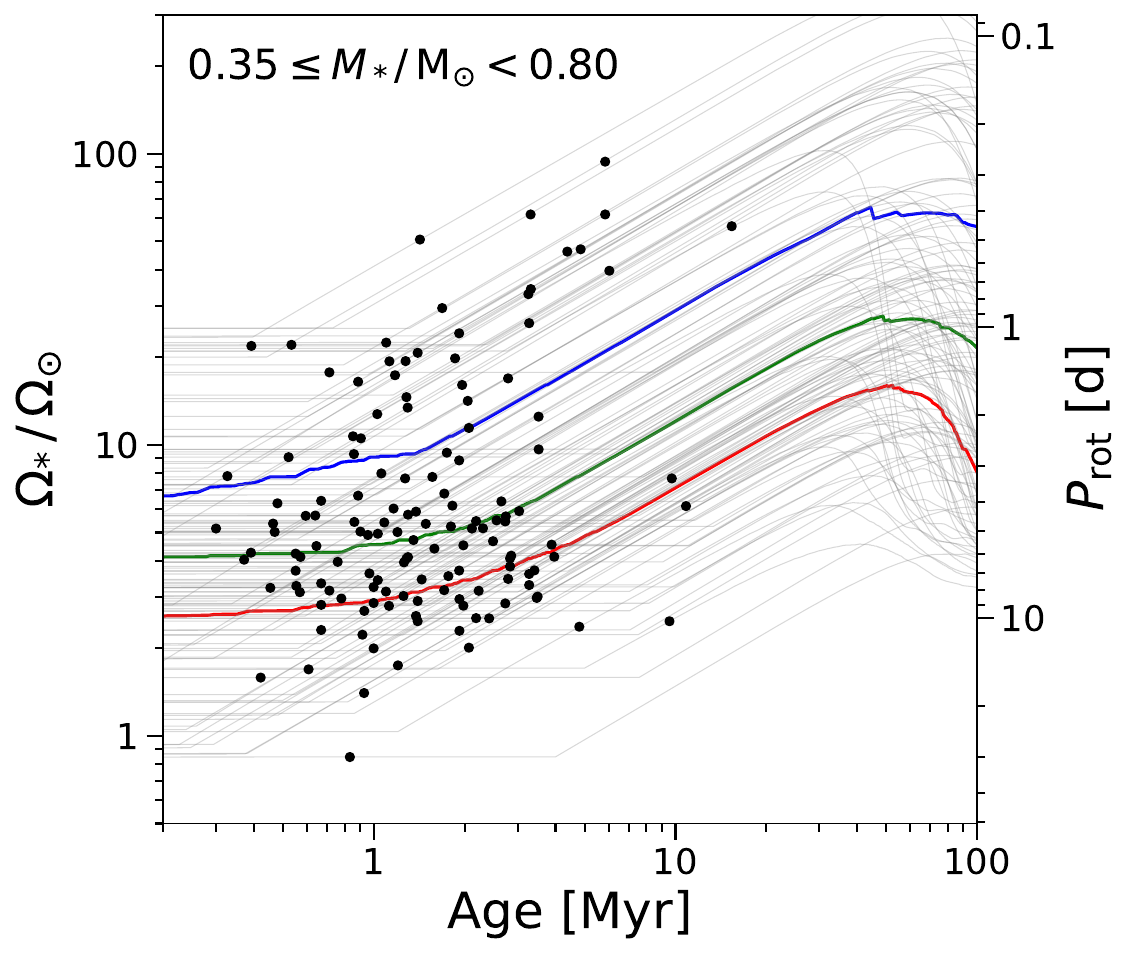}
 \includegraphics[scale=0.4]{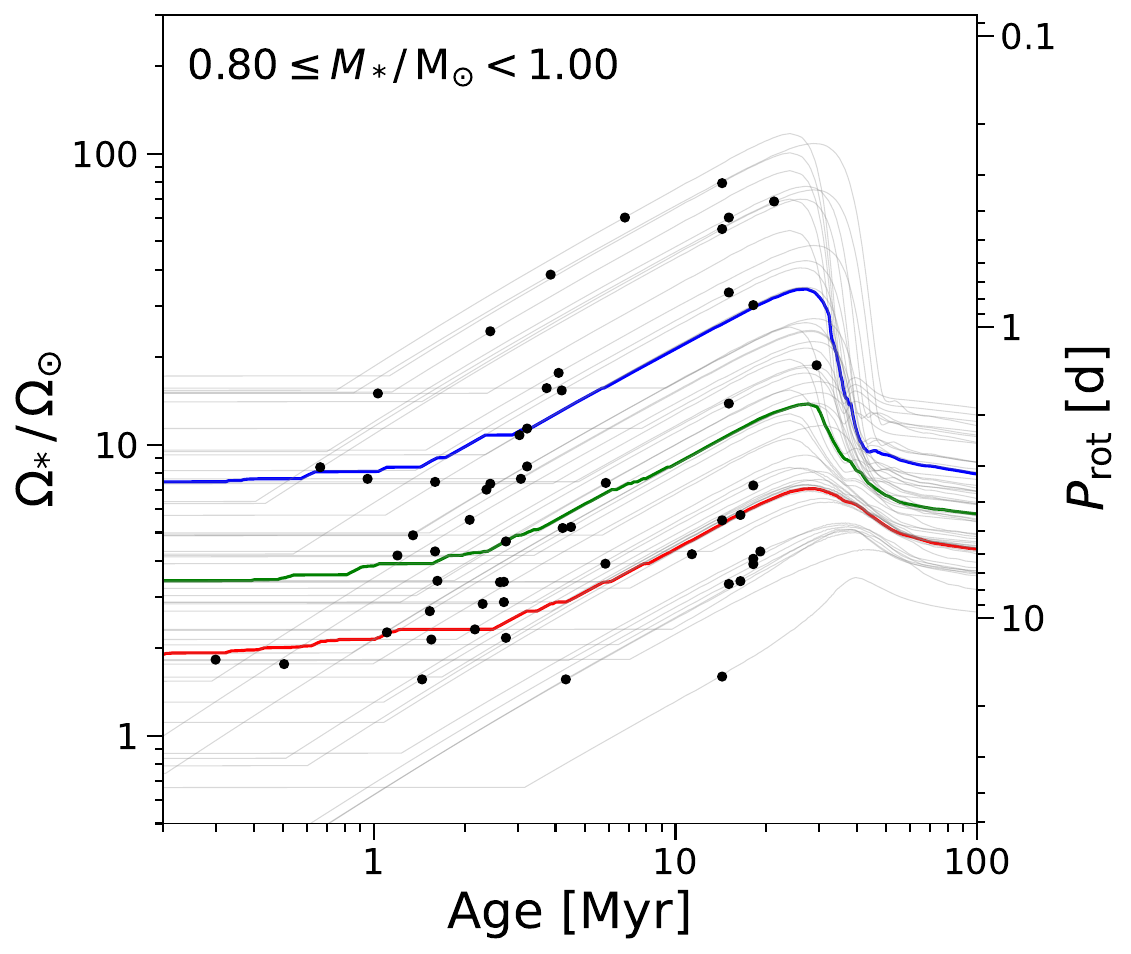}
 \includegraphics[scale=0.4]{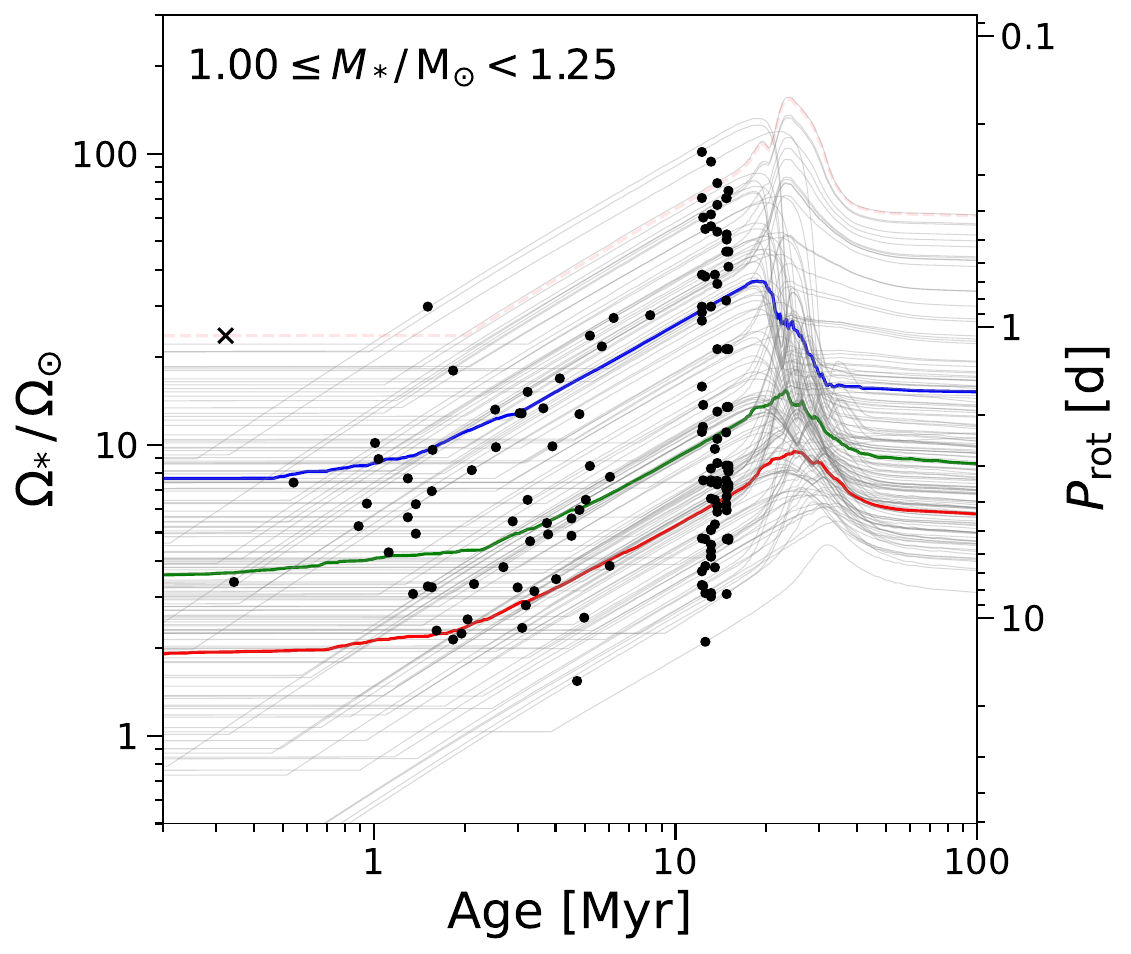}
 \caption{Rotation rate tracks of the PMS star sample for one simulation, separated into different mass ranges as labelled. Stars only up to the mass limit where stars become fully radiative are shown. Thin grey lines represent individual stellar rotational evolution tracks, with those that are dashed red indicating that the star's break-up rotation rate has been exceeded at some age. The thick red, green and blue solid lines represent the 25th, 50th and 75th percentiles of the rotation rate distribution, respectively. Black circles indicate the initial observed rotational rates of the PMS stars. Crosses indicate stars whose rotation rate always exceeds the break-up rotation rate when modelled.}
 \label{fig: massbinned rot evo}
\end{figure*}
%

\begin{figure}

\includegraphics[width=\columnwidth]{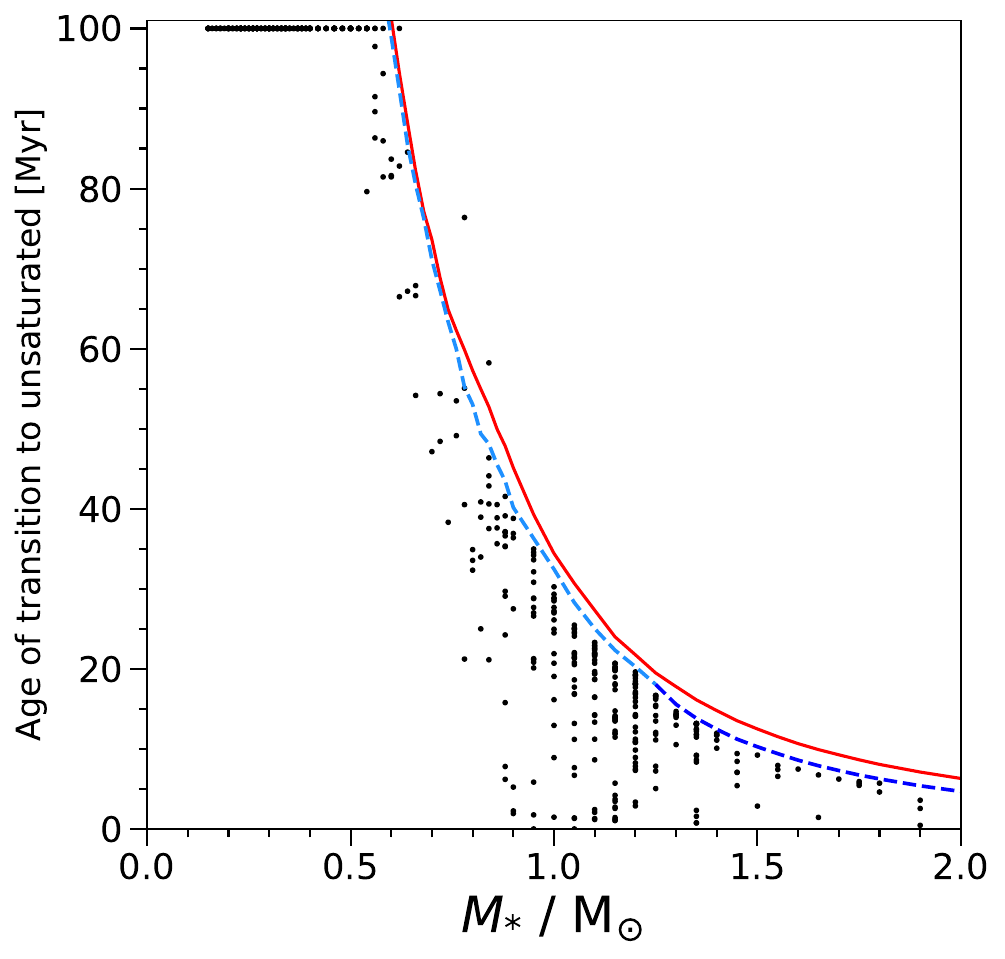}
\caption{Age at which stars leave the saturated regime during our 100$\,{\rm Myr}$ simulation as a function of stellar mass, for one simulation run. Points at 100\,Myr indicate stars that are still saturated by the end of the simulation. The dark blue dashed line indicates the age at which a star loses its outer convective envelope. (The Rossby number of a star is undefinable beyond this age.) The light blue dashed line indicates the age a star reaches its minimum convective turnover time during the PMS. The red line indicates the ZAMS age. It is notable that PMS stars (above 0.7\,M$_{\sun}$) have evolved from the saturated to the unsaturated regime of the rotation--activity relation before they reach the ZAMS. There are only two exceptions, the two stars above the red line, that remain exceptions for all 100 simulation runs.}
\label{fig: unsaturation ages}
\end{figure}
%

\begin{figure}

 \includegraphics[width=\columnwidth]{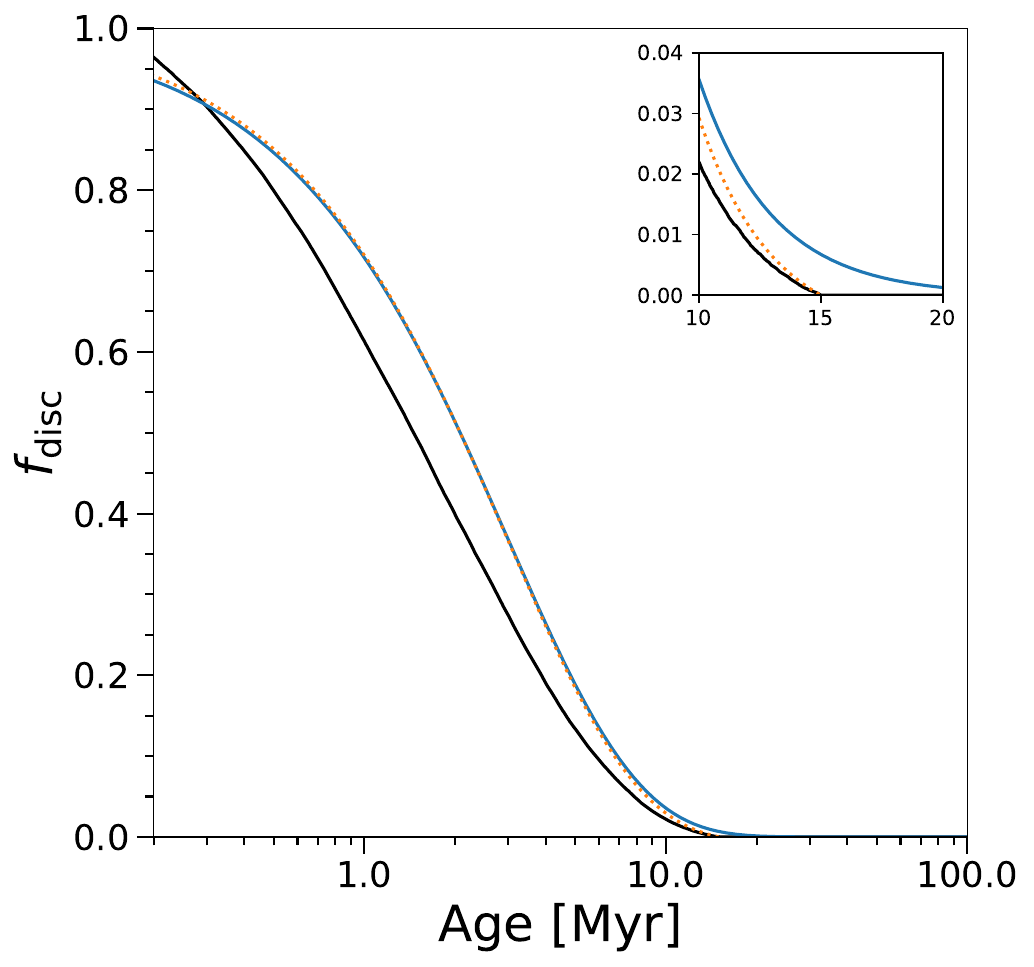}
 \caption{Fraction of stars in the sample with discs versus age. The blue line represents the expected distribution based on equation~(\ref{Eq: Disc fraction function}) for $t_{\rm{disc}}$ = 3\,Myr. The dotted orange line is the expected disc fraction evolution used in our models, which impose a maximum (15$\,{\rm Myr}$) and minimum (0.15$\,{\rm Myr}$) disc lifetime -- see Section \ref{sec: Rot Evo}. The black line represents the disc fraction averaged across all 100 of our simulation runs. Top right of the plot shows a zoomed-in portion of the plot from 10--20\,Myr. }
 \label{fig: disc fractions}

\end{figure}

The observed rotation rate distribution with age for our PMS star sample is displayed in Fig.~\ref{fig: init ProtvAge}. Our sample shows agreement with disc locking, as stars with discs show a lower average rotation rate than the general distribution, including those without discs or unknown disc status. The maximum observed rotation rate increases with age as stars have become free of discs and spin up during contraction. 

Fig.~\ref{fig: Omega&Ro evo} is an illustrative example of our rotational evolution model and the corresponding values of Rossby numbers for different stellar masses. For this example, we have chosen initial rotation periods of 6\,d and disc lifetimes of 3\,Myr. The example rotation tracks show key features, such as higher-mass stars having a short convective zone lifetime and, thus, having little time to evolve beyond spin-up from contraction (or not at all if the disc lifetime is long enough). Stars near the fully radiative limit ($M_* = 1.2$\,M${_{\sun}}$) do not have a significant spin-down torque and stay fast rotating on long timescales. The lowest mass stars have not yet finished contracting after 100\,Myr, so their rotation rates are still increasing. Only at late main-sequence ages does the rotation rate of the solar mass stars typically exceed that of lower mass stars. The typical age at which a star reaches the unsaturated regime of the rotation--activity relation (see Fig.~\ref{fig: Omega&Ro evo}, right panel) increases with stellar mass, with the lowest mass stars remaining saturated for 1\,Gyr. For partially convective stars, the switch from saturated to unsaturated Rossby numbers occurs before the ZAMS due to the rapid change in convective turnover time as the radiative core develops. For stars on the main sequence (that are partially convective), the Rossby number at a given age for a star increases with stellar mass for the chosen initial conditions. The Rossby number of a fully convective star decreases significantly as the star spins up and is dictated less by the change in convective turnover time and more by the timescale of the spin-up and spin-down of the star.

We ran our rotational evolution model for the PMS stars, simulating 100 times the evolution backwards and forwards in time for each PMS star in our dataset. In Fig.~\ref{fig: 4 star Rotevo examples}, we show four stars and how their rotation rates vary for the 100 simulations. The plotted stars have been selected as examples to highlight the different disc scenarios: observed with and without a disc, an undetermined observed disc status, and a limitation on disc lifetime due to break-up rotation rate.  We display the rotation rate evolution of the PMS stars for the first 100\,Myr in Fig.~\ref{fig: massbinned rot evo}, showing individual tracks for each star over one of the 100 simulations,  chosen at random for illustrative purposes. Plots of the rotational evolution are separated into different mass ranges (only up to the mass limit where stars become fully radiative) and show percentile markers of the rotational velocity distribution for each mass range of stars. Our models correctly predict no or few stars with rotation periods > 10\,d by around 30\,Myr, which agrees with period observations of NGC~2547 \citep{Jeffries_2011,Godoy-Rivera_2021}. In our model, disc lifetimes are restricted to avoid exceeding break-up rotation rates before the observed age of a star, which for our model (beginning at 0.15\,Myr) requires 85 of our stars to have lost their disc in the past such that they previously spun slower. Two of these 85 stars currently have discs, so the rotation rate will always exceed the break-up rotation rate at younger than observed ages. An additional five stars always exceed the break-up rotation rate at some point older than the age we determined from their H-R diagram position. These five stars are all less than $0.35\,\rm{M}_{\sun}$, and all have observed rotation rates close to break-up, so they lose discs early but eventually exceed break-up rotation rate during their long spin-up period as they continue their PMS contraction. Any time a star in our sample exceeds the break-up rotation rate, we exclude it from our subsequent analysis.

The rotational evolution behaviour in Fig.~\ref{fig: massbinned rot evo} is similar to that shown in the illustrative example in Fig.~\ref{fig: Omega&Ro evo}. Specifically, the typical rotation rate of lower-mass stars increases as they spin up, reaching a peak rotation rate larger than for higher-mass stars. Furthermore, stars that become partially convective have stopped spinning up by 100\,Myr, meaning the median rotation rate after 100\,Myr is considerably higher for lower mass stars. The spread in rotation rates for higher-mass stars after 100\,Myr is also smaller, which is driven in part due to rotational convergence being faster and reached within 100\,Myr for higher-mass stars \citep{Barnes_2003,Boyle_2023}. 

Considering the Rossby number evolution of the stars, we find that stars of mass $M_* \geq 0.7$\,M${_{\sun}}$ have left the saturated regime by 100\,Myr. This agrees with the expectation from the average rotator models of Fig.~\ref{fig: Omega&Ro evo}, where Rossby numbers dramatically increase as stars develop radiative cores. In Fig.~\ref{fig: unsaturation ages}, we display the ages at which stars leave the saturated regime from one simulation randomly selected from our 100 simulations (the following results hold for all of our simulation runs). The typical upper age limit of stars leaving the unsaturated regime correlates strongly with the ZAMS age for a given stellar mass, with only two stars above 0.7\,M$_{\sun}$ remaining saturated beyond the ZAMS. After $\sim$80\,Myr, no star above 0.7\,M$_{\sun}$ is saturated. All stars bar one with mass $M_* < 0.5$\,M${_{\sun}}$ remain in the saturated regime for the whole 100\,Myr length of our simulations, all of which are still contracting.

The disc lifetimes are capped for many stars in the sample to prevent the rotation rates from exceeding break-up before their observed age - an example of this is shown in the bottom right plot of Fig.~\ref{fig: 4 star Rotevo examples}. This cap contributes to our disc fraction being slightly lower than expected from the disc fraction function described by equation~(\ref{Eq: Disc fraction function}), which is visualised in Fig.~\ref{fig: disc fractions}. The need for stars to require shorter than expected disc lifetimes could arise if the age of a star has been estimated to be too young. Spurious, published rotation periods may also contribute to incorrectly identified rapid rotators (such as when a period alias has been reported in the literature). Alternatively, the problem of required short disc lifetimes may be resolved by considering a mechanism of moderate angular momentum loss during the star-disc interaction phase, as suggested by \cite{Lamm_2005}. Stellar winds driven by accretion during the disc phase could result in this angular momentum loss \citep[see][]{Matt_2005}, which has been implemented into recent models \citep[e.g.][]{Gallet_2019,Gehrig_2023}. This mechanism results in many rapid rotators having lower initial rotation rates at young ages $<$1\,Myr, where break-up rotation periods reach 1\,d when PMS stars have much larger radii -- see Fig.~\ref{fig: OmegabuVsAge}. As a result, this allows a star to have longer disc lifetimes without exceeding its break-up rotation rate.

\subsection{Rotation--activity evolution}
\label{sec: ARR results} 

\begin{figure*}
\includegraphics[scale=0.33]{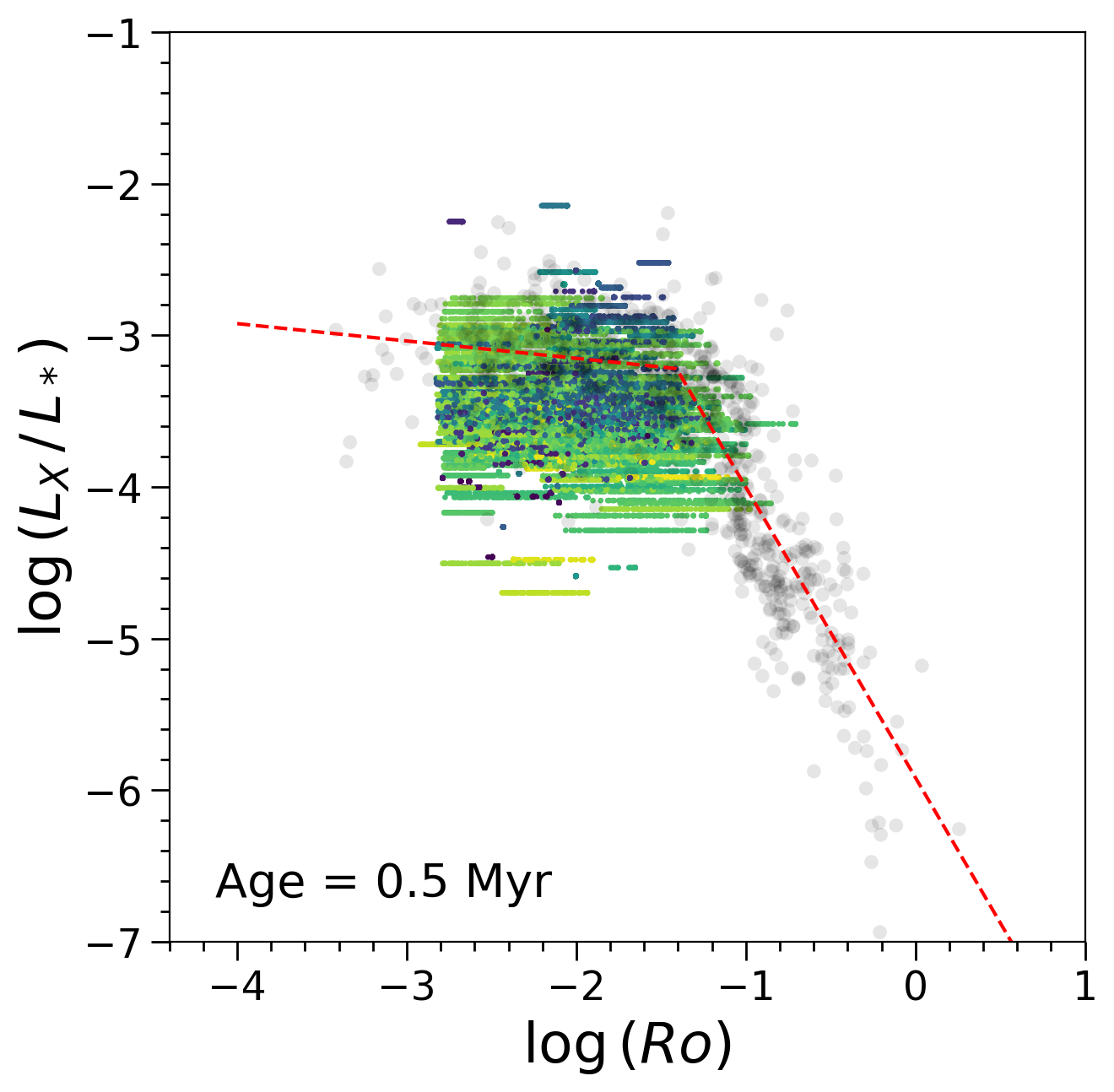}
\includegraphics[scale=0.33]{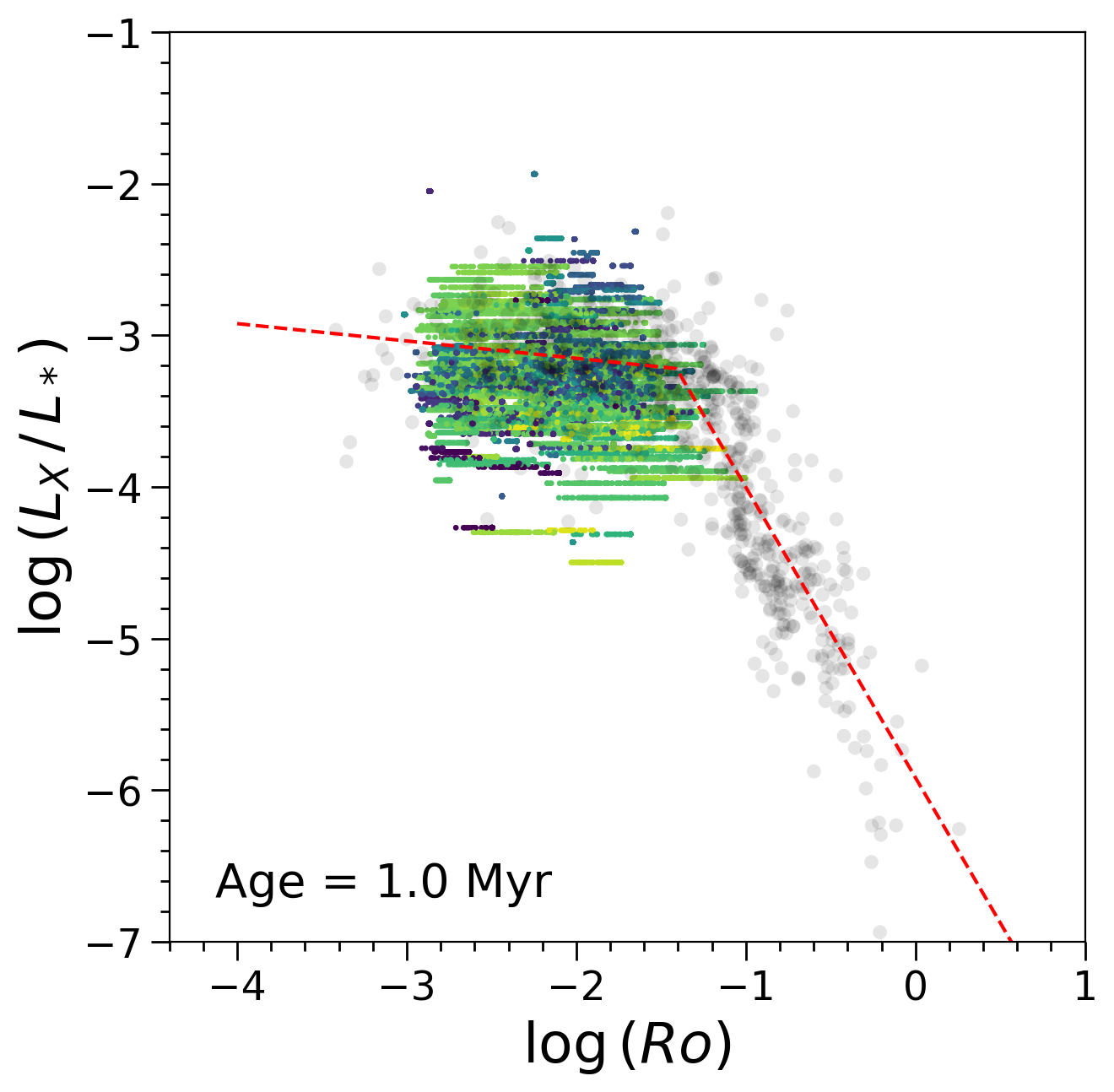}
\includegraphics[scale=0.33]{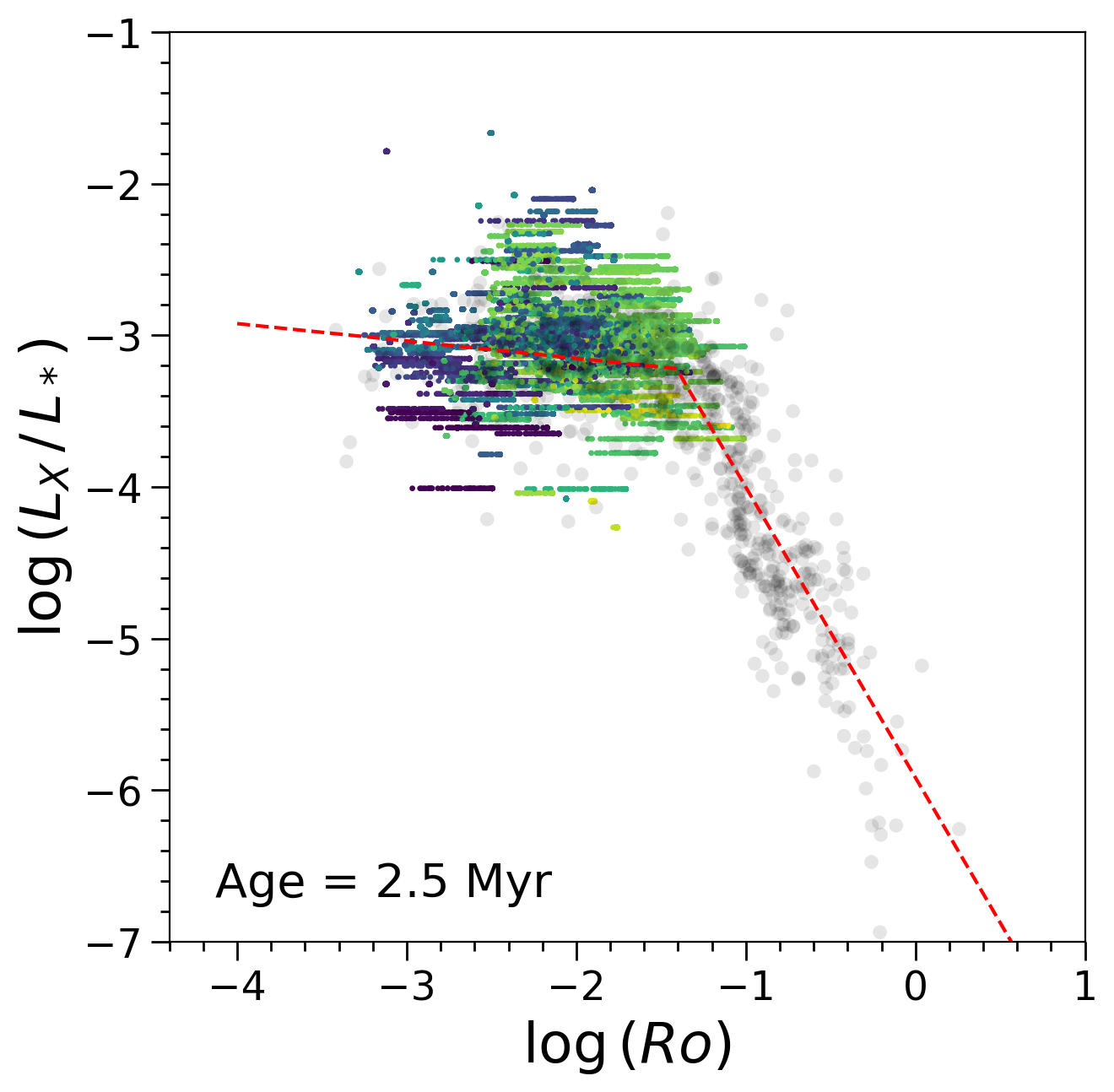}
\includegraphics[scale=0.33]{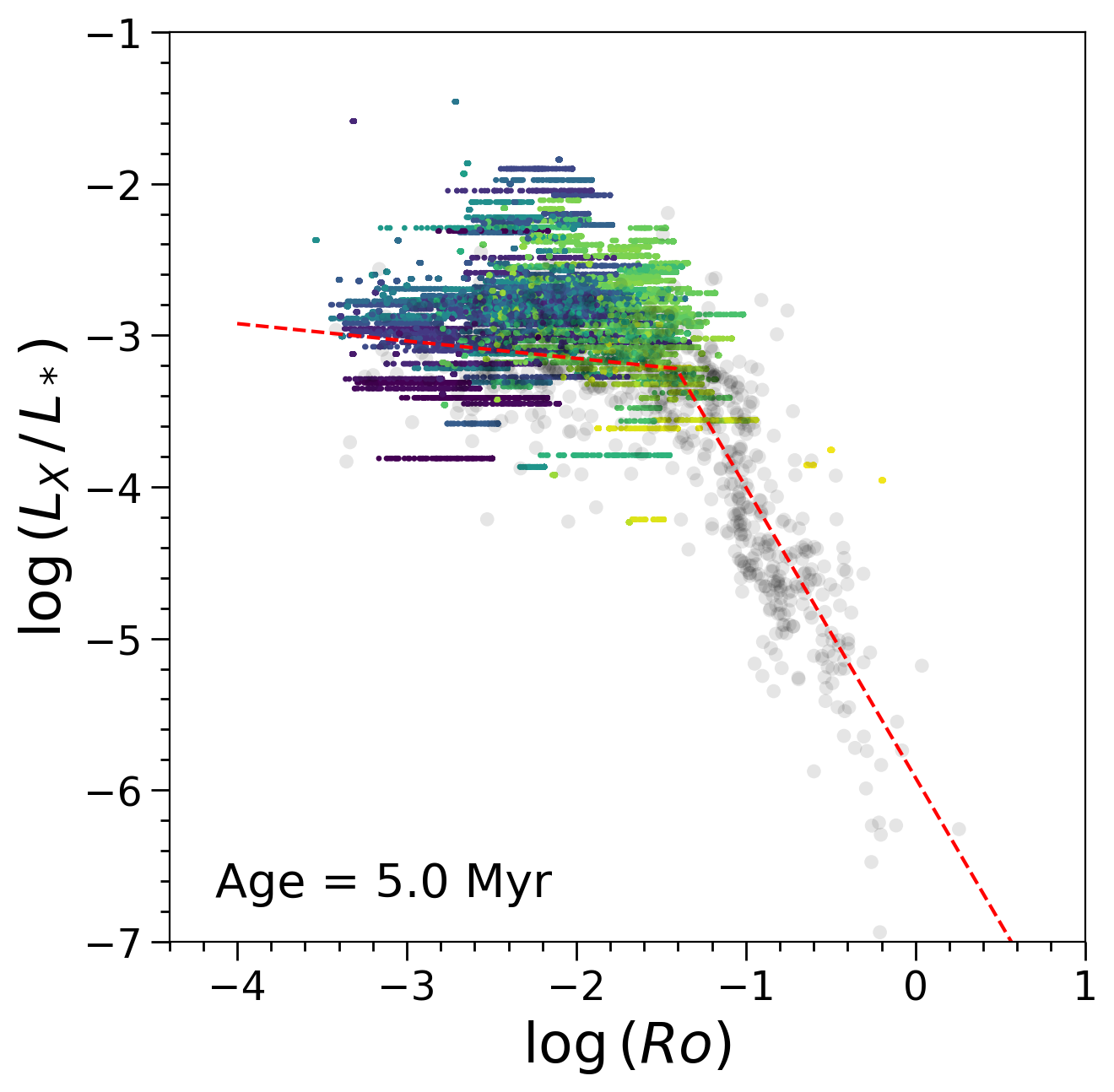}
\includegraphics[scale=0.33]{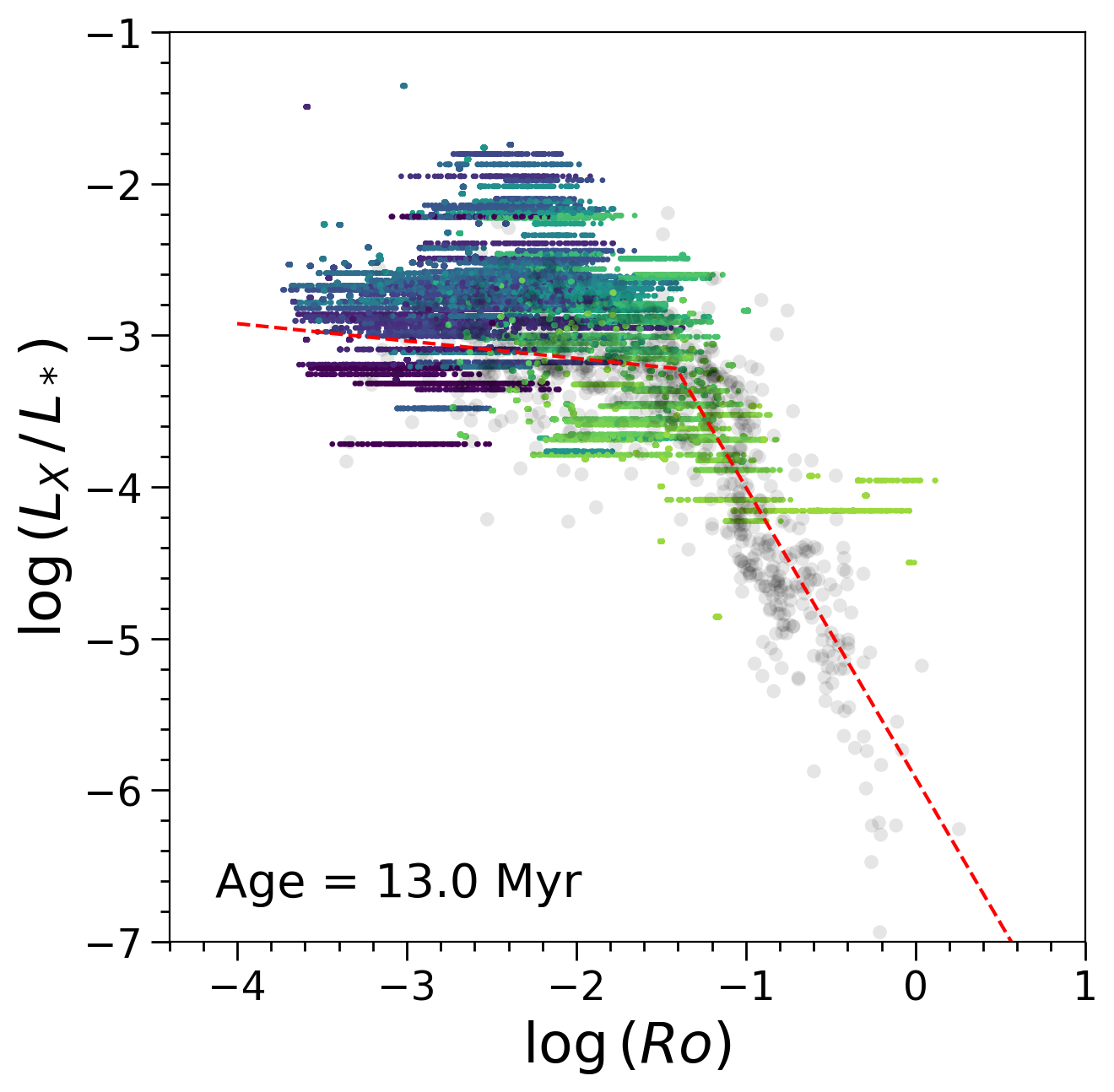}
\includegraphics[scale=0.33]{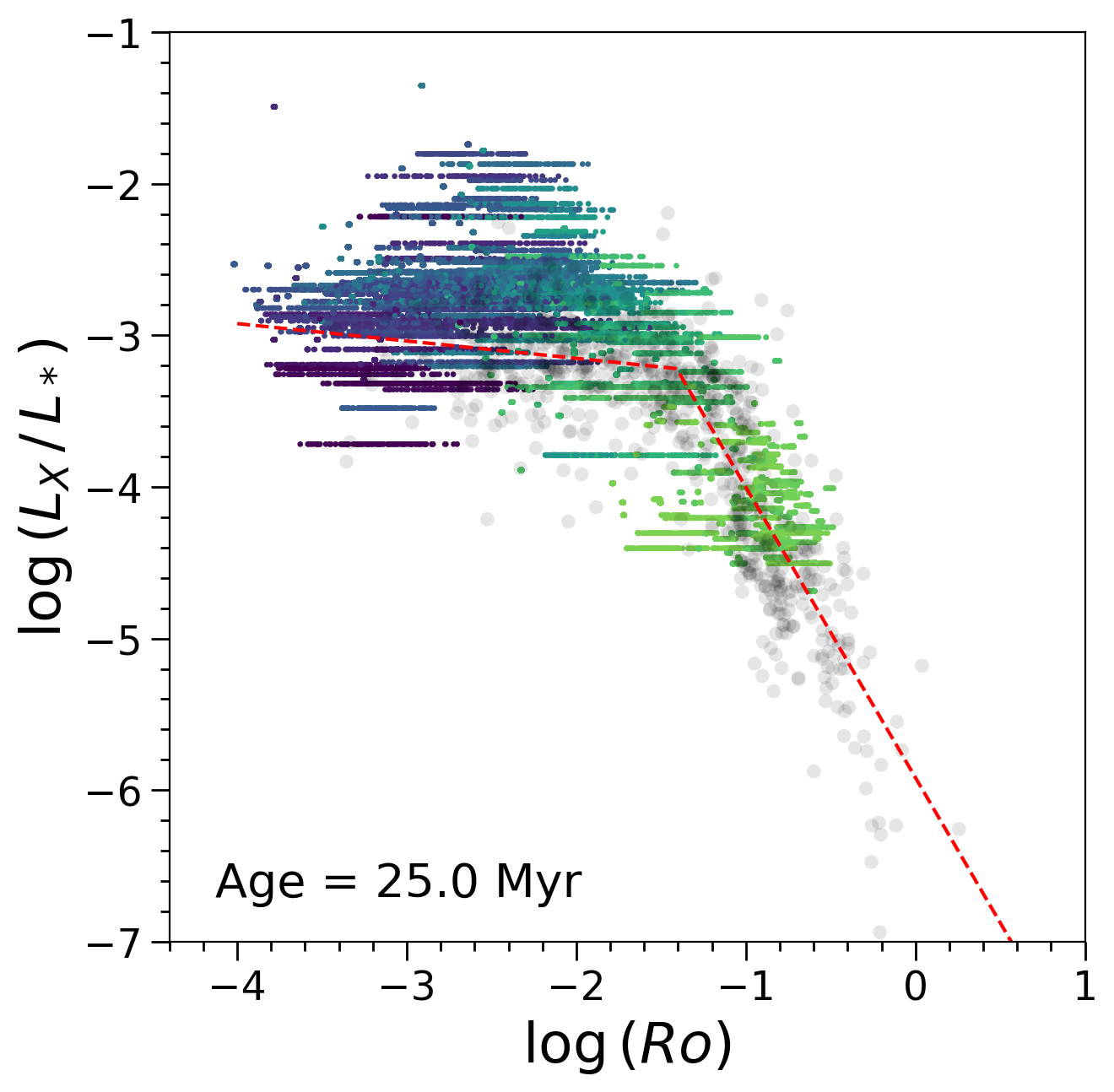}
\includegraphics[scale=0.33]{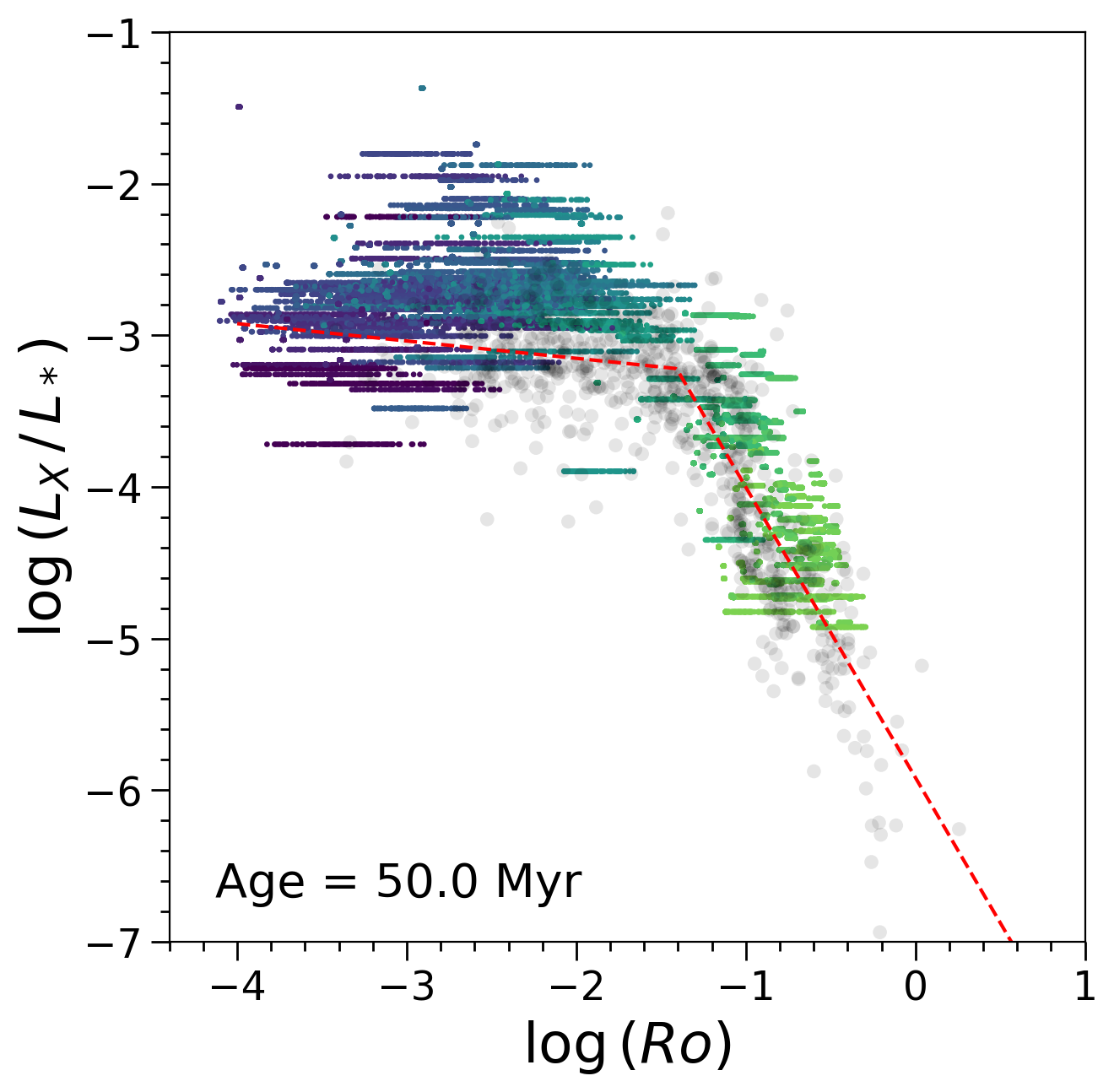}
\includegraphics[scale=0.33]{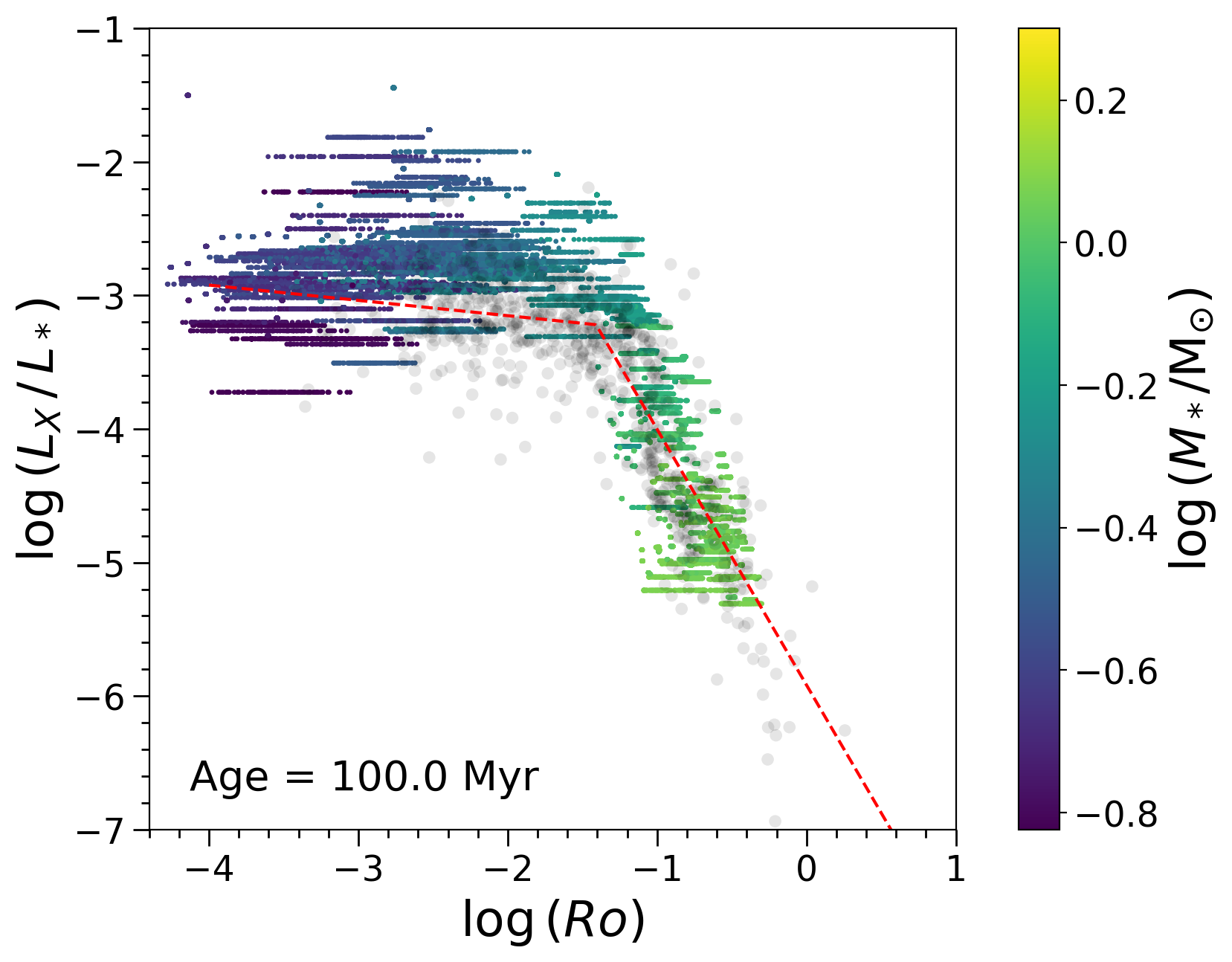}
\caption{Evolution of PMS stars in the rotation--activity plane, plotted for our entire sample. The grey dots represent main sequence stars from \protect\cite{Wright_2011}. The red dashed line represents our dual power-law fit to the main sequence sample. Stellar mass is represented by the colour scale at the bottom right. }

\label{fig: ARR evolution plots}
\end{figure*}
%

\begin{figure*}

\includegraphics[scale=0.33]{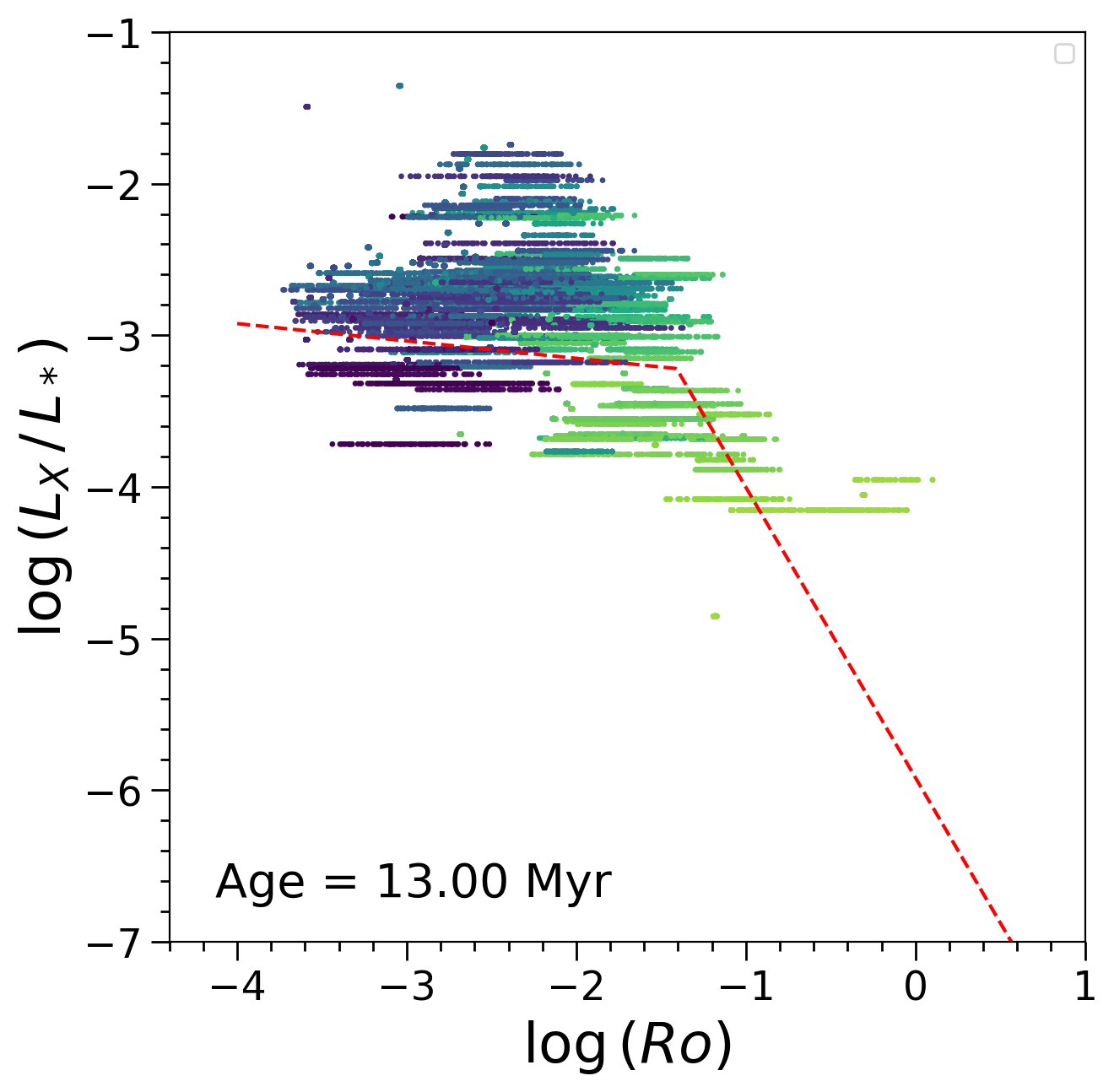}
\includegraphics[scale=0.33]{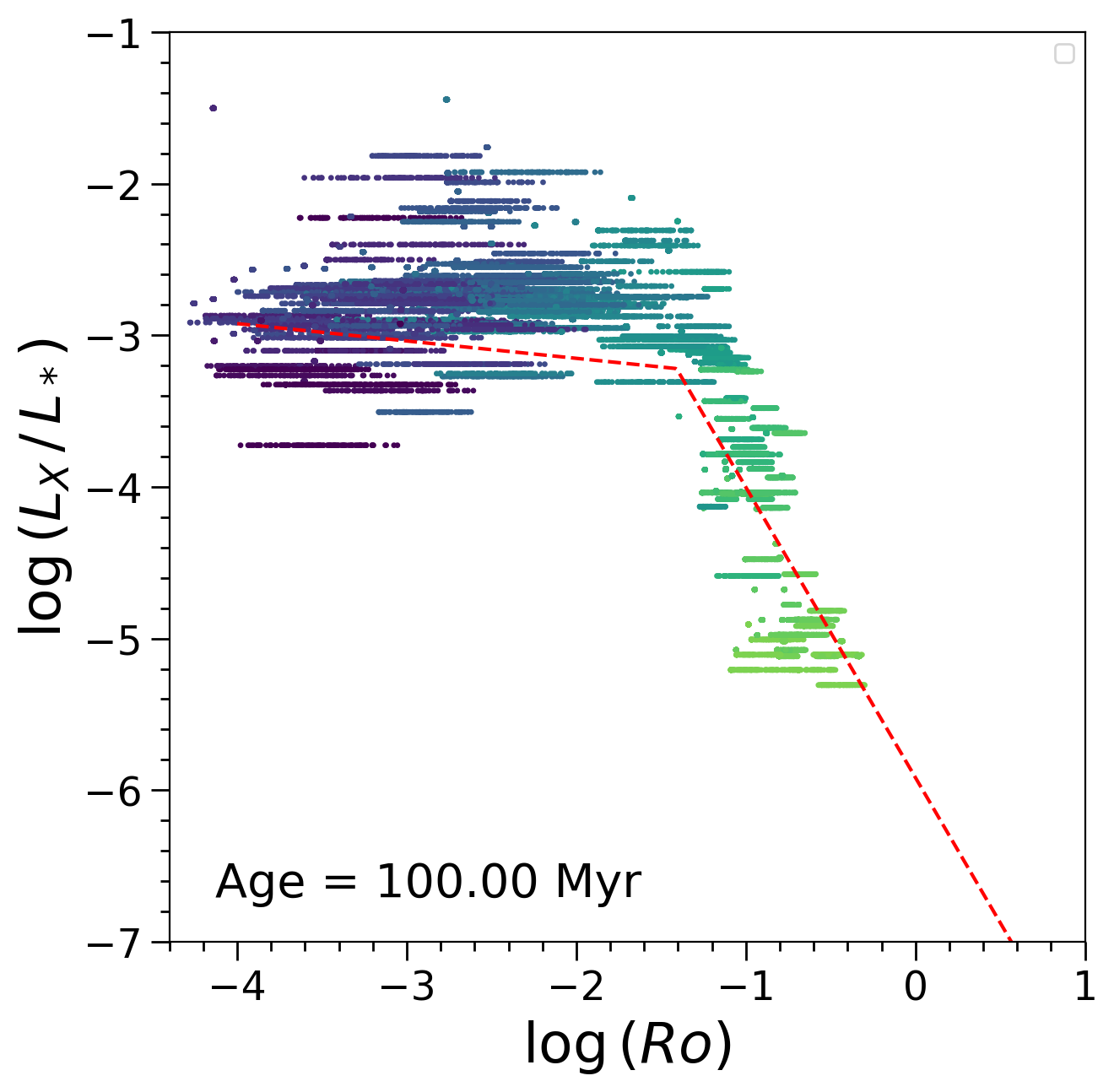}
\caption{Selected rotation--activity plots from Fig.~\ref{fig: ARR evolution plots} without stars from h~Persei. Colour represents stellar mass as indicated.}
\label{fig: no hPer ARR evolution plots}

\end{figure*}
%

\begin{figure}

\includegraphics[width=\columnwidth]{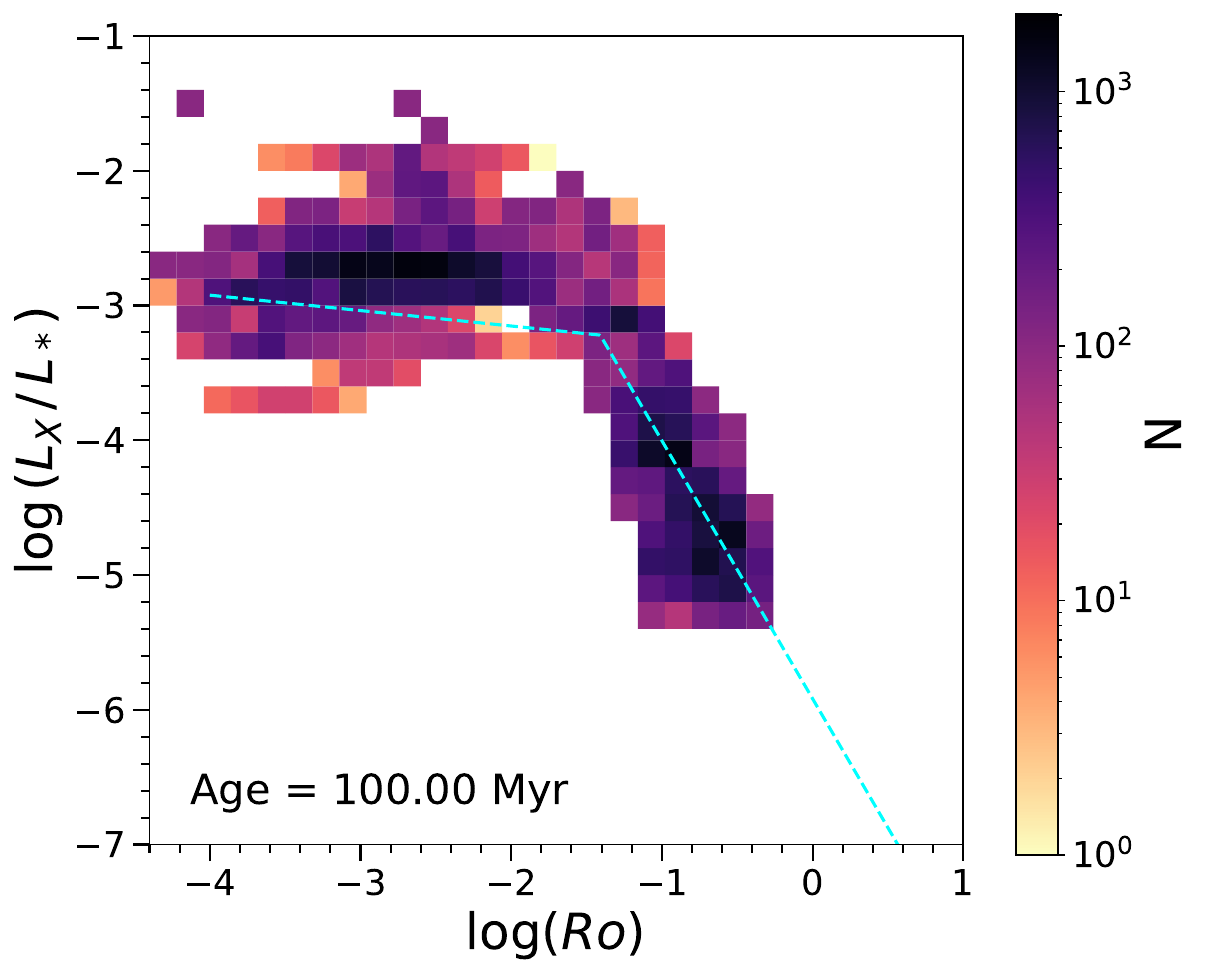}
\caption{The same rotation--activity plot at 100\,Myr, as shown in the bottom right panel of Fig.~\ref{fig: ARR evolution plots} but in the form of a 2D histogram and for only PMS stars. This shows the density of stars in the rotation-activity plane. The colour scale of the 2D bins is logarithmic. The majority of PMS stars are located in the darker regions. The blue dashed line is the best-fit main-sequence rotation-activity relation and is the same as shown in Fig.~\ref{fig: ARR evolution plots}.}
\label{fig: ARR density plot}

\end{figure}

We display the results of the evolution of the rotation--activity relation for the PMS star sample for the 100 simulations, showing snapshots at assorted ages in Fig.~\ref{fig: ARR evolution plots}. Notice that data points appear to form horizontal bands of $\log(L_\textrm{X}/L_*)$ across a range of Rossby numbers. Each band corresponds to a single star and is a result of a star's rotation period profile differing over the 100 simulations due to the randomly allocated disc lifetimes - see the outlined examples in Fig.~\ref{fig: 4 star Rotevo examples}.

At ages $\lesssim$ 2\,Myr, there appears to be a lower limit to the Rossby number that stars can have. This is caused by requiring that stellar rotation rates do not exceed break-up. At young ages, stars at this lower limit are mainly the higher-mass PMS stars as they have the lowest Rossby numbers due to their larger convective turnover times. This lower limit to $Ro$ is only apparent as we have simulated many initial rotation rates, some of which reach near break-up values. Plus, such a minimum is not observed in the individual PMS clusters as stars in clusters older than ~2\,Myr can obscure the minimum. Note also that the highest mass stars in our sample lose their outer convective envelopes before the end of our simulations, and when they do so, they no longer appear on the rotation activity plots. 

By $\sim$10\,Myr, the unsaturated regime is starting to become apparent for the highest-mass stars. \citet{Argiroffi_2016} reported similar from their observational analysis of h~Persei ($\sim$13\,Myr). Furthermore, an apparent mass stratification in the relation is evident by this age, where the Rossby number increases with increasing stellar mass - the highest mass stars in the unsaturated regime and the lowest mass stars remaining saturated. Such results are expected from the illustrative examples in Fig.~\ref{fig: Omega&Ro evo}. To determine if the emergence of the unsaturated regime by 10--13\,Myr is exclusively due to the inclusion of h~Persei in our sample, we also considered rotation--activity plots without h~Persei stars. Fig.~\ref{fig: no hPer ARR evolution plots} shows selected rotation--activity plots without h~Persei stars included. We find that even when only the younger PMS cluster stars are considered, we obtain the same results -- a clear mass stratification by 13\,Myr with stars starting to form the unsaturated regime.

By the end of our simulations at 100\,Myr, one can see that stars form a clear unsaturated regime with a gradient similar to that found for main sequence stars (which we see both with and without h~Persei stars being considered). The unsaturated regime is present at earlier ages too: it is clear by 25\,Myr, and it is well established by 50\,Myr, which we would expect from young open clusters of this age \citep{Randich_1996}. 

Upon visual inspection, we see that stars in the saturated regime begin the PMS with typically low fractional X-ray luminosities (many have $\log(L_\textrm{X}/L_*)\approx -4$), which increases with age (see Fig.~\ref{fig: RxVsAge_examples}). By $\sim$10\,Myr, the low-mass stars that remain in the saturated regime have, on average, fractional X-ray luminosities that are typically higher than the main-sequence average for saturated regime stars. A notable amount of stars reach values of $\log(L_\textrm{X}/L_*) > -2.4$ and remain high until 100\,Myr; levels which we do not see in our main-sequence sample (we discuss this further in Sec.~\ref{sec: discussion}). However, the overwhelming majority of our saturated stars have $\log(L_\textrm{X}/L_*) \approx -2.8$ and fall in the region of  $\log(L_\textrm{X}/L_*)$ of $-3.5$ to $-2.5$ as seen from main-sequence stars. This is reinforced in Fig.~\ref{fig: ARR density plot} where the simulated rotation--activity plot at 100\,Myr is plotted to indicate the density of data points.

\subsection{PMS sample age evolution cases}
\label{sec: age cases}

To analyse the fractional X-ray luminosity statistics, we consider the effect of stars in a PMS cluster having different ages. We consider two cases, described as follows.

\subsubsection{Case A}
\label{sec: case A}
In case A, we consider fractional X-ray luminosities when all stars are at the same simulated age. To achieve this, we remind readers that our models allow us to evolve individual stars both forwards and backwards in age (see Fig.~\ref{fig: 4 star Rotevo examples}). For case A, we evolve all of the stars in a simulation run to the same age. For example, if we wish to consider the X-ray properties of the sample at an age of 4\,Myr, some stars are evolved forwards in age while others are evolved backwards in age, thus ensuring that the entire sample is the same age. This is how the age evolution has been treated so far in our work, for example in Fig~\ref{fig: ARR evolution plots}. In case A, we can track statistics, for example the median absolute deviation of $\log(L_\textrm{X}/L_*)$, using all stars in our sample from our earliest to our oldest simulation ages (0.15--100\,Myr).

\subsubsection{Case B}
\label{sec: case B}

In case B, we evolve the stars in unison, beginning from their observed age. Thus, in case B, the sample has an age spread. We evolve the stars (both forwards and backwards in age) and consider the median age of the sample and how quantities such as $\log(L_\textrm{X}/L_*)$ evolve with this median age. In case B, the spread in the observed ages is conserved as the median age of the stars is increased or decreased; thus, the minimum median age from which the statistics can be tracked is limited by the age of the youngest star in the sample. This case is more akin to observing a set of stars from a PMS cluster which will have an age spread with a particular median age.

\subsection{The scatter in $\log(L_\textrm{X}/L_*)$}
\label{sec: Rx spread}

\begin{figure*}

\includegraphics[scale=0.4]{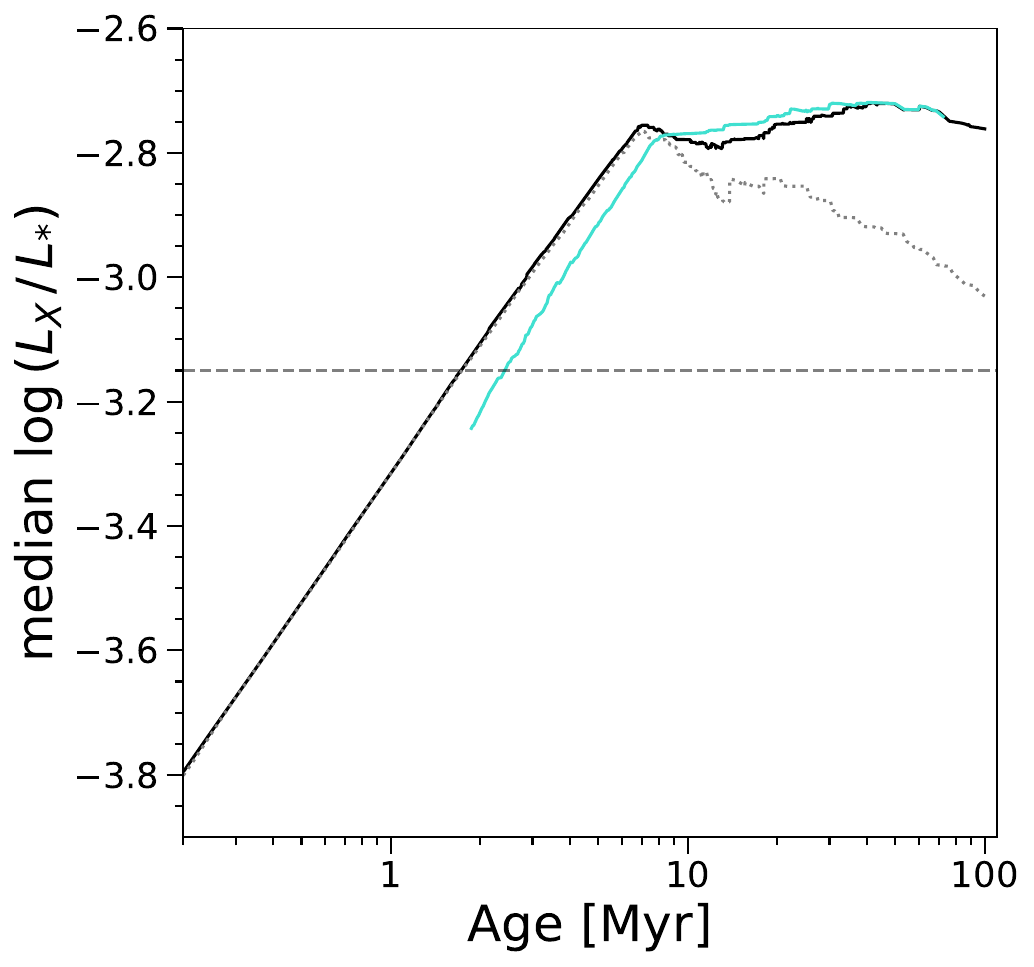}
\includegraphics[scale=0.4]{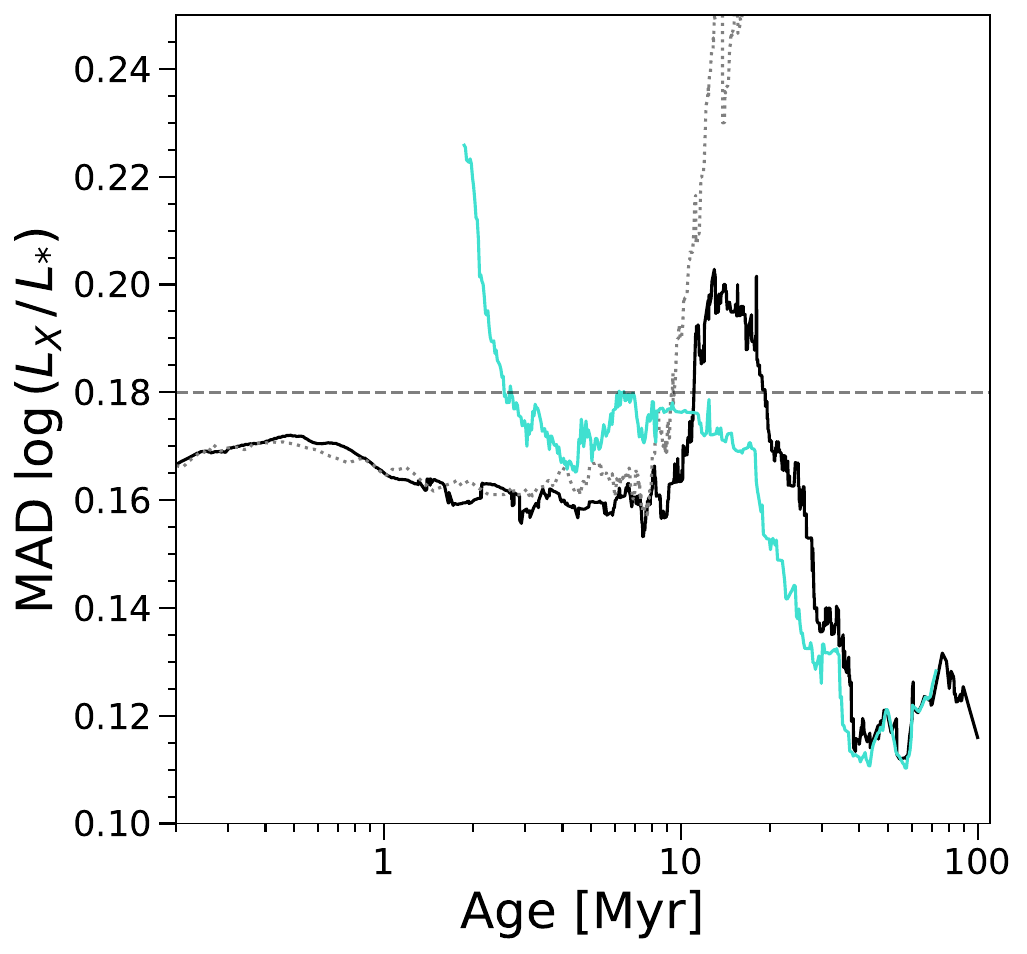}
\caption{Evolution of the median (left) and the median absolute deviation (right) of the fractional X-ray luminosity. The black lines indicate the values for saturated stars in case A (see Sec.~\ref{sec: case A}), and the dotted grey line shows the median fractional X-ray luminosity for all stars on the rotation--activity relation in case A. The light blue lines indicate the values for saturated stars in case B (see Sec.~\ref{sec: case B}). The grey dashed horizontal lines represent the expected main-sequence values for saturated stars as determined from the sample taken from \protect\cite{Wright_2011}.}
\label{fig: MEDMADRX whole sample}

\end{figure*}
%

\begin{figure*}

\includegraphics[scale=0.33]{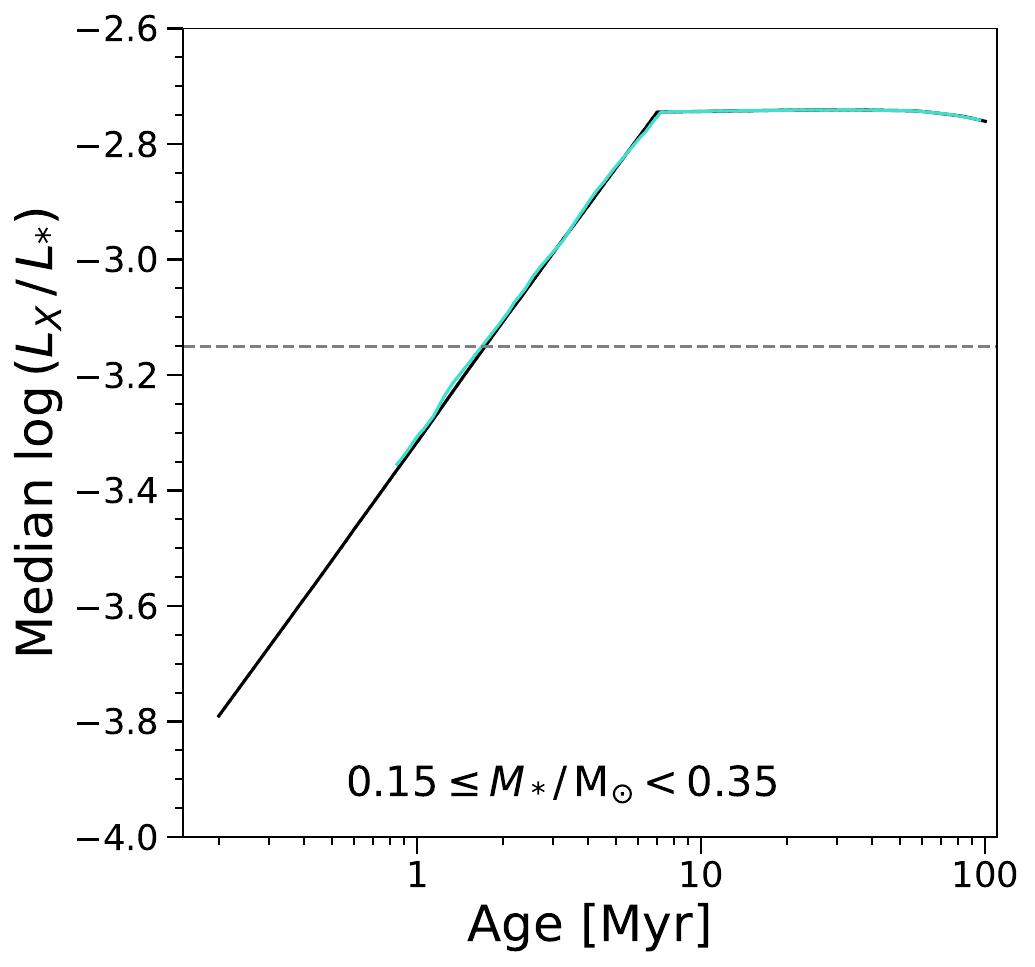}
\includegraphics[scale=0.33]{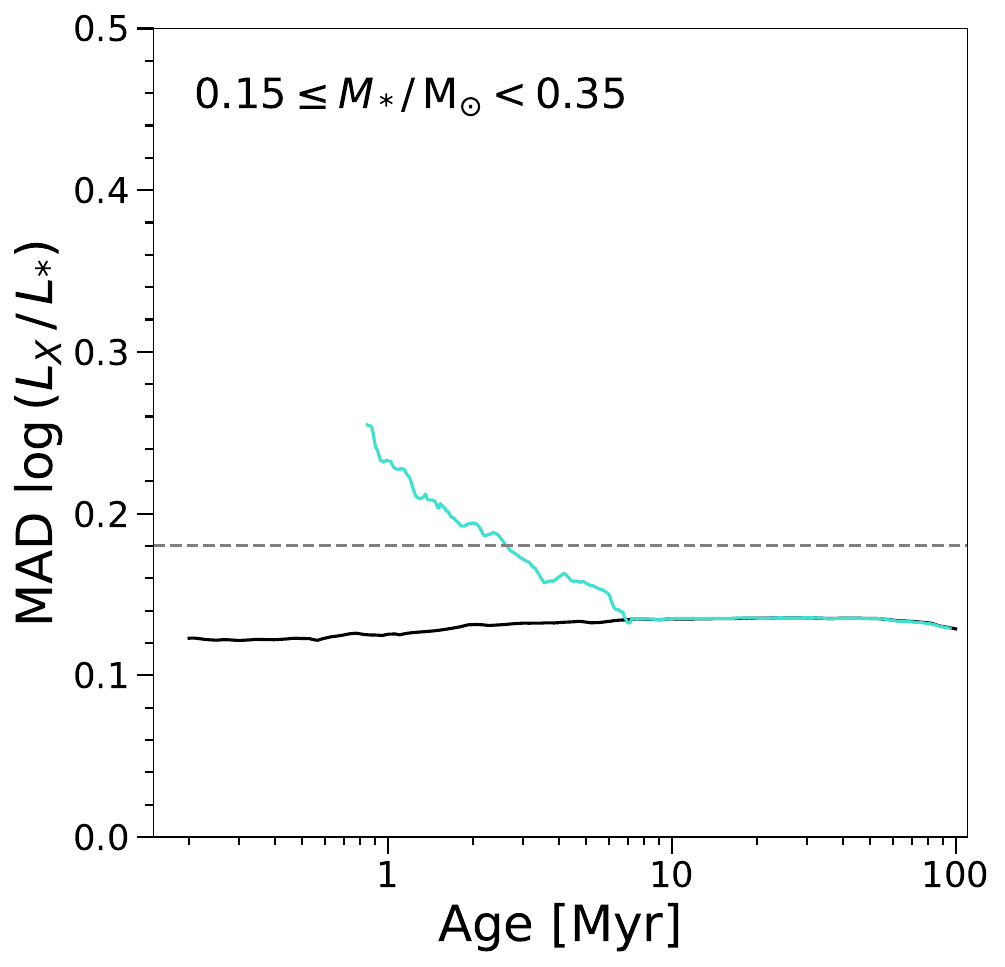}

\includegraphics[scale=0.33]{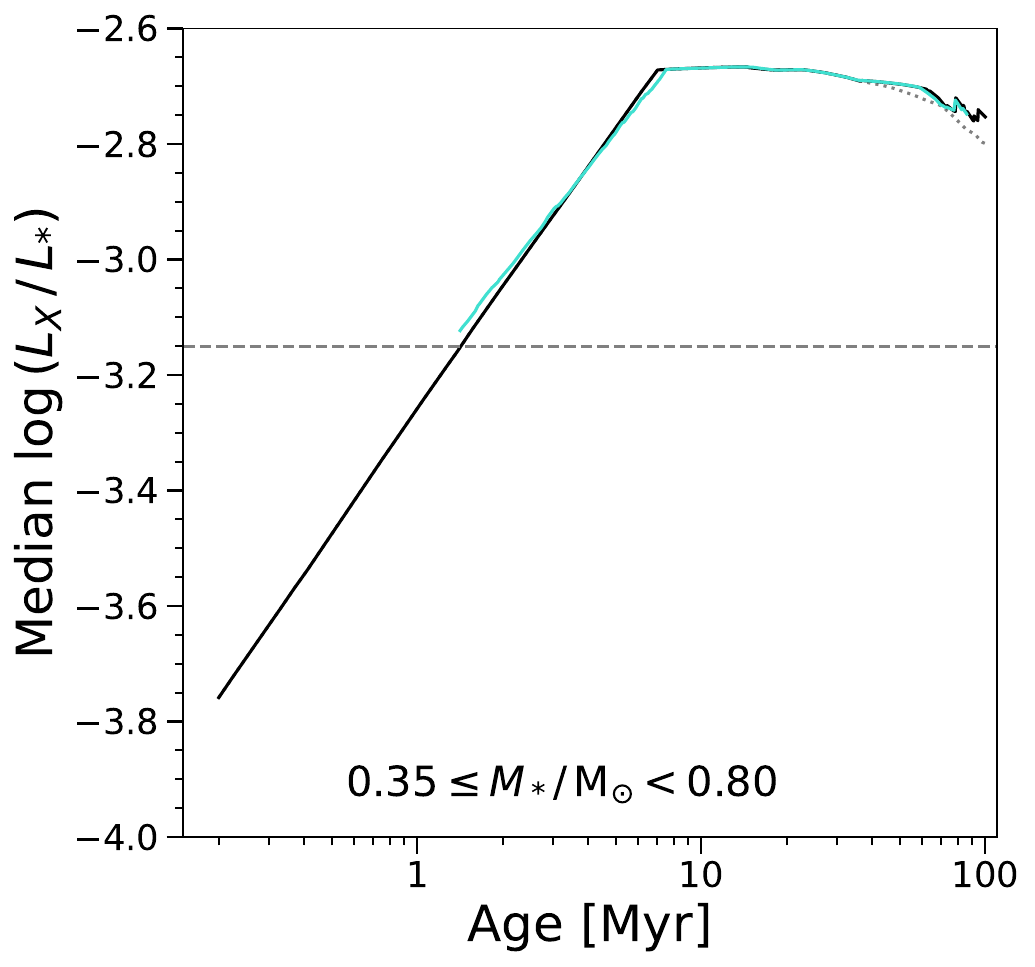}
\includegraphics[scale=0.33]{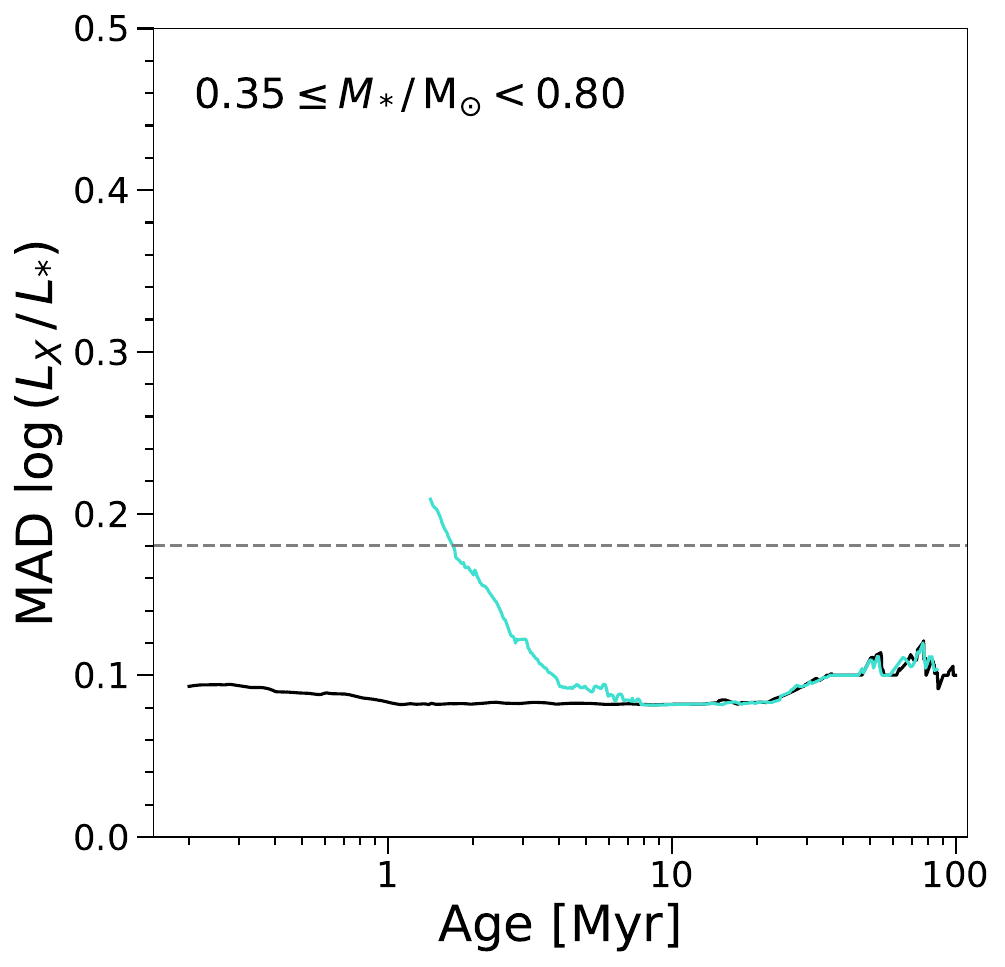}

\includegraphics[scale=0.33]{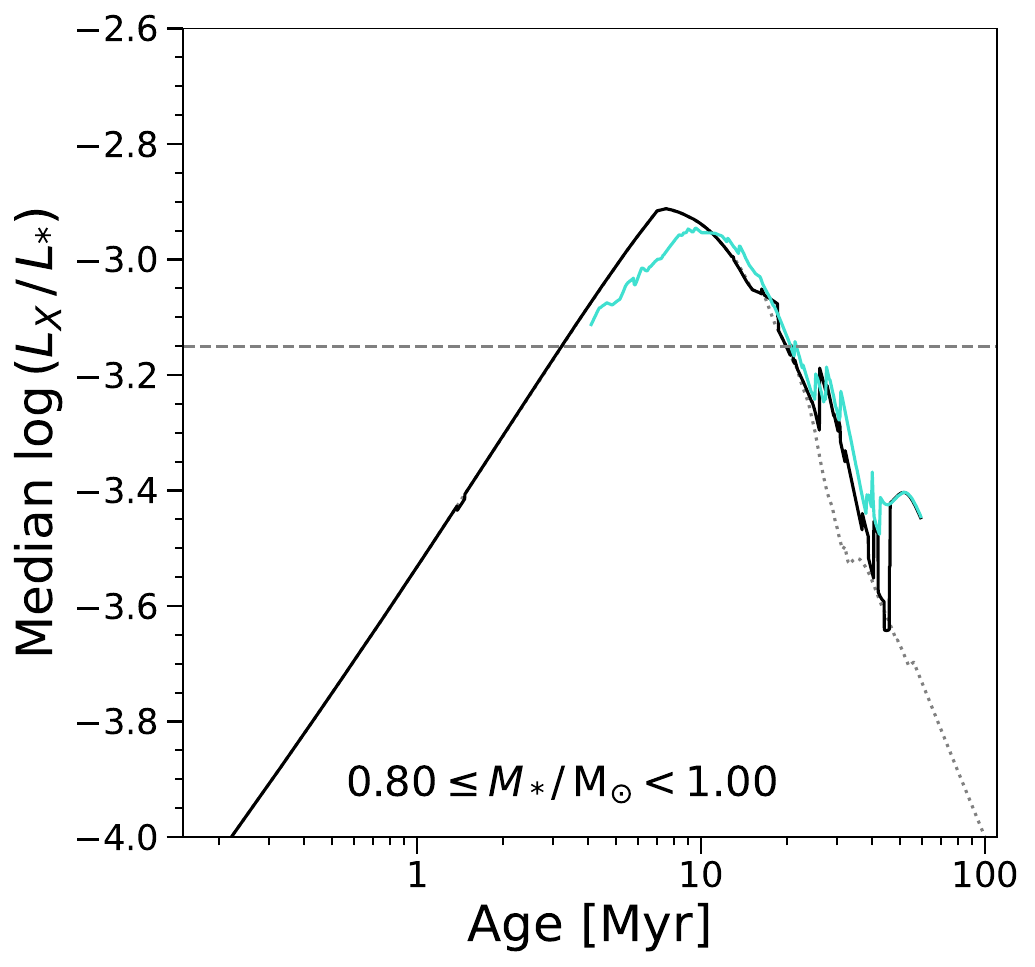}
\includegraphics[scale=0.33]{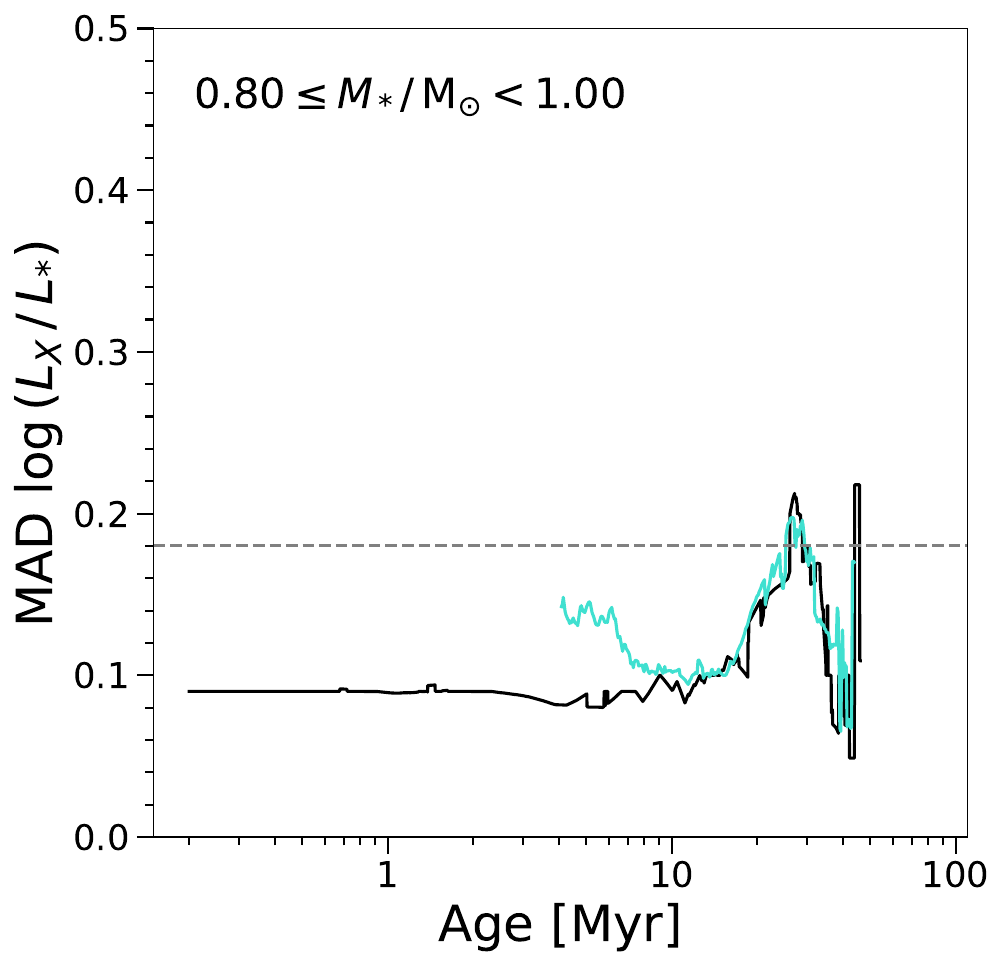}

\includegraphics[scale=0.33]{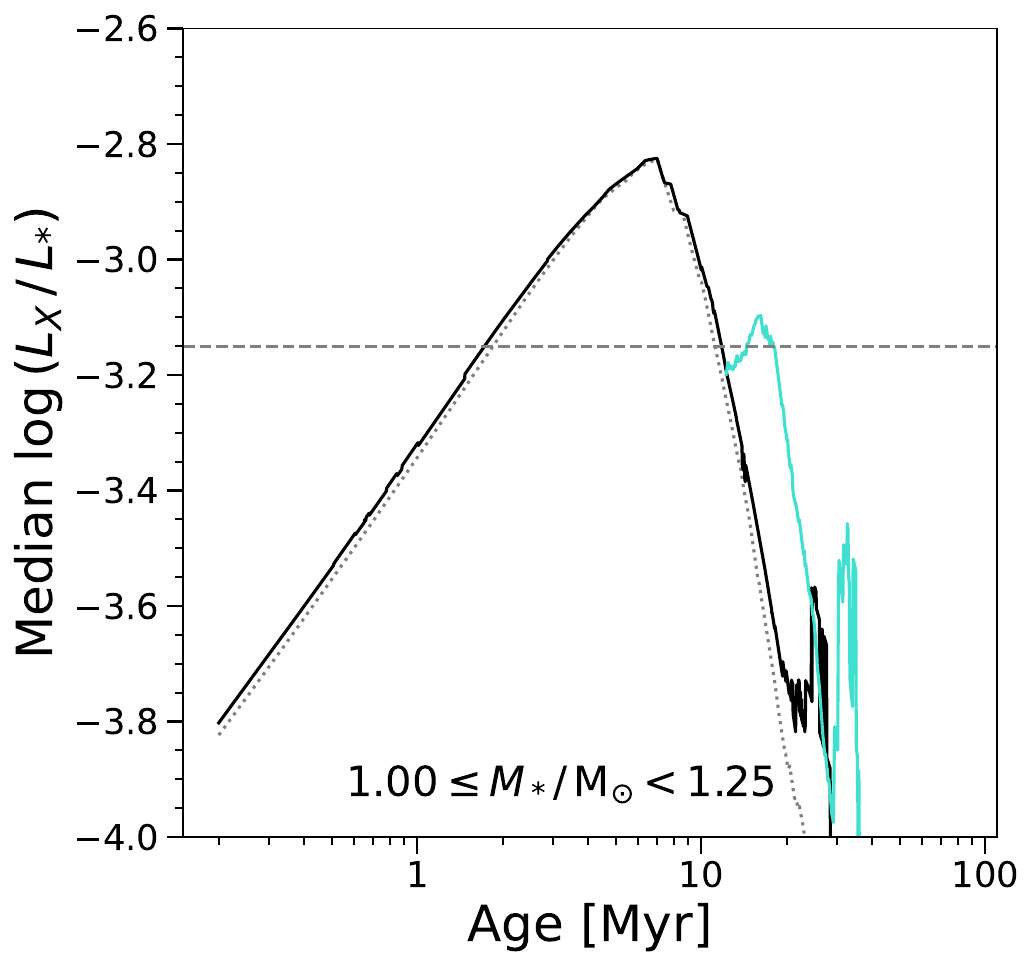}
\includegraphics[scale=0.33]{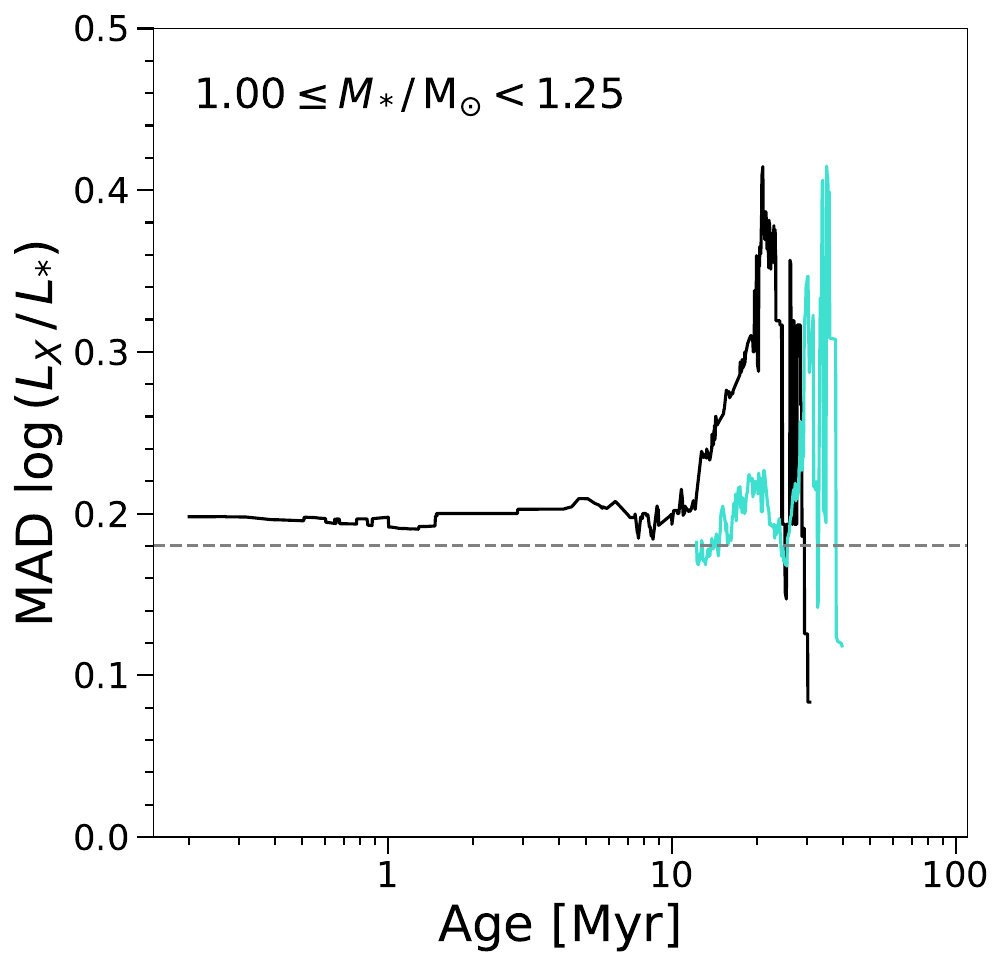}

\caption{Evolution of median (left) and median absolute deviation (right) of the fractional X-ray luminosity stars in selected mass bins as labelled. The black lines indicate the values for saturated stars in case A (see Sec.~\ref{sec: case A}), and the dotted grey line shows the median fractional X-ray luminosity for all stars in the particular mass bin on the rotation--activity relation in case A. The light blue lines indicate the values for saturated stars in case B (see Sec.~\ref{sec: case B}). The grey dashed horizontal lines represent the expected main-sequence values for saturated stars as determined from the sample taken from \protect\cite{Wright_2011}.}
\label{fig: MEDMADRX mass bins}

\end{figure*}
%

\begin{figure*}

\includegraphics[scale=0.4]{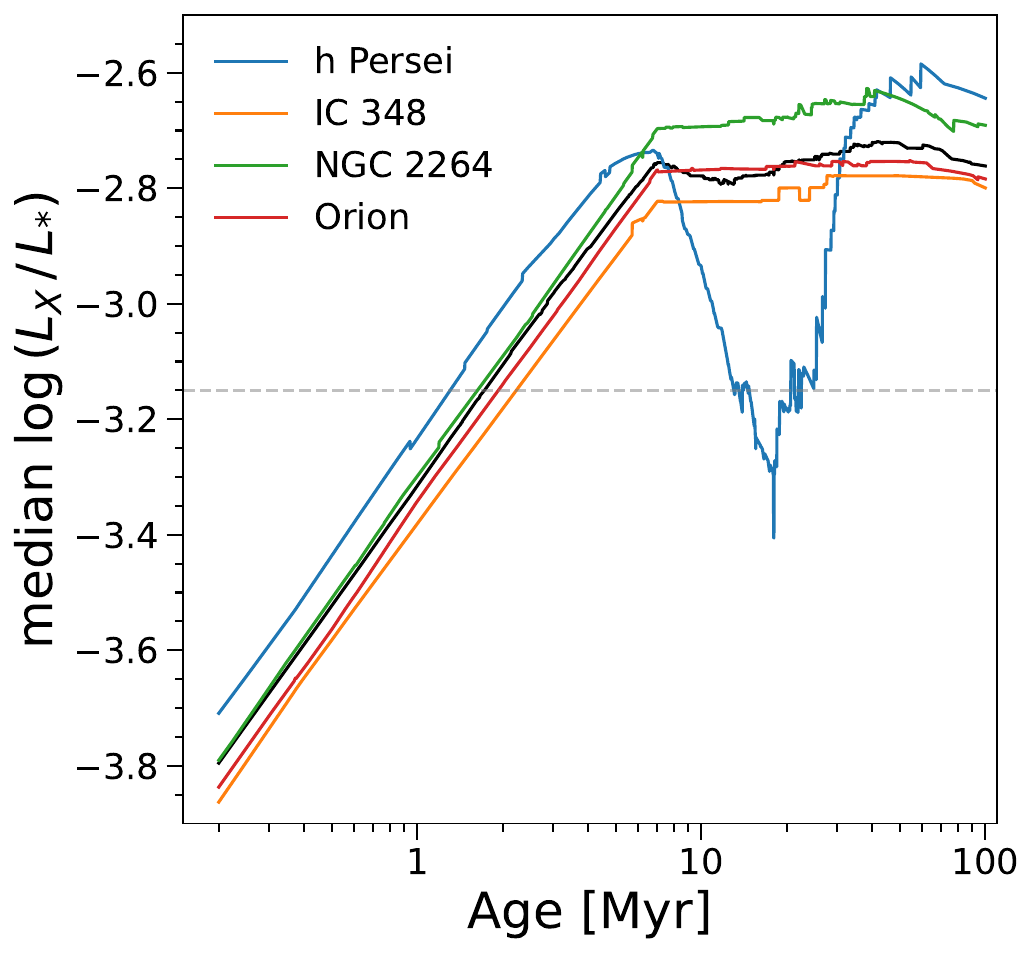}
\includegraphics[scale=0.4]{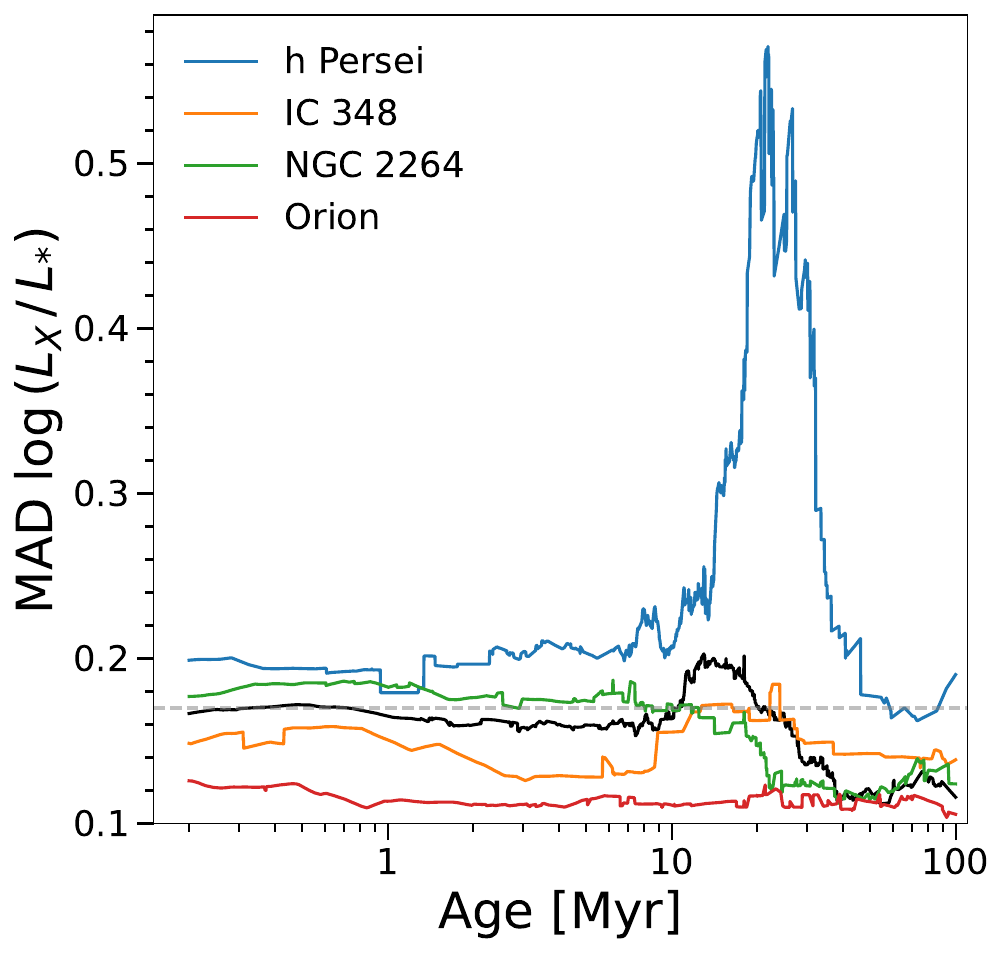}

\includegraphics[scale=0.4]{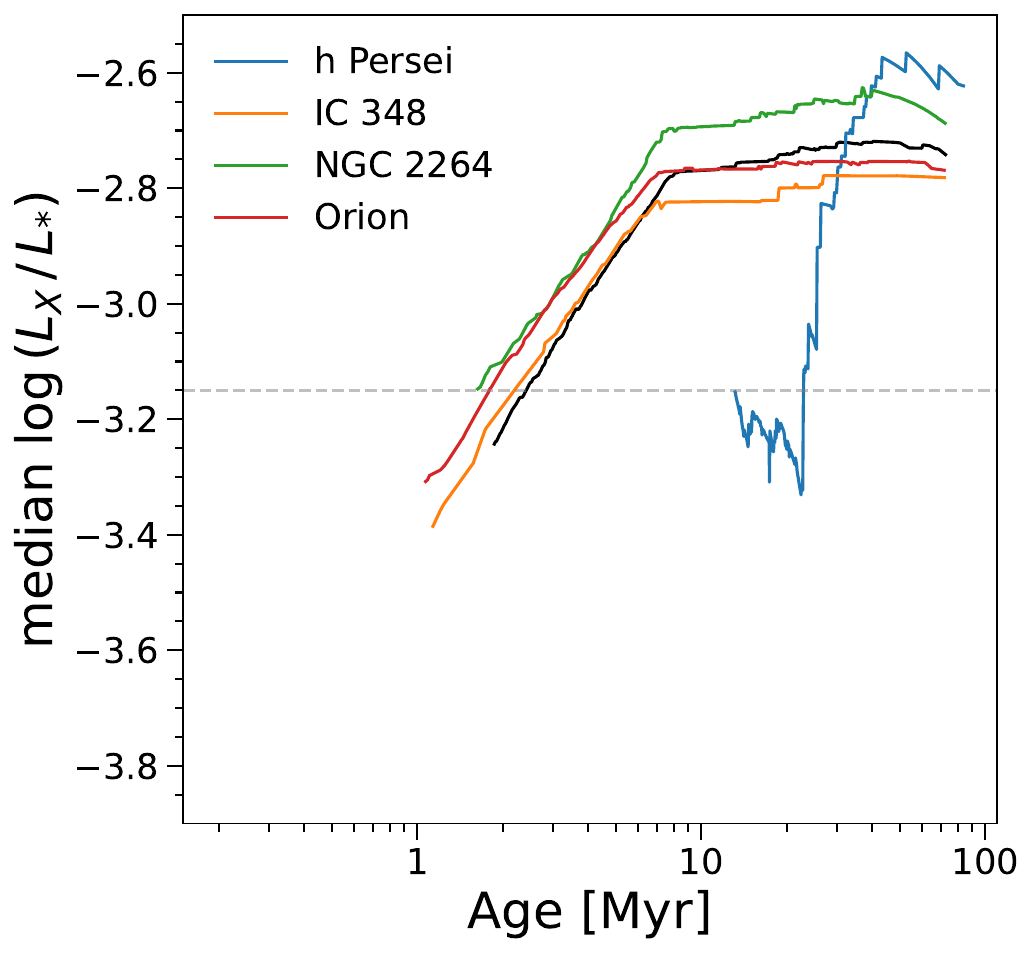}
\includegraphics[scale=0.4]{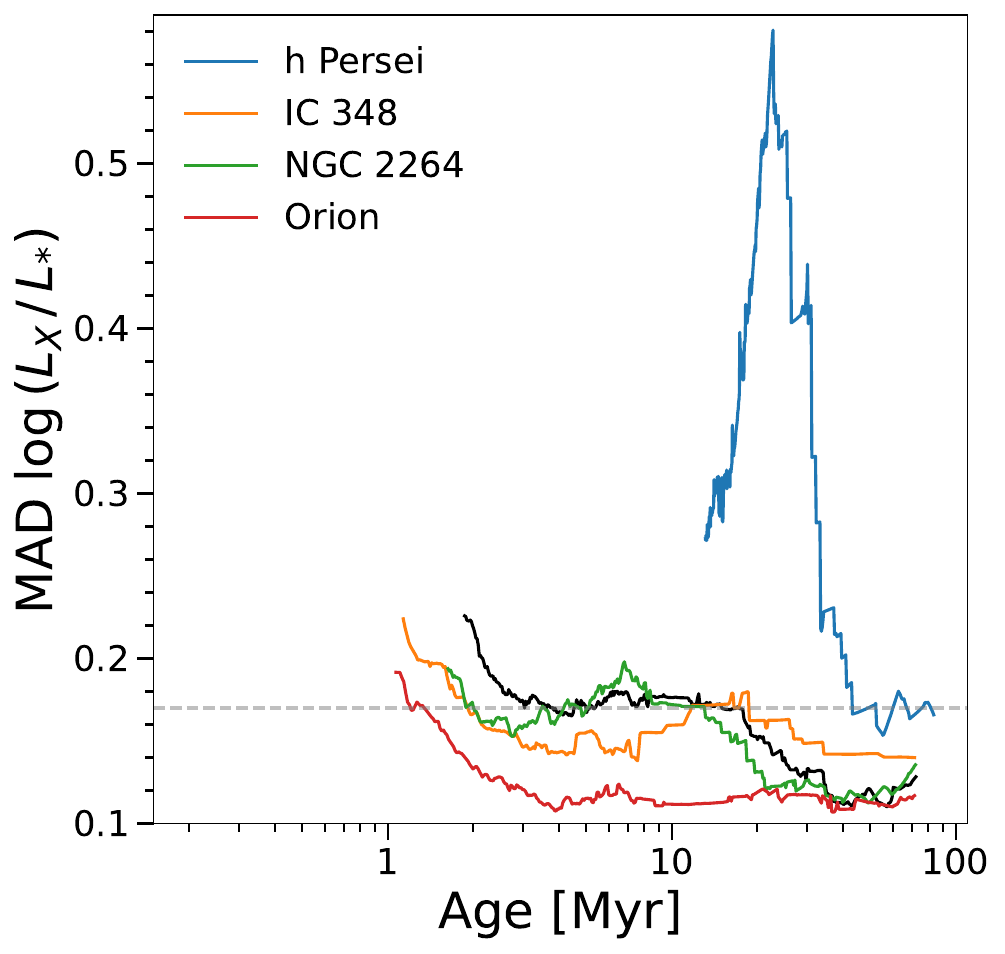}

\caption{Evolution of median (left column) and median absolute deviation (right column) of the fractional X-ray luminosity for saturated regime stars in individual PMS clusters. The top and bottom rows are for case A and case B, respectively (see Sec.~\ref{sec: age cases}). The values for the all the clusters combined are shown by the black lines. The dashed grey lines indicate the expected main-sequence values as determined from the sample taken from \protect\cite{Wright_2011}. }
\label{fig: MEDMADRX indiv clusters}

\end{figure*}

To analyse how the typical fractional X-ray luminosity and its spread change as PMS stars evolve, we track the median and MAD value of $\log(L_\textrm{X}/L_*)$ with age, using case A and case B. We focus on the median and MAD for saturated stars only as unsaturated regime stars are not expected to scatter around a median value of $L_\textrm{X}/L_*$. The evolution of the median and the MAD of the fractional X-ray luminosity versus age for our whole PMS cluster sample is shown in Fig.~\ref{fig: MEDMADRX whole sample}, for set mass bins in Fig.~\ref{fig: MEDMADRX mass bins}, and for individual clusters in Fig.~\ref{fig: MEDMADRX indiv clusters}. 

In case A, the median $\log(L_\textrm{X}/L_*)$ of the PMS stars starts low at around $-3.8$. The median initially increases at young ages, which is the expected behaviour while $L_{\rm{X}}$ remains constant for the first 7\,Myr (see the $L_\textrm{X}/L_*$ evolution examples in Fig.~\ref{fig: RxVsAge_examples}). By ages where X-ray luminosities begin decreasing, we see that the median fractional X-ray luminosity decreases and falls to near the main-sequence value by 100\,Myr, but only when considering both unsaturated and saturated stars. The median value for saturated stars stays approximately constant at just above  $\log(L_\textrm{X}/L_*) = -2.8$, a value higher than the best fit found for saturated main-sequence stars. 

It can be seen from Fig.~\ref{fig: RxVsAge_examples} that $\log(L_\textrm{X}/L_*)$ is approximately constant after 7\,Myr for the lowest mass PMS stars.
Similarly, from Fig.~\ref{fig: MEDMADRX mass bins}, it can be seen that the median $\log(L_\textrm{X}/L_*)$ value for the lowest mass stars in our sample is also approximately constant after 7\,Myr to just before 100\,Myr. For higher-mass stars, the fractional X-ray luminosity drops with increasing age (after 7\,Myr), and such stars leave the saturated regime and reach the ZAMS well before the end of our simulation time. However, for the sample as a whole, the median $\log(L_\textrm{X}/L_*)$ for saturated regime stars remains approximately constant (see left panel of Fig.~\ref{fig: MEDMADRX whole sample}), similar to the behaviour in lower-mass stars. This is because significantly more lower-mass stars are in the sample and those with higher mass leave the saturated regime.

There is a noticeable difference in the median and MAD $\log(L_\textrm{X}/L_*)$ evolution with age when comparing stars in the highest and lowest mass bins (bottom and top plots of Fig.~\ref{fig: MEDMADRX mass bins} respectively), particularly after 20\,Myr. This is caused by the highest mass stars reaching the ZAMS and leaving the saturated regime of the rotation--activity relation. By 50\,Myr stars above a solar mass are all in the unsaturated regime, and have all completed their PMS contraction, while stars below 0.5\,$\rm{M}_{\sun}$ all remain saturated for the 100\,Myr of the simulation and on the PMS. The results of the median $\log(L_\textrm{X}/L_*)$ evolution for the mass ranges do agree well with what is expected from \citetalias{Getman_2022}, especially for the comparable mass range of $0.8\rm\,{M}_{\sun} \leq M_* < 1.\,\rm{M}_{\sun}$ (see their figure 10).

The median $\log(L_\textrm{X}/L_*)$ evolution is similar among individual clusters, see Fig.~\ref{fig: MEDMADRX indiv clusters}, and they behave similarly to the overall sample of stars in the lower-mass bins. The exception is h~Persei. This is because the h~Persei sample consists mostly of stars above a solar mass. In contrast, the other clusters have large populations of fully convective stars (see Fig.~\ref{fig: init Mass hists}). Thus, for h~Persei, the evolution of the median $\log(L_\textrm{X}/L_*)$ follows similar behaviour to the overall sample of stars in the higher-mass bins initially. Then around $\sim$20\,Myr, when most of the higher-mass stars have become unsaturated, the median value in h~Persei increases to levels similar to the other clusters. For the evolution of the median $\log(L_\textrm{X}/L_*)$ in case B, the behaviour is very similar to case A in all cases.

If we consider stars that are in the saturated regime only (right panel Fig.~\ref{fig: MEDMADRX whole sample}), by 100\,Myr the MAD of $\log(L_\textrm{X}/L_*)$ is notably lower than expected for the sample of main-sequence stars. In case A, the initial decrease in the MAD over the first 10\,Myr is insignificant. The initial MAD $\log(L_\textrm{X}/L_*)$ is below main sequence levels, and eventually reduces further until about 30\,Myr. Conversely, for case B, the MAD of $\log{(L_\textrm{X}/L_*)}$ decreases significantly within the first 10\,Myr, from above to below main-sequence levels. \citet{Alexander_2012} considered X-ray observations of three PMS clusters and reported that the MAD of $\log(L_\textrm{X}/L_*)$ decreased to a level comparable to that of saturated main-sequence stars in the rotation--activity plane by $\sim$30\,Myr. We find the scatter decreases over a similar timescale but that the scatter for saturated PMS stars is below the main sequence level. This is likely because \citet{Alexander_2012} have included in their analysis some NGC~2547 stars (a cluster of age $\sim$30\,Myr) that have already left the saturated regime.

The clear difference in the evolution of the MAD of $\log(L_\textrm{X}/L_*)$ for cases A and B (see Fig.~\ref{fig: MEDMADRX whole sample}, right panel) over the first 10\,Myr can be explained as follows. In case A, for the first 7\,Myr, $L_\textrm{X}/L_*$ increases at a rate set by the $L_{*}\propto t^{-2/3}$ relation for Hayashi track evolution as $L_\textrm{X}$ is approximately constant. In case B, where there is an age spread in the sample, there is a larger possible range of bolometric luminosities, and in turn the overall spread in $(L_\textrm{X}/L_*)$ is higher. Additionally, some stars in case B will be significantly older than the average cluster age and will be decreasing in X-ray luminosity (and decreasing in fractional X-ray luminosity) while others are still increasing.

When considering individual mass bins, the MAD of $\log(L_\textrm{X}/L_*)$ is low and remains relatively constant at young ages for stars in case A. This is expected as stars of a given mass have the same $L_\textrm{X}/L_*$ evolution and have not yet left the saturated regime. The MAD of $\log(L_\textrm{X}/L_*)$ is the largest for stars in the highest mass bin. This is because the highest-mass bin stars have a greater spread in bolometric luminosity than stars in the lower-mass bins. For the higher-mass bins, the MAD also increases at around 10\,Myr as stars leave the saturated regime. When this occurs, fractional X-ray luminosities begin to drop at different rates, due to the mass dependent rate of decrease of $L_{\rm{X}}$ (see equation \ref{Eq: LXvt Getman} and \ref{Eq: LXvt Gudel}) and a difference in age when bolometric luminosity becomes constant as the stars move onto the main-sequence. The significant increase in the MAD of $\log(L_\textrm{X}/L_*)$ for high-mass stars contributes to the increasing MAD around 10\,Myr when considering the whole PMS sample.

In case B, where stars are not all the same age, there are significant decreases in the MAD of $\log(L_\textrm{X}/L_*)$ for the low-mass bins before 10\,Myr. We do not see such a decrease in the higher-mass bins as higher-mass stars in our sample typically have older ages. This can be seen in the age distributions of our observed rotation rate distributions for different masses as plotted in Fig.~\ref{fig: massbinned rot evo}. This occurs for two reasons. Firstly, higher-mass stars are typically found to be older and vice versa from their H--R diagram positions (see Sec.~\ref{sec: interpolation results}). Secondly, a large sample of our higher-mass stars are from our oldest cluster h~Persei. This affects the minimum age at which we can trace the statistics in case B as reflected in Fig.~\ref{fig: MEDMADRX mass bins} where the starting point of the light blue lines increases in age for higher-mass bins. Thus, the lack of a significant decrease before 10\,Myr in the MAD of $\log(L_\textrm{X}/L_*)$ for the stars in the highest-mass bins is because the higher-mass stars are mostly found to be older than ages where we may have expected to see a significant decrease in the MAD (at <10\,Myr). This is also reflected in the MAD[$\log(L_\textrm{X}/L_*)$] evolution in individual clusters, see Fig.~\ref{fig: MEDMADRX indiv clusters} bottom right panel, where only the younger PMS clusters (not h~Persei) show an initial decrease in the MAD (and those clusters consist of primarily younger and lower-mass stars).

Finally, h~Persei stars show a significant increase in the MAD of $\log (L_\textrm{X}/L_*)$ as they evolve towards 20\,Myr due to the dominant population of high-mass stars in the sample. IC~348 and NGC~2264 have a noticeable, but not nearly as pronounced, increase in the MAD after an initial decrease at early ages -- with NGC~2264 having a larger increase than IC~348. The level that the MAD increases appears to correlate with the fraction of high-mass stars in the cluster, with NGC~2264 having the second highest fraction after h~Persei (see the mass histograms in Fig.~\ref{fig: init Mass hists}). IC~348 has a moderate amount while Orion is dominated by lower-mass, fully convective stars and has little to no increase in MAD[$\log (L_\textrm{X}/L_*)$] during our simulation.


\section{discussion}
\label{sec: discussion}
We have considered $\sim$600 PMS stars across four star-forming regions and modelled their evolution across the rotation-activity plane. We find that the unsaturated regime of the rotation-activity relation has started to emerge by approximately 10\,Myr, with the gradient of the unsaturated regime approaching that found of main sequence stars by approximately 25\,Myr. However, we have made various assumptions in our rotational and X-ray evolution models that we now discuss. 

In our work, we have not assumed that every PMS star in a given cluster is the same age. There is, therefore, an age spread within our sample, both intra and inter-cluster, based on the H--R diagram position of each star. There is a large range of bolometric luminosities within our sample. As stars evolve across the H--R diagram, this significantly contributes to the large scatter in fractional X-ray luminosity. A spread in $\log(L_\textrm{X}/L_*)$, quantified by the MAD, is reduced if we instead evolve each star (forwards or backwards in age) to the same simulated age (case A). Thus, the age spread of stars within a cluster is a key driver of the amount of scatter in $L_\textrm{X}/L_*$. However, if stars in a PMS cluster really are coeval, then the large scatter in $L_\textrm{X}/L_*$ from the observational data cannot be attributed to the effects of intra-cluster age spread. It is of contention whether it is appropriate to approximate individual ages of stars in a PMS cluster instead of assigning a single age to all stars based on the best-fitting isochronal age of the cluster. Determining individual ages from H--R diagram position using stellar evolution models is expected to generate an age spread, even if stars are all of the same age, due to observational uncertainties \citep{Hillenbrand_2008,Preibisch_2012}. However, the intra-cluster age spread in our sample exceeds that expected by such errors, and later studies found that observational uncertainties cannot account for intra-cluster age spreads alone \citep{Jeffries_2017}. For example, considering our fitted ages for stars in Orion, 95\% of stars of solar mass or less are in the age range 0.23 -- 4.74\,Myr. This is a larger age spread than expected from observational uncertainty in the H--R diagram position alone. This age spread is slightly higher than that reported by \cite{Jeffries_2011b}, but our value agrees well for the M dwarf stars (< 0.7\,$\rm{M}_{\sun}$). Evidence of age spreads is backed up by age gradients of $\sim$1\,Myr\,pc$^{-1}$ in young PMS clusters \citep{Getman_2018} -- with stars on the periphery of clusters older than those in the central regions -- along with varying ages in spatial/kinematic substructures of clusters \citep{Kounkel_2018}.

Additionally, on the subject of stellar age, there is a bias in the YaPSI stellar evolution models whereby, in a given PMS cluster, lower-mass stars are found to be typically younger than higher-mass stars. This may be due to the discrepancies in low-mass stellar radii, being found to be around 3\% larger than in evolutionary models  \citep{Spada_2013}; which may be reconciled by including the effect of magnetic fields on convection \citep{Feiden_2016}. Furthermore, the YaPSI models assume a non-rotating star. Rotation plays a role in heat transport and, thus, the convective turnover time. The effects of varying surface rotation rates and the resulting internal differential rotation rates on convective turnover time have been investigated in other evolutionary models \citep{Amard_2019}. 

In our stellar rotational evolution model, fitting the wind torque constant (see Appendix~\ref{App: Kw calibration}) relies on the rotation rate of stars of a given mass converging on timescales on gigayear timescales, which is a common assumption in previous rotation rate models \citep[e.g.][]{Bouvier_1997,Irwin_2011}. This is grounded in rotation rate observations of partially convective stars having converged and become dependent on mass for a given age by at least $\sim$700\,Myr and earlier for higher-mass stars \citep{Barnes_2003,Boyle_2023,Douglas_2019}. However, many fully convective stars are observed to be rotating much faster than expected at $\sim$1\,Gyr by gyrochronology relations \citep{Irwin_2011,Douglas_2019}. Additionally, stars near the fully convective limit may not follow the gyrochronology relations because of the sudden braking of rotation rates that can occur due to the $^3\rm{He}$ instability in the cores \citep{Chiti_2024}. In our simulations, the lowest mass stars follow such that they spin down as fast as expected from gyrochronology relations. If the spin-down of such stars were to occur over longer timescales, they would not leave the saturated regime of the rotation--activity relation until much older ages. One could consider a model that considers multiple fast and slow regime rotational evolution tracks by adjusting the wind torque for fully convective stars to replicate the spread of rotation rates at old ages. Ultimately, this change would have little to no effect on analysing our fractional X-ray luminosity statistics for stars in the saturated regime. This is due to all fully convective stars remaining in the saturated regime over our 100\,Myr simulated period, despite our use of the lower limit of rotation rates expected following gyrochronology. 

\begin{figure}

\includegraphics[width=\columnwidth]{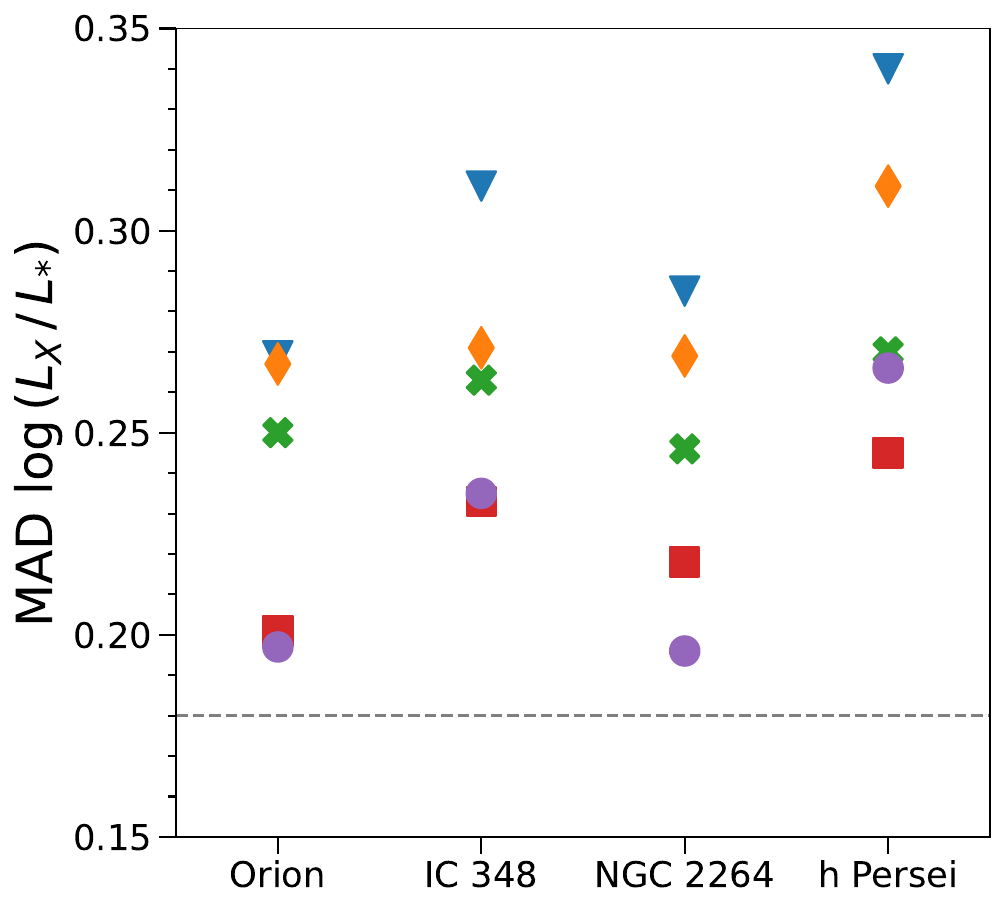}
\caption{The median absolute deviation of $\log(L_\textrm{X}/L_*)$ for observations of individual PMS clusters, ordered left-to-right by increasing median cluster age. Blue triangles represent all stars in a cluster with X-ray luminosity data. Orange diamonds represent values for stars with bolometric luminosity data from which we could determine mass and age (see Sec.~\ref{sec: ARR interpolation}). Green crosses represent the sample represented by the orange diamonds, but only for stellar masses of $M_{*} \leq 2\,$M$_{\odot}$. Red squares represent stars that also have literature rotation periods. Purple circles represent stars in the saturated regime of the rotation--activity plane, as seen in Fig.~\ref{fig: PMS_init_ARRplot}. The dashed horizontal line represents the MAD value seen for MS saturated stars in the \protect{\citet{Wright_2011}} sample.}
\label{fig: MADRX cluster sampling}

\end{figure}

In our analysis of the evolution of the MAD of $\log{(L_\textrm{X}/L_*)}$, we are restricted to stars that we can position on the rotation--activity plane. Thus, we lose a large sample of stars with X-ray luminosity but no known rotation period (we cannot determine if they are in the saturated or unsaturated regime for such stars). We find the scatter in $L_\textrm{X}/L_*$ in all clusters is larger if we consider all stars with X-ray luminosities compared to the subset of stars which also have a measured rotation period, as shown in Fig.~\ref{fig: MADRX cluster sampling}. The decrease in MAD of $\log(L_\textrm{X}/L_*)$ when considering PMS stars with known rotation periods is driven in part by no stars in that sample with $\log(L_\textrm{X}/L_*) < -4.7$ (which is far from the median). This is primarily caused by higher-mass PMS stars, which typically have lower $\log(L_\textrm{X}/L_*)$ -- see \citet{Getman_2022}, \citet{Gregory_2016}, \citet{Rebull_2006} -- not having literature rotation period estimates. Note, for example, in Fig.~\ref{fig: MADRX cluster sampling}, we see MAD[$\log(L_\textrm{X}/L_*)$] decreases when we remove all stars above 2 solar masses (with or without rotation periods) from the samples. The lack of literature rotation periods in less active, higher-mass PMS stars could be because they have stellar surfaces covered in more numerous and smaller cool spots, which give rise to less apparent rotationally modulated variability in their light curves compared to active, lower-mass PMS stars \citep{Saunders_2009}. 
 
In our simulations, we find that low-mass stars that remain in the saturated regime after 100\,Myr have a median $L_\textrm{X}/L_*$ that is higher than found for main-sequence stars. Studies of main-sequence stars rarely find values of $\log(L_\textrm{X}/L_*) > -2.5$ and none above $-2$ \citep{Nunez_2024,Wright_2018}: stars exceed this value in our model. However, we only evolve the stars in our sample to 100\,Myr. In contrast, the sample of main-sequence stars that we are comparing to includes substantially older stars (e.g. from the Praesepe and Hyades clusters of $\sim$600--700\,Myr in age). If we assume that the decay of $L_{\rm{X}}$ with age that we have used for stars of age 25--100\,Myr holds for older stars [see equation (\ref{Eq: LXvt Gudel})], then the median $\log(L_\textrm{X}/L_*)$ for saturated stars would drop below the main sequence value at around $\sim$400\,Myr. 

While our models successfully reproduce the rotation--activity relation regarding the emergence of the unsaturated regime, we have not considered mechanisms such as coronal stripping. Coronal stripping, where X-ray emitting plasma is centrifugally stripped from the extended corona due to the stars spinning rapidly enough that the co-rotation radius comes within the closed emitting regions [see \citet{Jardine_1999}], would reduce the X-ray luminosity. The inclusion of coronal stripping and exploring supersaturation is beyond the scope of our models and is reserved for future work. Had we included this effect in our models, this may have limited some of the peak fractional X-ray luminosities from our simulations seen by 100\,Myr. 

We have assumed that X-ray luminosity is constant for the first 7\,Myr of evolution, following the trends reported by \citetalias{Getman_2022}. For Hayashi track PMS stars, whose bolometric luminosity is decreasing, this means that $\log(L_\textrm{X}/L_*)$ is increasing for the first 7\,Myr of evolution (see Fig.~\ref{fig: RxVsAge_examples}). This may also explain why we find higher than expected fractional X-ray luminosities by 100\,Myr for saturated regime stars. We investigated the effect of changing the age up to which $L_{\rm{X}}$ is assumed constant in the range of 3--8\,Myr. We find that the median of $\log(L_\textrm{X}/L_*)$ at 100\,Myr decreases as the age at which $L_{\rm{X}}$ is kept constant decreases, with a change of $\sim$0.3\,dex over the range of 3--8\,Myr. The change in fractional X-ray luminosities is dependent on the value of $\beta_{X}$ from equation (\ref{Eq: LXvt Getman}), $L_{\rm X}\propto t^{\beta_{\rm X}}$. As a result, any change in the median $\log(L_\textrm{X}/L_*)$ value at 100\,Myr reflects only the change for the $<0.75$\,M$_{\sun}$ sample (as only these stars are left unsaturated). Furthermore, the gradient of the unsaturated regime in the rotation-activity plane becomes noticeably steeper by decreasing the age at which $L_{\rm{X}}$ is kept constant as the higher-mass ($>1.0$\,M$_{\sun}$) stars have a greater change in $\log(L_\textrm{X}/L_*)$ due to the more negative value of $\beta_{\rm{X}}$. When all stars are the same age in our simulation, low values of when $L_{\rm{X}}$ ends being constant ($\leq 5$\,Myr) lead to MAD $\log(L_\textrm{X}/L_*)$ values that are initially higher than main-sequence values. When stars evolve in unison from observed ages, a change in when $L_{\rm{X}}$ is kept constant has an insignificant effect on MAD $\log(L_\textrm{X}/L_*)$.

While we have adopted the \citetalias{Getman_2022} X-ray luminosity relations for the decay of $L_{\rm X}$ with age (with $L_{\rm X}$ constant to 7\,Myr), their analysis focuses on stars of mass $\ge$0.75\,M$_{\sun}$. The trends may not be appropriate for lower-mass PMS stars, particularly the prescription of constant $L_{\rm X}$ for the first few Myr of evolution. More analysis of X-ray data from clusters around 5\,Myr of age would help refine our understanding of X-ray evolution.

%


\section{Conclusions}
\label{sec: conclusions}

We have developed a model that allows us to use observational data of PMS stars and evolve their positions on the X-ray rotation--activity plane both backwards and forwards in age. To achieve this, we used the YaPSI stellar evolutionary models, a coupled core-envelope rotational evolution model, and observed trends of X-ray luminosity with age. 

We simulated the emergence and evolution of the rotation--activity relation, up to 100\,Myr, using stars from four PMS clusters. Our model successfully produced the emergence of the main-sequence rotation--activity relation. We find that the unsaturated regime begins to emerge after approximately 10\,Myr, which is in good agreement with the observational study of h~Persei by \citet{Argiroffi_2016} who find that the higher-mass PMS stars have started to form the unsaturated regime by $\sim$13\,Myr. While our analysis also includes h~Persei, we still find that the unsaturated regime emerges after approximately 10\,Myr when h~Persei stars are excluded. 

By $\sim$25\,Myr, the slope of the unsaturated regime is comparable to that seen for main-sequence stars and is well established by 50\,Myr, as expected by observations of the $\alpha$~Persei cluster of such age \citep{Randich_1996}. As we evolve the stars in age, the unsaturated regime emerges as a result of higher-mass stars developing partially convective interiors and, in turn, increasing in Rossby number while decreasing in fractional X-ray luminosity. Notably, the emergence of the unsaturated regime is found without an a priori assumption of the shape of the main-sequence rotation--activity relation or any connection between rotation rate and typical X-ray luminosities. 

Our rotational evolution model reveals that the rotation rate of stars of mass $M_* \geq$ 0.6\,M${_{\sun}}$ become slow enough, and therefore their Rossby numbers high enough, that they typically leave the saturated regime before reaching the main-sequence. No star $M_* \geq$ 0.7\,M${_{\sun}}$ remains saturated after 100\,Myr.  Fully convective stars and stars of mass $M_* < 0.5$\,M${_{\sun}}$ all remain in the saturated regime (bar one) over the first 100\,Myr of our simulations.

For saturated regime stars, the median $\log(L_\textrm{X}/L_*)$ initially increases as our model predicts increasing fractional X-ray luminosities for Hayashi track stars until $L_{\rm X}$ starts decreasing. After 10\,Myr, the median $\log(L_\textrm{X}/L_*)$ stays approximately constant for saturated stars up to 100\,Myr. A decrease is seen in the scatter of $\log(L_\textrm{X}/L_*)$, quantified by the MAD, for saturated stars as they age. The scatter is below what is expected from main-sequence stars by 100\,Myr. The decrease in the MAD of $\log(L_\textrm{X}/L_*)$ is only significant if we treat the evolution of the stars in unison and evolve them forwards in age, beginning from their age spread found from their H--R locations. \citet{Alexander_2012} reported similar from X-ray observation, with their MAD[$\log(L_\textrm{X}/L_*)$] for saturated stars reducing when comparing older to younger PMS clusters. When we instead evolved all stars, either forward or backwards in age such that the entire sample was the same age, and then evolved forwards to 100\,Myr, we find initial MAD[$\log(L_\textrm{X}/L_*)$] below main sequence levels. This indicates that the observed large scatter in $\log(L_\textrm{X}/L_*)$ observed in PMS clusters is influenced by the age spread within PMS clusters [either a real intra-cluster age spread \citep[e.g.][]{Getman_2018,Kounkel_2018} or an inferred age spread based on ages determined from H--R diagram locations]. The age spread means that stars in the sample cover different stages of PMS contraction, increasing the range of bolometric luminosities and the initial values of the fractional X-ray luminosities in the sample. The exact nature of the evolution of the scatter in $\log(L_\textrm{X}/L_*)$ is highly influenced by the mass distribution of the sample. The MAD of $\log(L_\textrm{X}/L_*)$ increases again after 10\,Myr if the sample is dominated by higher-mass, partially convective stars that are leaving the saturated regime. Meanwhile, a sample of stars that are dominantly low-mass, fully convective stars will have a lower MAD[$\log(L_\textrm{X}/L_*)$], which remains low.

While our model successfully produces the emergence of the unsaturated regime on the rotation-activity relation, the median fractional X-ray luminosity by 100\,Myr is above what we expect compared to the sample of main-sequence stars. Our work suggests that the observed $L_{\rm X}$ trends with age from \citetalias{Getman_2022}, which are based on stars of $M_* \ge 0.75$\,M${_{\sun}}$ and assume that $L_{\rm X}$ remain constant for the first few Myr of evolution, cannot be readily applied to lower-mass PMS stars. In addition, if $L_{\rm X}$ is assumed to continue to decay at the same rate as we have adopted for 25-100\,Myr stars, from the work of \citet{Gudel_2004}, then the median fractional X-ray luminosity in our model becomes too low by $\sim$400\,Myr. More X-ray studies of fully convective PMS stars are needed, especially between 5\,Myr and up to at least 100\,Myr where the lowest mass stars are still on the Hayashi tracks. 

To model the emergence and evolution of the X-ray rotation--activity relation, we did not use established relations between $L_{\rm{X}}$ and rotation rate/period for main sequence stars (e.g., \citealt{Pizzolato_2003}). However, these two parameters are intrinsically connected. The rotation rate influences the stellar dynamo behaviour, which generates magnetic fields that permeate into the coronae and dictate the nature of coronal X-ray emission. Future works could consider linking the X-ray luminosity to the magnetic field using trends of magnetic field strength with rotation periods / Rossby number \citep{Vidotto_2014}. By assigning a typical magnetic field geometry to a star at a given age, one could construct coronal models of PMS stars to calculate the coronal X-ray emission \citep[e.g.][]{Jardine_2006}. The decay of PMS star coronal X-ray emission could be driven by the increasing magnetic complexity as stars develop radiative cores \citep{Gregory_2016,Stuart_2023}. Such models would allow the implementation of mechanisms for supersaturation of X-ray emission, which may reduce the large values of fractional X-ray luminosities that we see in the model presented in this paper. Furthermore, by considering the evolution of the magnetic field, a better description of mass loss and accretion rates with age would allow a more detailed modelling of wind and disc torques \citep{Johnstone_2021,Gallet_2019}. Our understanding of the detailed long-term evolutionary behaviour of magnetic field geometries currently limit such an approach.


\section*{Acknowledgements}
KAS acknowledges support from STFC via a Doctoral Training Partnership grant (ST/W507404/1, project reference 2647716). The authors thank the reviewer for their comments which have improved our work. 

\section*{Data Availability}

The data underlying this article will be shared on reasonable request to the corresponding author.



\bibliographystyle{mnras}
\bibliography{References}


\appendix

\section{Calibration of the wind torque constant}
\label{App: Kw calibration}

For our rotational evolution model, the wind torque constant $K_{\rm{w}}$ in equation (\ref{Eq: wind torque tauw}) must be determined. To do so, we use established gyrochronology relations. By ages of at least $\sim$700\,Myr, the rotation rates of partially convective stars have converged and become dependent on mass for a given age \citep{Barnes_2003,Douglas_2019}. The form of the relationship between a star's $B-V$ colour, age $t$ in Myr, and rotation period $P_{\rm rot}$ in days was identified by \cite{Barnes_2007}, and updated by  \citet{Mamajek_2008} and \citet{Meibom_2009}, and remains consistent with later studies of period distributions for large samples of field stars \citep{McQuillan_2014}. The relation takes the form 
\begin{equation}
P_{\rm rot}(B-V,t) = a[(B-V)-c]^b t^n.
\label{Eq: Barnes Prot}
\end{equation}
We use the value of constants determined by \cite{Meibom_2009}: $a=0.77$, $b=0.553$, $c=0.472$, and $n=0.52$. For a given stellar mass, $K_{\rm{w}}$ is determined by enforcing that the period is equivalent to the period given by this relation at the age of 2\,Gyr. The $B-V$ colours for a given mass are retrieved using the 2\,Gyr YaPSI model isochrone.

The above method to fit $K_{\rm{w}}$ is only applicable for stars that remain fully or partially convective ($M_{*}$ $\leq$ $1.2\,\rm{M}_{\sun}$).
Thus, to model the rotational evolution of higher mass PMS stars before they become fully radiative, we assume that a fixed value of $K_{\rm{w}}$ matching that fitted for a $1.2\,\rm{M}_{\sun}$ star is appropriate for all higher masses. Precise refinement of the wind torque constant for these higher-mass PMS stars is not required, as we only track their rotational evolution while they have an outer convective envelope. In the YaPSI models, stars of mass $\ge1.25\,\rm{M}_{\sun}$ become fully radiative before they reach the ZAMS, and the torques from the expansion/contraction of the stellar radius or disc locking are dominant over the torque exerted by the stellar wind.


\bsp	
\label{lastpage}

\end{document}